\documentclass[a4paper,12pt]{amsart}
\usepackage{amsfonts,amsmath,amssymb}
\usepackage[english]{babel}
\usepackage{color}
\usepackage{cite}
\numberwithin{equation}{section}

\let\p\partial
\let\bb\mathbb

\def\phi{\varphi}
\let\bsy\boldsymbol
\let\ge\geqslant
\let\le\leqslant
\let\t\tilde
\newcommand{\ord}{\text{\rm ord}\,}

\let\ds\displaystyle
\newcommand{\eqdef}{\overset{\rm def}{=}}

\def\be{\begin{equation}}
\def\ee{\end{equation}}
\def\ba{\begin{aligned}} 
\def\ea{\end{aligned}}

\def\const{\mbox{const}}
\def\n{\nonumber}

\def\N{\bb N}

\newcounter{theo}
\newcommand{\theo}{\addtocounter{theo}{1}\textbf{Theorem \thetheo.} }
\newcounter{lem}
\newcommand{\lem}{\addtocounter{lem}{1}\textbf{Lemma \thelem.} }
\newcounter{prop}
\newcommand{\prop}{\addtocounter{prop}{1}\textbf{Proposition \theprop.} }
\newcounter{rem}
\newcommand{\rem}{\addtocounter{rem}{1}\textbf{Remark \therem.} }
\newcounter{defi}

\newcounter{examp}
\newcommand{\examp}{\addtocounter{examp}{1}\textbf{Example \theexamp.} }

\numberwithin{equation}{section}

\begin{document}

\allowdisplaybreaks

\thispagestyle{empty}

\title{\bf Integrable evolution equations with constant separant}

\author{\bf A.G. Meshkov, V.V. Sokolov}

\address{A.G. Meshkov. Higher Mathematics Department, State University -- UNPK, 29 Naugorskoe st.,  Orel, Russia.}
\address{V.V. Sokolov. Landau Institute for Theoretical Physics, Kosygina 2, 119334, Moscow, Russia.}
\address{E-mails: a\_meshkov@orel.ru, vsokolov@landau.ac.ru}

\maketitle
\begin{quote}
\noindent{\bf Abstract. }
The survey provides classification results for integrable one-field evolution equations of orders 2, 3 and 5 with the constant separant. The classification is based on necessary integrability conditions following from the existence of the formal recursion operator for integrable equations.
Recurrent formulas for the whole infinite sequence of necessary conditions are presented for the first time. The most of the classification statements can be found in papers by S.I. Svinilupov and V.V. Sokolov but the proofs have never been published before. The result concerning the fifth order equations is stronger than obtained before.

\medskip

\medskip
\end{quote}

\tableofcontents

\section*{Introduction}

This survey is devoted to the classification of integrable evolution equations
\begin{equation}
u_t=u_n+F(x,u, \, u_x, \, u_{xx}, \dots, u_{n-1}), \qquad u_i=\frac{\partial^i u}{\partial x^i}. \label{scalar}
\end{equation}
The equations with such dependence of the highest $x$-derivative are often referred to as equations with a constant separant.

Let us specify what we mean by the integrability in the present paper. Unfortunately, at present there exists no unified rigorous definition for the integrability of differential equations (for various approaches see, for instance, \cite{int1,int2,nu}). However, for some types of differential equations there are efficient criteria of the integrability, which can be not only checked for these equations, but also allow one to find  all the equations from this class satisfying this criterion.

For  evolution equations (\ref{scalar}) with one temporary variable and one spatial variable the most effective integrability criterion is the existence of generalized local symmetries. In the works \cite{ibshab,ss} a way of ``excluding a symmetry'' from this relation and obtaining necessary conditions for the existence of symmetries only in terms of the right hand side of the equation was suggested. These conditions which we call integrability conditions are written as so-called canonical conservation laws. Their main advantages are the independence of the conditions on the order of symmetry and their invariancy w.r.t. all point transformations not leading out of the class of equations (\ref{scalar}).

It was shown in the papers \cite{ibshab,soksvin1, soksvin2,ss} that necessary integrability conditions are implied by the existence of an infinite series of generalized symmetries or conservation laws for  equation (\ref{scalar}). In more details the technique of obtaining the conditions is discussed in the reviews \cite{umn,MSS}. Here we do not deal with it. We note that there exists an alternative way \cite{CLL, IP} for calculating canonical conservation laws by the logarithmic derivative of the formal eigenfunction of the linearization operator for equation (\ref{scalar}) (see Appendix 3). The equivalency of these two approaches for the scalar equations follows from Theorem~2.9 in   survey \cite{drsok}.

Let us describe the results of the work. In Chapter 1 by the simplest examples we show how canonical conservation laws look like and how one can classify integrable equations by employing them. In particular, in this chapter the problem of the classification for  equations (\ref{scalar}) with $n=2$ is solved. General second order integrable evolution equations were classified in \cite{sv4}. In \cite{C-C} the results of the latter work were generalized for the case of weakly nonlocal symmetries.

In Chapter 2 we provide the solution of the classification problem for the integrable equations of the form
\begin{equation}\label{kdv}
u_t=u_3+F(x, u,\, u_1, \, u_2).
\end{equation}
The famous Korteweg-de Vries equation
\begin{equation} \label{kdF}
u_t=u_3+u u_1
\end{equation}
belongs to this class. The case when the function $F$ is independent on $u_2$ and $x$ (see Section 1.2) was considered in \cite{ibshab, fokas}. The results of Chapter 2 were announced in \cite{soksvin1,soksvin2}, but the proof is published now for the first time. We also present for the first time a recurrent formula describing all infinite series of canonical densities. In the works \cite{soksvin1,soksvin2} only 4 first densities were written down explicitly which were indeed used then in the classification. Third order evolution integrable equations more general than (\ref{kdv}) were studied in
 \cite{MSS,hss,her}.

In Chapter 3 we consider a computationally complicated problem on the classification of integrable equations of the form
\begin{equation}\label{eq0}
u_t=u_{5}+F(u, u_1, u_2, u_3, u_4).
\end{equation}
In the note \cite{DSS} a solution to this problem was announced under an additional assumptions that even canonical densities are trivial (see Remark 2). However, not only the proof but also any complete list of the found equations is absent in \cite{DSS}.  For the first time the list of
equations (\ref{eq0}) possessing generalized conservation laws was published in \cite{MSS}. In the present work the condition of the triviality of even canonical densities is not employed and we solve thus a technically more complicated problem on the classification of   equations (\ref{eq0}) possessing generalized symmetries. The answer coincides in essence  with the list in \cite{MSS}. As in the case of the third order equations, a general formula for the whole infinite series of canonical densities is published for the first time in the present paper.

The results of the works \cite{soksvin1,soksvin2,DSS,MSS} were obtained by hard calculations made ``by hand''. This is why it was a non-zero probability of errors which could lead to losing integrable equations. Once computer systems like Maple, Mathematica, etc. appeared, an opportunity to automate partially the calculations rose. The results of the present paper were obtained by the program package Jet written by the first author. It was found no essential errors in the lists of the integrable equations but we found and corrected several misprints in \cite{MSS}.

At first glance, the problem of the classification of integrable equations (\ref{scalar}) with arbitrary $n$ seems to be far from the complete solution. This is not quite so. Each integrable equation together with all its symmetries form a so-called hierarchy of integrable equations. For the equations integrable by the inverse scattering problem method \cite{Z} all the equations of the hierarchy possess the same $L$-operator. This fact lies in the basis of the commutativity of the flows in hierarchies (each equation of the hierarchy is a symmetry for all others). A general statement on ``almost'' commutativity of the symmetries for   equation (\ref{scalar}) is contained in \cite{sok1}.

Assuming that the right hand side of   equation (\ref{scalar}) is polynomial and homogenous, it was proven in the works
\cite{sw,ow} that the hierarchy of any such  equation contains an equation of second, third, or fifth order. This statement looks very credible also without any additional restrictions for the right hand side of the equation. The proof in the general case is absent and this statement has a status of the conjecture well-known for experts. No counterexamples to this conjecture are known.

Up to this conjecture, in the survey we describe all the hierarchies of the integrable equations of the form
(\ref{scalar}). In other words, any integrable equation of order $4$ or $>5$ is equivalent to a generalized symmetry of one of the equations given in this survey. We note that the calculation of symmetries for given equation is a linear problem, and there are several effective computer programs for solving it. Moreover, the generalized symmetries can be found by the use of quasilocal recursion operators (see \cite{tur} and the references therein).

Various results on the classification of integrable systems of evolution equations can be found in
\cite{umn,sow,Mmikh,svin1,svin2,svin3,svinsok1,soksv11,habsokyam,os1,sokwolf,meshbal1,meshbal2,meshbal3,bal,M2,meshsok,meshsok1}. Further references are contained, for instance, in the survey \cite{miksok}.

A separate difficult problem is the classification of integrable hyperbolic equations and systems \cite{zibshab1,M5,M6,zibshab2, zib, zibsok, M3,M4,meshsok2}.

\vskip.3cm
\noindent
{\bf Acknowledgments.}
The authors are grateful to D.I. Borisov, A.V. Mikhailov, S.I. Svinolupov, and A.B. Shabat for numerous useful discussions. V.S. is grateful to Max Planck Institute (Bonn) for the hospitality. The research is partially supported by RFBR grant
 11-01-00341-a, the grant for supporting leading scientific school 6501.2010.2, and the grant of Ministry of Education and Science of Russia (project 1.2.11).

\section{Simplest classification problems}

All necessary integrability conditions we shall use below are given in the form of local conservation laws.
We remind \cite{olver} that  a pair of functions $\rho$ and $\theta$ depending on a finite number of the variables
$x,u,u_1,\dots$ such that
\begin{equation} \label{conslaw} \frac{d}{dt}(\rho)=\frac{d}{dx}(\theta) \end{equation}
is called a local conservation law for equation (\ref{scalar}).
Here
\begin{equation}\label{D}
\frac{d}{dx}=\frac{\partial}{\partial x}+ u_1\frac{\partial}{\partial u_{0}}+
u_2\frac{\partial} {\partial u_{1}}+ u_3\frac{\partial}{\partial
u_{2}}+\cdots\, , \qquad u_0=u,
\end{equation}
\[
\frac{d}{dt}=\frac{\partial}{\partial
t}+K_0\frac{\partial}{\partial u_{0}}+ K_1\frac{\partial}
{\partial u_{1}}+K_2\frac{\partial}{\partial u_{2}}+ \cdots\, ,
\]
 where
\[ K_i= \frac{d^i}{dx^i}\Big(u_n+F(x,u_0, \, u_1, \, u_{2}, \dots, u_{n-1})\Big).\]
the operators $\ds\frac{d}{d x}$ and $\ds\frac{d}{d t}$ are often referred to as the total derivative w.r.t. $x$ and the total derivative w.r.t. $t$ in virtue of  equation (\ref{scalar}). The function $\rho$ is called a density, and $\theta$ a flux of the conservation law.

Relation (\ref{conslaw}) is called a conservation law due to the following reason. Consider, for instance, the Korteweg-de Vries equation $u_t=u_3+u u_1$. It is known that it possesses an infinite number of conservation laws. In particular, since the equation can be rewritten as
$$
u_t=(u_2+\frac{1}{2} u^2)_x,
$$
the function $u$ is the density of the conservation law. Suppose the solution $u(x,t)$ decays as $\vert x \vert\to \infty.$ Then we have
$$
\frac{d}{dt} \int_{-\infty}^{+\infty} u\, dx=0,
$$
i.e., the area under the graph of the solution is independent of $t$. In the same way, the integrals of others densities of the conservation laws are conserved.

It is clear that if $\rho$ is a density of a conservation law, then ${\rho_{1}=\ds\rho+\frac{d}{dx}(h)}$ is also a density for any function $h$. We call two such densities equivalent and write $\rho \sim \rho_{1}$. A conservation law is called trivial if $\rho \sim 0$.

The order of the higher derivative, on which the function $f(x,u,u_1,\dots,u_k)$ depends, is called {\bf differential} order of this function. The differential order is usually indicated as $\ord f=k$. The minimum of the {\bf differential} orders of equivalent densities is called the order of the conservation law.

The deduction of necessary integrability conditions as an infinite series of so-called canonical conservation laws was discussed in details in \cite{umn,MSS,IP,CLL}; for an alternative version see Appendix 3. In this paper we often give appropriate formulas without proofs. But on the other hand we dwell on how to retrieve the complete list of integrable equations of the form (\ref{kdv}) using these necessary conditions, and we describe point transformations necessary to reduce an arbitrary integrable equation to one of the canonical forms.

\subsection{Integrable Burgers type equations}

Consider second order evolution equations
\begin{equation}
u_t=u_{2}+f(x, \,u, \, u_1). \label{scal2}
\end{equation}
The canonical densities for this equation are defined by the  recurrent formula
\begin{align}\label{cd1}
2\rho_{n+1}=\theta_n+\sum_{i=0}^{n} \rho_{n-i}\rho_i-\frac{\p f}{\p u_1}\rho_n+\frac{\p f}{\p u_1}\delta_{n,-1}+\frac{\p f}{\p u}\delta_{n0}-\frac{d}{dx}\rho_n,\ \ \ n\ge-1.
\end{align}
Here $\rho_{-1}=0$, $\delta_{ij}$ is the Kronecker delta. One of the ways of obtaining similar formulas is described in Appendix 3. The fluxes associated with these densities are calculated consequently in the process of classification. At that, the obstacles to their existence pose the restrictions for the right hand side of equation (\ref{kdv}) that finally allow us to find all integrable equations (\ref{scal2}).

Letting $n=-1,0$ in (\ref{cd1}), we find two first canonical conservation laws,
\begin{align}
&\frac{d}{dt}\frac{\partial f}{\partial u_{1}}=\frac{d}{dx}\sigma_1, \label{rho1}\\
 &\frac{d}{dt}\left(\sigma_1+2\frac{\p f}{\p u}-\frac{1}{2}\left(\frac{\p f}{\p u_1}\right)^2\right)=\frac{d}{dx}\sigma_2 , \label{rho2}
\end{align}
where $\sigma_1=2\,\theta _0$ and $\ds \sigma_2=4\,\theta_1+\frac{d}{dx}\sigma_1.$

The former of these formulas means that for each integrable equation (\ref{scal2}) the partial derivative w.r.t. $u_1$ of its right hand side is a density of a conservation law. For instance, for Burgers equation $u_t=u_2+u u_1$ this formula yields a density $\rho=u$. In this case the function $\sigma_1$ is calculated easily,
$$
\sigma_1=u_2+\frac{1}{2} u^2.
$$
A general algorithm of calculating the flux for a given density is given below (see Remark 4).

Let us demonstrate the main modes for working with the conditions like (\ref{rho1}), (\ref{rho2}). In order to determine the character of the dependence of the right hand side on
$u_1$, the most simplest way is to exclude the unknown function $\sigma_1$ in (\ref{rho1}). For this we apply the Euler operator $$
\frac{\delta}{\delta u}=\frac{\partial}{\partial u}-\frac{d}{dx}\circ \frac{\partial}{\partial u_1}+
\frac{d^2}{dx^2}\circ \frac{\partial}{\partial u_2}- \cdots
$$
to both sides of (\ref{rho1}). It is well known
 \cite{olver} that
$$
\frac{\delta}{\delta u}\circ \frac{d}{dx}=0,
$$
and therefore,
\begin{equation} \label{condn1}
0=\frac{\delta}{\delta u}\frac{d}{dt}\left(\frac{\partial f}{\partial u_{1}} \right) =-2u_4\frac{\p^3 f}{\p u_1^3}-4u_3\frac{d}{dx}\frac{\p^3 f}{\p u_1^3}+O(2),
\end{equation}
where the symbol $O(2)$ indicates terms whose order w.r.t. the derivatives are at most two. The latter identity must hold true for each solution (\ref{scal2}). Since there exists no ordinary differential equation in $x$ satisfied by all the solutions to equation (\ref{scal2}),   relation (\ref{condn1}) must hold identically w.r.t. the variables $u,u_1,\dots,u_4$. Equating the coefficient at $u_4$ to zero, we find that the equation reads as
 \begin{equation}\label{eqv}
u_t=u_2+ A(x,u) u_1^2+B(x,u) u_1+C(x,u).
\end{equation}
Thus, each integrable equation (\ref{scal2}) is quadratic in $u_1$. It can be checked that for   equation (\ref{eqv})  condition (\ref{condn1}) is equivalent to  two equations
$$
(C\phi )_u=(B\phi -\phi _x)_x,\qquad \phi _u=A\phi ,
$$
where $\phi =B_u-2A_x$.

Taking into consideration that the integrability of any differential equation is preserved under point transformations, we simplify  equation (\ref{eqv}) by a point transformation $u=\psi(x,v)$ before we proceed to study the integrability conditions. Simple calculations lead us to the evolution equation for $v$
$$
v_t=v_2+v_1^2\left(\frac{\p\psi }{\p v}\right)^{-1}\left(\frac{\p^2\psi }{\p v^2}+A(x,\psi )\left(\frac{\p \psi }{\p v}\right)^2\right)+\bar B(x,v) v_1+\bar C(x,v).
$$
It is obvious that the equation
$$
\frac{\p^2\psi }{\p v^2}+A(x,\psi )\left(\frac{\p \psi }{\p v}\right)^2=0
$$
has a solution depending on $v$ forany function $A$. This is why by a point transformation one can vanish the function $A$ in  equation (\ref{eqv}). This transformation is the first step in
reducing any integrable equation to one of the canonical forms.

Condition (\ref{rho1}) for the equation
\begin{equation}\label{eqv1}
u_t=u_2+ B(x,u) u_1+C(x,u)
\end{equation}
becomes
\begin{equation} \label{condn1red}
 B_u\big(u_2+ B(x,u) u_1+C(x,u)\big)=\frac{d}{dx}\sigma_1.
\end{equation}

Since to use  condition (\ref{rho2}) we need to know completely
or partially the function $\sigma_1$, instead of applying the variational derivative to both sides of (\ref{condn1red}) we employ an alternative approach which is a separation of a total $x$-derivative in the left hand side of (\ref{condn1red}).
This approach is completely algortihmical and can be programmed in any language of symbolic computations (see Remark on page \pageref{D-1}).

We have
$$
\begin{aligned}
&B_uu_2+ B_uB u_1+B_uC=\frac{d}{dx}\left( B_uu_1+ \frac{1}{2}B^2\right)-u_1(B_{uu}u_1+B_{ux})-BB_x+B_uC=\\
&\qquad=\frac{d}{dx}\left( B_uu_1+ \frac{1}{2}B^2-B_x\right)-B_{uu}u_1^2+B_{xx}-BB_x+B_uC.
\end{aligned}
$$
Substituting the last expression in (\ref{condn1red}), we obtain
$$
-B_{uu}u_1^2+B_{xx}-BB_x+B_uC=\frac{d}{dx}\left(\sigma_1- B_uu_1- \frac{1}{2}B^2+B_x\right)\equiv \frac{d\psi }{dx}.
$$
Since the left hand side depends only on $x$, $u$, $u_1$, then the function $\psi$ may depend only on $x$ and $u$. Substituting $\ds\frac{d\psi }{dx}
=\psi _x+\psi _uu_1$ and equating the coefficients at $u_1^2$ and $u_1$, we get
$$B_{uu}=0,\qquad \psi _u=0, \qquad B_{xx}-BB_x+B_uC=\psi_x.$$
Letting
$B=\alpha (x)u+\beta (x)$, we find that each integrable equation (\ref{eqv1}) reads as
\begin{equation}
u_t=u_2+ (\alpha (x)u+\beta (x)\big)u_1+C(x,u), \label{eqq1}
\end{equation}
where
\begin{equation}
\alpha C(x,u)-\alpha \alpha'u^2+(\alpha''-\alpha \beta'-\alpha' \beta )u=\psi'+\beta \beta'-\beta''.
\label{eqq2}
\end{equation}
At that,
\begin{equation}
\sigma _1=\psi +\alpha u_1+\frac{1}{2}(\alpha u+\beta )^2-\alpha'u-\beta'. \label{eqq3}
\end{equation}

If $\alpha\ne 0$, then by (\ref{eqq2}) one can determine the function $C$. In this case  equation (\ref{eqq1}) can be simplified by a point transformation $u\to u\,f_1(x)+f_2(x)$. By taking $f_1=1/\alpha $ ${ f_2=2\alpha'/\alpha ^2-\beta /\alpha }$ we get $\alpha =1,\,\beta =0$. At that, equation (\ref{eqq1}) casts into the form
\begin{equation}\label{eqq4}
u_t=u_{xx}+uu_x+\psi'(x).
\end{equation}
For this equation conditions (\ref{rho1}), (\ref{rho2}), as well as all other necessary integrability conditions, hold true. The Burgers equation (\ref{eqq4}) is reduced to the linear equation
$$
v_t=v_{xx}+\phi (x)v_x,
$$
by the Cole-Hopf substitution $u=2v_x/v+\phi (x)$, where $\phi$ and $\psi$ are related by the identity ${\phi''+\phi\phi'=-\psi'}$.

In the case $\alpha=0$ the left hand side of   equation (\ref{condn1red}) vanishes and this is why $\sigma_1$ is constant. Then   condition (\ref{rho2}) implies
$$
\frac{\delta }{\delta u} \left(\sigma_1+2\frac{\p f}{\p u}-\frac{1}{2}\left(\frac{\p f}{\p u_1}\right)^2\right)_t=0,
$$
that is equivalent to the system of equations
$$
C_{uuu}=0,\ \ \ CC_{uu}+C_{xuu} -(\beta C_u)_x+\psi'(x)=0.
$$
It yields $C=p(x)u+q(x)$, and we arrive at the linear equation
\begin{equation}\label{eqqq4}
u_t=u_{xx}+\beta (x)u_x+p(x)u+q(x).
\end{equation}
For this equation all necessary integrability conditions hold true.

\rem Among obtained second order integrable equations (\ref{eqq4}) and (\ref{eqqq4}) there is no the potential Burgers equation $u_t=u_{xx}+u_x^2$. The reason is that this equation is linearized by the point transformataion $u=\ln v$. This transformation is a particular case of the point transformation that has to be applied to  equation (\ref{eqv}) for eliminating the function $A$.

\subsection{Integrable KdV type equations}

The list of integrable equations obtained in the previous section is quite poor. Let us consider a more substantial classification problem. Let us find all integrable evolution equations of the form
\begin{equation} \label{kdvt}
u_{t}=u_{3}+f(u_1, u).\ \ \ \
\end{equation}
It turns out (see Section 2.1) that for each such integrable equation
\begin{equation} \label{con1}
\frac{d}{dt}\left(\frac{\partial f}{\partial u_{1}}\right)=\frac{d}{dx}(\sigma_1),
\end{equation}
where $\sigma_1$ is a function depending on $u$, $u_x$, \ldots, $u_3$.

\examp For the mKdV equation $u_t=u_3+u^2 u_1$   conservation law (\ref{con1}) reads as
$$
(u^2)_t=(2 u u_2-u_1^2+\frac{1}{2} u^4)_x. \qquad \square
$$

Applying the Euler operator to both sides of (\ref{con1}), we obtain
\begin{equation} \label{conn1}
0=\frac{\delta}{\delta u}\left(\frac{\partial f}{\partial u_{1}} \right)_t =
3 u_4 \left(u_2 \, \frac{\partial^4 f}{\partial u_{1}^4}+u_1 \, \frac{\partial^4 f}{\partial u_{1}^3 \partial u}\right)+O(3).
\end{equation}
The last identity must hold true for each solution of (\ref{kdvt}) and this is why it must be identity in the variables $u,u_1,\dots,u_4$. Equating the coefficient at $u_4$ to zero and employing that $f$ is independent of $u_2$, we get
$$
f(u_1,u)=\mu u_1^3+A(u) u_1^2+B(u) u_1+C(u)
$$
with some constant $\mu$. It is easy to check that for each such function $f$  condition (\ref{conn1}) is equivalent to the  system of ODEs
$$
\mu A'=0, \quad \qquad B'''+8 \mu B'=0, \quad \qquad (B'C)'=0, \qquad \quad A B'+6 \mu C'=0.
$$

The next necessary integrability condition reads as
$$
\frac{d}{dt}\left(\frac{\partial f}{\partial u}\right)=\frac{d}{dx}(\sigma_2)
$$
that yields
\begin{equation} \label{conn2}
\frac{\delta}{\delta u}\frac{d}{dt}\left(\frac{\partial f}{\partial u} \right)=0.
\end{equation}
The latter condition leads to additional equations
$$
A'=0, \qquad \quad AC''=0, \qquad \quad (C'''+2\mu C')'=0, \qquad \quad (CC'')'=0.
$$

In the case $\mu\ne 0$ the obtained equations are sufficient to determine completely the functions $A$, $B$, and $C$. As a result, up to a scaling $u\to const \,u,$ we arrive at the equations
\begin{equation}
u_t=u_{xxx}-\frac{1}{2} u_{x}^3+(c_1 e^{2u}+c_2 e^{-2u}+c_3)\,u_{x}\, \label{e1}
\end{equation}
and
\begin{equation}
u_t=u_{xxx}+c_1 u_x^3+c_2 u_x^2+c_3 u_x+c_4, \label{e2}
\end{equation}
where $c_i$ are arbitrary constants.

If $\mu=0$, then solving the above system of ODEs for the functions $A,B,C$, we obtain that the equation reads as
$$
u_t=u_{xxx}+c_0 u_x^2 +(c_1 u^2+c_2 u +c_3) u_x+c_4u+c_5,
$$
where
$$
c_0c_1=0,\qquad c_0c_2=0,\qquad c_4c_1=0,\qquad c_4c_2=0,\qquad c_1c_5=0.
$$
By the third integrability condition (see Section 2) we find additional relations,
$$
c_0c_4=0,\qquad c_2c_5=0.
$$

In the case $c_0\ne0$ we arrive at a particular case of  equation (\ref{e2}). If $c_0=0$, two cases are possible; a) $c_1\ne0$ or $c_2\ne0$,
$c_4=c_5=0$ and b) $c_1=c_2=0.$ They lead us to two equations
\begin{align}
u_t&=u_{xxx}+(c_1 u^2+c_2 u +c_3) u_x, \label{e3}\\
u_t&=u_{xxx}+c_3u_x+c_4u+c_5. \label{e4}
\end{align}

The experts in nonlinear equation regard each linear equation as exactly integrable. Equations (\ref{e1}), (\ref{e2}), and (\ref{e3}) have been found by necessary integrability conditions. This is why it should be discussed independently in which exactly sense they are integrable. It is well-known that to all of these equations the method of the inverse scattering problem is applicable. Moreover, all of them are related with the Korteweg-de Vries equation $u_t=u_3+u u_1$ by Miura type differential substitutions \cite{cvsokyam}.

\rem Conditions (\ref{conn1}), (\ref{conn2}) hold true for the equations (\ref{kdvt}) possessing generalized symmetries. If the equation possesses generalized conservation laws (at that, the existence of the symmetries is not assumed),   condition (\ref{conn1}) still holds, and   condition (\ref{conn2}) can be strengthen,
$$
\frac{\delta}{\delta u}\left(\frac{\partial f}{\partial u} \right)=0.
$$
It is implied by the general statement \cite{soksvin1} in accordance to which for the equations with generalized conservation laws the canonical densities with even indices are trivial.

\subsection{On admissible point transformations}

In the process of classification of integrable equations, as a rule, we use point transformations reducing integrable equation to one or another canonical form. For instance, in Section 1.1 we employed point transformations while reducing   equation (\ref{eqv}) to the form (\ref{eqv1}), and also for normalizing the functions $\alpha (x)$ and $\beta(x)$ in   equation (\ref{eqq1}).

Let us describe point transformations we use in the classification of   equations (\ref{scalar}).

Each equation of the form
 (\ref{scalar}) admits the transformations
\begin{equation}
 \tilde u = \phi(u,x). \label{t1}
\end{equation}
Hereinafter, once  transformation rules for some of the variables $t,$ $x$, or $u$ are not indicated in the formulas, this means that the corresponding variables are not changed. The scalings
\begin{equation}
\tilde x = ax,\qquad \ \tilde t = a^nt \label{t2}
\end{equation}
are also admitted. Under such transformations
$$
\qquad F(x,u, u_1, u_{2},\dots)\to a^{-n}F(a^{-1}x,u, au_1, a^2u_{2},\dots).
$$

For some subclasses of  equations (\ref{scalar}) additional transformations depending on $t$ are admitted. In particular,
if $F(x,\lambda u, \lambda u_1, \dots,\lambda u_{n-1}) =\lambda F(x,u, u_1, \dots,u_{n-1})$, then for arbitrary constants $a$ and $b$ the transformation
\begin{equation}
\tilde u = u \exp(at + bx) \label{t6}
\end{equation}
is applicable. Under this transformation $ u_n \to (\p_x - b)^nu, \quad F \to F + au.$

If, as in Section 1.2, it is assumed that the right hand side $F$ of equation (\ref{scalar}) is independent on the variable $x$, then the class of admissible transformations changes. Among (\ref{t1}), only the transformations
\begin{equation}\label{t4}
 \tilde u = \phi(u)
\end{equation}
are admitted. At the same time additional point transformations appear. In particular, the Galilean transformation
\begin{equation}
 \tilde x = x + ct \label{t3}
\end{equation}
is always  admissible; under this transformation $F\to F - cu_1.$
If the function $F$ is independent of $u$ and $x$, then the transformation
\begin{align}
\tilde u = u + c_1x + c_2t\label{t5}
\end{align}
is admissible. Under such transformation
$$
F(u_1, u_2,u_3,\dots)\to F(u_1- c_1, u_2,u_3,\dots) + c_2.
$$

The equations related by aforementioned transformations are called
{\it equivalent}. It is important to note that our classification is pure algebraic. We are not interesting in such properties of the solutions to the studied equations as being real. This is why the functions and constants being involved in formulas (\ref{t1})--(\ref{t6}) can be both real and complex. For instance, the equations $u_t=u_3-u_1^3$ and
$u_t=u_3+u_1^3$ are regarded as equivalent.

Integrable equations can involve arbitrary constants which can be eliminated by one or another transformation. Consider as an example  equation (\ref{e2}), where $c_1\ne 0$. By (possible complex) scaling $u\to \lambda u$ we fix a normalization $c_1=1$. Then the transformation
$u\to u + \alpha x + \beta t$ leads us to the equation
$$
u_t+\beta =u_{xxx}+ (u_x+\alpha )^3+c_2 (u_x+\alpha )^2+c_3 (u_x+\alpha) +c_4.
$$
It is easy to see that taking $\alpha=-c_2/3$ and $\beta=c_4+\alpha ^3+c_2\alpha^2+c_3\alpha$, we obtain $c_2=0,$ $c_4=0$. The constant $c_3$ is eliminated by the Galilean transformation, and we obtain the potential modified Korteweg-de Vries equation,
$$
u_t =u_{xxx}+u_x^3.
$$
Similarly, the parameters in   equation (\ref{e3}) are inessential.

\section{Third order equations with constant separant}

\subsection{Integrability conditions}

For the equations of the form (\ref{kdv}) an infinite chain of the canonical conservation laws
\begin{equation}
\label{law1}
\frac{d}{dt}(\rho_{n})=\frac{d}{dx}(\theta_{n}), \quad n=0,1,\dots ,
\end{equation}
can be defined by the formulas (for the deduction see Appendix 3),
\begin{align}
\rho_{n+2}&=\frac{1}{3}\bigg[\theta_n-\delta_{n,0}F_{u} -F_{u_1}\rho_{n}-
F_{u_2}\Big(\frac{d}{dx}\rho_{n} + 2\rho_{n+1}+\sum_{s=0}^{n} \rho_{s}\,\rho_{n-s}\Big)\bigg]
-\sum_{s=0}^{n+1} \rho_{s}\,\rho_{n+1-s}
 \nonumber\\[2mm]
&-\frac{1}{3}\sum_{0\le s+k\le n}\rho_{s}\,\rho_{k}\,\rho_{n-s-k}
-\frac{d}{dx}\biggl[\rho_{n+1}+\frac{1}{2}\sum_{s=0}^{n}\rho_{s}\,\rho_{n-s}+
\frac{1}{3}\frac{d}{dx}\rho_{n} \biggr], \ \ n\ge0, \label{rekkur_sc}
\end{align}
where the first two elements of the sequence $\rho_i$ read as
$$
\rho_0=-\frac{1}{3}F_{u_2},\qquad \rho_1=\frac{1}{9}F_{u_2}^2-\frac{1}{3}F_{u_1}+\frac{1}{3}\frac{d}{dx} F_{u_2}.
$$
Here $\delta_{i,j}$ is the Kronecker delta, $F_{u_i}=\p F/\p u_i$, where $i=0,1,2.$ The fluxes $\theta_n$ are calculated consequently in the process of classification. At that, the obstacles for its existence lead to differential equations, which must be satisfied by the right hand of  integrable equation (\ref{kdv}).

It is easy to check that first four conditions in this series are equivalent to the conditions
\begin{align}
& \frac{d}{dt} \frac{\partial F}{\partial u_{2}}=\frac{d}{dx}\sigma_{0}, \label{cond1} \\[2mm]
& \frac{d}{dt} \left(3 \frac{\p F}{\p u_{1}}- \left(\frac{\p F}{\p u_{2}}\right)^{2}\right)=\frac{d}{dx}\sigma_{1}, \label{cond2} \\[2mm]
&\frac{d}{dt} \left(9 \sigma_{0}+2 \left(\frac{\p F}{\p u_{2}}\right)^{3} -9\left(\frac{\p F}{\p u_{2}}\right)\, \left(\frac{\p F}{\p
u_{1}}\right) +27 \frac{\p F}{\p u}\right)=\frac{d}{dx}\sigma_{2}, \label{cond3} \\[2mm]
& \frac{d}{dt} \sigma_{1}=\frac{d}{dx}\sigma_{3} \label{cond4}
\end{align}
given in \cite{soksvin2}. At that, $\ds\sigma_0=-3\theta_0,\ \sigma_1=3\frac{d}{dx}\sigma_0 -9\theta_1,\dots$
As it will be shown below, these four conditions are ``almost'' sufficient to obtain the complete list of  integrable equations (\ref{kdv}).

In order to employ efficiently the canonical series for the classification, it is useful to study first a possible structure of the densities of local conservation laws of small orders for the considered class of equations.

\lem {\it If a density $\rho$ of a conservation law for  equation {(\ref{kdv})} has the differential order $\ord\rho =2$, then
\begin{equation}\label{ro_0}
\rho =f_1u_2^2+f_2u_2+f_3,
\end{equation}
 where $f_i$ are some functions in $x,u$, $u_1$, and}
\begin{equation}\label{ro_01}
\frac{d}{dx} f_1=\frac{2}{3}f_1\frac{\p F}{\p u_2}.
\end{equation}

{\bf Proof}. Eliminating the terms $u_5$ and $u_4$ by the subtraction of total $x$-derivatives, we find that
\begin{align}\label{rho0}
\frac{d}{dt}\rho&=\frac{\partial \rho}{\partial u} (u_3+F)+
\frac{\partial \rho}{\partial u_{1}}\left(u_4+\frac{d}{dx}F\right)+\frac{\partial \rho}{\partial u_{2}}\left(u_5+\frac{d^2}{dx^2} F\right)\sim \nonumber\\
&\sim \frac{u_{3}^{3}}{2} \frac{\partial^{3} \rho}{\partial u_{2}^{3}}+\frac{3}{2} u_{3}^{2}\left(\frac{\p^3\rho }
{\p u_2^2\p u_1}u_2+\frac{\p^3\rho }{\p u_2^2\p u}u_1+\frac{\p^3\rho }{\p u_2^2\p x}- \frac23\frac{\p F}{\p u_2}\frac{\p^2\rho }{\p u_2^2}\right) +\cdots,
\end{align}
where the dots indicate a linear in $u_3$ expression. By the definition of the conservation law, the last expression should read
$\ds\frac{d}{dx}\sigma$. It is clear that the function $\sigma$ can not depend on the derivatives higher than $u_{2},$ and the degree in $u_3$ of the function $\ds\frac{d}{dx}\sigma$ is at most one. Hence, equating the coefficients at $u_3^3$ and $u_3^2$ to zero, we obtain (\ref{ro_0}) and (\ref{ro_01}).
$\square$

\rem We observe that in (\ref{ro_0}) the identity $f_1=0$ is possible. It concerns also other similar lemmas.

\rem In the proof of Lemma 1 we used the following algorithm of checking whether \label{D-1} a given function $S(x,u, u_{1}, \dots , u_{n})$ is a complete derivative
w.r.t. $x$ (i.e., whether it belongs to ${\rm Im}\, \ds\frac{d}{dx}$). At first, $S$ must be linear in the highest derivative $u_{n}$. If it holds true, then as one can see easily, we can subtract a total derivative from $S$ so that the difference has the order less than $n$. Repeating this order lowering procedure, we either arrive at the situation when the function is nonlinear in its highest derivative, or we get zero.

Let us show how one can employ formulas (\ref{ro_0}) and (\ref{ro_01}) in the classification of  equations (\ref{kdv}).

\lem {\it Let for  equation} (\ref{kdv}) {\it the first integrability condition } (\ref{cond1})
{\it holds. Then $F$ is a polynomial in $u_2$ of at most second degree.}

{\bf Proof}. According to condition (\ref{cond1}), the function $\ds\frac{\partial F}{\partial u_{2}}$ should be a density of a conservation law. Applying Lemma~1 to it, we write  equations (\ref{ro_0}) and (\ref{ro_01}),
\[
 \frac{\partial F}{\partial u_{2}}=f_1 u_2^2+f_2 u_2+f_3,
\]
$$
\frac{\p f_1}{\p x}+\frac{\p f_1}{\p u_0}u_1+\frac{\p f_1}{\p u_1}u_2=\frac{2}{3}f_1(f_1 u_2^2+f_2u_2+f_3).
$$
Since $f_i$ are independent of $u_2$, then equating the coefficients at $u_{2}^{2}$, we obtain $f_1=0$. Integrating the equation $\ds\frac{\p F}{\p u_2}=f_2u_2+f_3$ w.r.t. $u_2$, we arrive at the desired result. $\square$

\subsection{List of integrable equations}

Our main aim is to prove the following statement \cite{soksvin2}.

\theo {\it Up to transformations of the form} (\ref{t1})--(\ref{t5}) {\it each equation} (\ref{kdv}) {\it satisfying   integrability conditions } (\ref{law1}), (\ref{rekkur_sc}) {\it for} $n=0,1,...,5,$
{\it belongs to the list
\begin{align}
&u_t=u_{xxx}+uu_x,\ \label{list1}\\
&u_t=u_{xxx}+u^2u_x, \label{list2}\\
&u_t=u_{xxx}+u_x^2, \label{list3}\\
&u_t=u_{xxx}-\frac{1}{2}u_x^3+(c_1e^{2u}+c_2e^{-2u})u_x, \label{list4}\\
&u_t=u_{xxx}-\frac{3u_xu_{xx}^2}{2(u_x^2+1)}+a_1(u_x^2+1)^{3/2}+a_2u_x^3, \label{list5}\\
&u_t=u_{xxx}-\frac{3u_{xx}^2}{2u_x}+\frac{1}{ u_x}-\frac{3}{2} \wp(u)u_x^3, \label{list7}\\
&u_t=u_{xxx}-\frac{3u_xu_{xx}^2}{2(u_x^2+1)}-\frac{3}{2} \wp(u)u_x(u_x^2+1), \label{list6}\\
&u_t=u_{xxx}-\frac{3u_{xx}^2}{2u_x}, \label{list8}\\
&u_t=u_{xxx}-\frac{3u_{xx}^2}{4u_x}+c_1u_x^{3/2}+c_2u_x^2,\ \ c_1\ne0\ \mbox{\rm or } c_2\ne0, \label{list9}\\
&u_{t}=u_{xxx}-\frac{3\,u_{xx}^{2}}{4\,u_x} +\alpha(x)u_x, \label{list10}\\
&\begin{aligned} \label{list11a}
u_{t}&=u_{xxx}-\frac{3\,u_{xx}^{2}}{4\,u_x} +\frac{3}{\xi} u_{xx}(\sqrt{\alpha'\,u_x}+u_x)+\frac{3\,u_x^3}{\xi^2}+\frac{6}{\xi^2}\,u_x^{5/2}\sqrt{\alpha'}\\
&\quad+\frac{3\,u_x^{3/2}}{\xi^2\sqrt{\alpha'}}\,(\xi\alpha''-2\alpha'^2)+f\,u_x+c_0+c_1\,u+c_2\,u^2,
\end{aligned}
\end{align}
\begin{align}
&\mbox{\rm where}\
\xi=\alpha(x)-u,\ \ \ f=-\frac{\alpha'''}{\alpha'}+\frac{3\alpha''^2}{4\alpha'^2}+3\frac{\alpha''}{\xi}-3\frac{\alpha'^2}{\xi^2}-\frac{c_0+c_1\,\alpha+c_2\,\alpha^2}{\alpha'},\n \\
&u_t=u_{xxx}+3\,u^2u_{xx}+9\,uu_x^2+3\,u^4u_x+u_x\alpha (x)+\frac{1}{2}\,u\alpha'(x), \label{list13}\\
&u_t=u_{xxx}+3\,uu_{xx}+3\,u_x^2+3\,u^2u_x+(u\gamma (x))_x+\beta(x), \label{list14}\\
&u_t=u_{xxx}+\alpha(x)u_x+ \beta(x)u. \label{list15}
\end{align}
Here $ ( \wp')^2=4 \wp^3-g_2 \wp-g_3$, $a_1,a_2,c_0,c_1,c_2, g_2, g_3$ are arbitrary constants, $\alpha$, $\beta$, and $\gamma $ are arbitrary functions.}

\rem Quite often instead of  equations (\ref{list7}) and (\ref{list6}) one considers point equivalent to them equations
\begin{equation}\label{kdv4}
u_t=u_{xxx}-\frac{3}{2}\,\frac{u^{2}_{xx}}{u_{x}}+\frac{Q}{u_{x}},
\end{equation}
and
\begin{equation}\label{kdv5}
u_{t}= u_{xxx}- \frac{3}{8}\frac{\big((Q+u_x^2)_{x}\big)^2}{u_x\,(Q+u_x^2)}+\frac{1}{2} Q''\,u_x .
\end{equation}
In both cases, $Q=c_{0}+c_{1} u+c_{2} u^{2}+c_{3} u^{3}+c_{4} u^{4}$ is an arbitrary polynomial.
If $Q'\ne0$, then one can make the substitution $u=f(v)$ in  equations (\ref{kdv4}) and (\ref{kdv5}), where $(f')^2= Q(f)$. Then for $v$ we get   equations (\ref{list7}) and (\ref{list6}), respectively. At that,
$$
g_2=\frac{4}{3} c_2^2-4 c_1 c_3+16 c_0 c_4, \qquad g_3=\frac{8}{27}c_2^3-\frac{4}{3}c_1c_2c_3-\frac{32}{3}c_0c_2c_4+4c_0c_3^2+4c_1^2c_4.
$$
We observe that under the linear fractional transformations
\begin{equation}\label{meb}
u=\frac{z_1 \t u+z_2}{z_3 \t u+z_4}
\end{equation}
the polynomial $Q$ changes by the law
$$
\t Q(\t u)=Q\left(\frac{z_1 \t u+z_2}{z_3 \t u+z_4}\right) (z_3 \t u+z_4)^4 (z_1 z_4-z_2 z_3)^{-2}.
$$
The expression $g_2, g_3$ are invariants of  transformations group (\ref{meb}). Subject to the structure of the multiple roots, by a transformation (\ref{meb}) and scalings of $x$ and $t$ the polynomial $Q$ can be reduced to one of following canonical forms, $Q(x)=x(x-1)(x-k)$, $Q(x)=x(x-1)$, $Q(x)=x^2$, $Q(x)=x$, $Q(x)=1$ and $Q(x)=0$.

\rem In  equation (\ref{list7}) the degenerate case $\wp=const$ is possible, and in the equation (\ref{list6}) the same degeneration leads to a special case of equation (\ref{list5}).

Let us prove Theorem 1. We note that the provided below proof contains the algorithm for reducing an arbitrary integrable equation (\ref{kdv}) to one of canonical forms (\ref{list1})--(\ref{list15}) by point transformations.

\begin{proof}
According to Lemma~2, each integrable equation can be written as follows,
\begin{equation}\label{eq0_3}
u_{t}=u_{xxx}+A_{2}(u_{x},u,x) u_{xx}^{2}+A_{1}(u_{x},u,x) u_{xx}+A_{0}(u_{x},u,x).
\end{equation}
It is easy to see that the density of conservation law (\ref{cond2}) reads as (\ref{ro_0}), where $f_1=3\,A_{2,u_1}-4\,A_2^2$. Relation (\ref{ro_01}) leads us to two equations
$$
\begin{aligned}
&9 \frac{\partial^{2 }A_{2}}{\partial u_1^{2}}-36\, A_{2} \frac{\partial A_{2}}{\partial u_1}+16\, A_{2}^{3}=0,\\
&24\,A_2\left(\frac{\p A_2}{\p u}u_1+\frac{\p A_2}{\p x}\right)+2\,A_1\left(3\frac{\p A_2}{\p u_1}-4\,A_2^2\right)-9\frac{\p^2 A_2}{\p x\p u_1}-9\frac{\p^2 A_2}{\p u\p u_1}u_1=0.
\end{aligned}
$$
The first of the equations has a solution of the form
\begin{equation}\label{A2}
A_{2}=-\frac{3}{4 B}\frac{\p B}{\p u_1}, \qquad {\mbox{\rm where}} \quad \frac{\p^3 B}{\p u_1^3}=0,
\end{equation}
at that, the second equation becomes
\begin{equation}\label{A1}
\left(2\,A_1B+3\,\frac{\p B}{\p x}+3\,u_1\frac{\p B}{\p u}\right) \frac{\p^2 B}{\p u_1^2}=3\,B\frac{d}{dx}\frac{\p^2 B}{\p u_1^2}.
\end{equation}

In view of the formula for the function $A_2$, it is clear that without loss of generality the leading coefficient of the polynomial $B(u_1)$ can be assumed to be one. This is why we have three cases,
$$
{\bf I}.\ B=u_1^{2}+B_{1}(x,u)u_1+B_{0}(x,u),\ \ \ {\bf II}.\ B=u_1+B_{0}(x,u),\ \ \ {\bf III}.\ B=1.
$$
Equation (\ref{A1}) holds identically in the cases {\bf II} and {\bf III}, and in the first case it determines the  function $A_1$,
$$
A_{1}=-\frac{3}{2 B}\left(\frac{\p B}{\p x}+u_1\frac{\p B}{\p u} \right).
$$

{\bf Case I.} Under the point transformation $u=\phi(x,v)$ the function $B$ changes by the rule
$$
\t B(x,v)=(\phi_vv_1+\phi_x)^2+B_1(x,\phi)(\phi_vv_1+\phi_x)+B_0(x,\phi).
$$
Therefore, taking for $\phi$ any solution to the equation $\phi_x=-\frac12 B_1(x,\phi)$, we reduce the issue to the case $B_1=0$.

Returning back to the study of  second integrability condition (\ref{cond2}), we find that
\begin{align}\label{kk1}
\frac{d}{dt}\rho _1\sim -\frac{u_2^4B_{0,x}}{4\,(u_1^2+B_0)^3}-\frac{u_2^3}{6}\left[\frac{\p^4 A_0}{\p u_1^4}+
\frac{3}{u_1^2+B_0}\left(u_1\frac{\p^3 A_0}{\p u_1^3}-\frac{\p^2 A_0}{\p u_1^2}\right)+\Phi (B_0,u,u_1)\right]+\\
+Z_2 u_2^2+Z_1 u_2+Z_0,\n
\end{align}
where the expression $\Phi$ depends on the derivatives of the functions $B_0$ and vanishes as $B_0$ is constant; $Z_i$ are some functions in $x,u,u_1$. Equating the coefficient at $u_2^4$ to zero, we find that $B_{0,x}=0$ and therefore, $B=u_1^{2}+B_{0}(u)$. By an appropriate point transformation
$u\rightarrow \phi(u)$ we convert $B_0$ to a constant $c_0$ being zero or one (case {\bf I.1}), or zero (case {\bf I.2}).

Equating the coefficient at $u_2^3$ in (\ref{kk1}), where $\Phi=0$, $B_0=c_0$,  to zero, we find the function $A_0$. As a result,   equation (\ref{eq0_3}) casts into the form,
\begin{equation}\label{eq0_31}
u_{t}=u_{xxx}-\frac{3\,u_{x}u_{xx}^{2}}{2(u_x^2+c_0)}+A_{0}(u_1,u,x),
\end{equation}
where $A_0$ is defined by one of the following formulas,
$$
\begin{aligned}
&{\bf I.1.}\ c_0=1,\ \ A_0=a_0(u_1^2+1)^{3/2}+ a_1u_1(u_1^2+1)+a_2u_1+a_3, \\
&{\bf I.2.}\ c_0=0,\ \ A_0=\frac{a_0}{u_1}+ a_1u_1^3+a_2u_1+a_3.
\end{aligned}
$$
In both cases $a_i=a_i(x,u)$.

In the case {\bf I.1} it follows from the further implications of the second integrability condition that $a_0$, $a_2$, and $a_3$ are constant and the function $a_1$ is independent of $u$. Moreover, ${a_1'''=-8a_1a_1',\ a_0a_1'= a_3a_1'=0}$. If $a_1'\ne0$, then
eliminating the constant $a_2$ by the Galilean transformation, we obtain  equation (\ref{list6}). If $a_1'=0$, then up to the Galilean transformation we get   equation (\ref{list5}).

In the case {\bf I.2}, equating the coefficient at $u_2^2$ in (\ref{kk1}) to zero, we find the equation
$$
5\frac{\p a_{1}}{\p x}\,u_1^4-4\frac{\p a_2}{\p u}\,u_1^3-\frac{\p a_2}{\p x}\,u_1^2+2\frac{\p a_3}{\p x}\,u_1+\frac{\p a_0}{\p x}=0.
$$
It yields that $a_2$ is constant and the functions $a_0$, $a_1$, and $a_3$ depend on $u$ only. The constant $a_2$ is eliminated by the Galilean transformation, and one of the functions $a_0$,$a_1$, or $a_3$ can be made constant by an appropriate transformation $u\rightarrow \phi(u)$.

{\bf I.2.1}. If $a_0\ne0$, then without loss of generality we can assume that $a_0=1$. In this case the second integrability condition is equivalent to three equations,
$a_3'=0$, $a_3a_1'=0$, $a_1'''+8a_1a_1'=0$. If $a_1'\ne0$, then letting $a_1=-3/2\wp$, we arrive at  equation (\ref{list7}). If $a_1'=0$, then the transformation
$u\to u+a_3t$ is admissible and it eliminates the constant $a_3$. In this case we get the equation coinciding with (\ref{list7}) for a constant function $\wp$.

{\bf I.2.2}. If $a_0=0$, then by a transformation $u\rightarrow \phi(u)$ one can simplify $a_3$ or $a_1$. If $a_3=0$, then by such a transformation one can eliminate $a_1$,
and we obtain  equation (\ref{list8}). If $a_3\ne0$, then by a transformation $u\rightarrow \phi(u)$ we make $a_3$ constant. Then   the second integrability condition implies $a_1'=0$ that allows us to employ the transformation ${u\to u+a_3t}$ eliminating $a_3$. That is,  we arrive at the case $a_3=0$ considered above.

In the case {\bf I} the complete classification has been obtained just by  conditions (\ref{cond1}) (Lemma 2) and (\ref{cond2}). It has happened to be possible because $\rho_1$ is the density of high (second) order.

{\bf Case II.} In this case $B=u_{x}+B_{0}(x,u).$ By a transformation $u\rightarrow \psi(x,u)$ one can eliminate the function $B_{0}$. Letting $B_0=0$, we find that
$$
\rho_0\sim A_1,\qquad \rho_2\sim \frac{u_2^2}{u_1^2}\left(2\frac{\p^2 A_1}{\p u_1^2}\,u_1^2-\frac{\p A_1}{\p u_1}\,u_1+A_1\right)+O(1).
$$
It is easy to check that
\begin{align}
&\frac{d}{dt}\rho_0\sim \frac{u_2^3}{4\,u_1}\left(2\,u_1\frac{\p^3 A_1}{\p u_1^3}+3\frac{\p^2 A_1}{\p u_1^2}\right)+h_2u_2^2+h_0, \label{r0}\\
&\frac{d}{dt}\rho_2\sim \frac{u_3^2u_2}{4\,u_1^3}\left(2u_1^3\frac{\p^3 A_1}{\p u_1^3}+u_1^2\frac{\p^2 A_1}{\p u_1^2}+u_1\frac{\p A_1}{\p u_1}-A_1\right)+
g u_3^2+O(2), \label{r2}
\end{align}
where $h_i$ and $g$ are some functions of $u_1,\,u,\,x$. Equating to zero the first terms in these expressions, we obtain a system being reduced to one second order equation
$$
2u_1^2\frac{\p^2 A_1}{\p u_1^2}-u_1\frac{\p A_1}{\p u_1}+A_1=0.
$$
It yields $A_1=a_1(x,u)u_1+a_2(x,u)\sqrt{u_1}$, and   equation (\ref{eq0_3}) reads as
\begin{equation}\label{eq0_32}
u_{t}=u_{xxx}-\frac{3\,u_{xx}^{2}}{4\,u_x} +(a_1u_x+a_2\sqrt{u_x}) u_{xx}+A_{0}(u_{x},u,x).
\end{equation}
Equating the term at $u_2^2$ in (\ref{r0}) to zero, we obtain  two relations
\begin{equation}\label{a1a2}
3\frac{\p a_2}{\p u}=a_1a_2,\qquad 3\frac{\p a_1}{\p x}+a_2^2=0.
\end{equation}

For the equations of the form (\ref{eq0_32}) we have
\begin{equation}\label{r1t}
\frac{d}{dt}\rho_1\sim Z_3 u_2^3+Z_2 u_2^2+Z_1 u_2+Z_0,
\end{equation}
where $Z_i=Z_i(u_1,u,x)$. Equating the expression $Z_3$ to zero, we obtain a linear inhomogeneous fourth order equation for $A_0$, which determines the dependence of the function $A_0$ on $u_1$,
$$
A_0=\frac19\,{{u_1}}^{3} \left( 6\,{\frac {\p {a_1}}{\p {u}}}+{{a_1}}^{2}\right) +\frac23\,{a_1}\,{a_2}\,{{u_1}}^{5/2}+{a_3}\,{{u_1}}^{2}+{a_4}\, {{u_1}}^{3/2}+{a_5}\,{u_1}+a_6,
$$
where $a_i=a_i(x,u)$. Now the dependence of all the coefficients $A_i$ on $u_1$ is determined and this is why one can split the equations w.r.t. $u_1$ under the integrability conditions. For instance, the coefficient $Z_2$ at $u_2^2$ in (\ref{r1t}) is linear in $u_1$, and this is why the identity $Z_2=0$ leads to two identities. These equations read as
\begin{equation}\label{a3a5}
3\frac{\p a_4}{\p u}=a_1a_4+2a_2\frac{\p a_1}{\p x},\qquad 6\frac{\p a_5}{\p u}-3a_2a_4-12 \frac{\p a_3}{\p x}+2\,a_2\frac{\p a_2}{\p x}=0.
\end{equation}

The splitting w.r.t. $u_1$ in  conditions (\ref{cond1})--(\ref{cond3}) in view of  equations (\ref{a1a2}) and (\ref{a3a5}) yields several additional equations. The simplest of them are
\begin{align}
&a_2a_3=0,\qquad a_2\left(3\frac{\p a_5}{\p u}-a_2a_4\right)=0, \label{a2-1}\\
&27\frac{\p^2 a_5}{\p u^2}-18a_1\frac{\p a_5}{\p u}+2\,a_2^4=0, \label{a2-2}\\
&\frac{\p a_6}{\p x}=0,\qquad 3\frac{\p a_3}{\p u}=2\,a_1a_3,\qquad a_2\left(2\frac{\p a_2}{\p x}-a_4\right)=0. \label{a2-3}
\end{align}

To analyze  equations (\ref{a1a2})--(\ref{a2-3}), it is natural to consider two cases,
{\bf II.1} $a_2=0$ or {\bf II.2} $a_2\ne0$.

{\bf II.1}. If $a_2=0$, then it follows from (\ref{a1a2}) that $a_{1}=a_1(u)$. The point transformation $u\to \phi(u)$,
where $\phi$ satisfies the equation $3\,\phi''+2\,(\phi')^2a_1(\phi)=0$, vanishes $a_1$. Taking into consideration  relations (\ref{a1a2})--(\ref{a2-3}), we can write  equation (\ref{eq0_32}) as
\begin{equation}\label{eq0_32a}
u_{t}=u_{xxx}-\frac{3\,u_{xx}^{2}}{4\,u_x} +a_3(x)u_x^2+a_4(x)u_x^{3/2}+a_5(x,u) u_x+a_6(u),
\end{equation}
where $a_5=\alpha (x) +2\,u\,a_3' (x)$. For this equation conditions (\ref{cond1})--(\ref{cond4}) are equivalent to  simple relations
$$
a_3=c_3,\ \ a_4=c_4,\ \ a_5=\alpha (x),\ \ a_6=c_1+c_2u,\ \ c_3c_2=c_4c_2=0,\ \ c_3\alpha'=c_4\alpha'=0,
$$
where $c_i$ are arbitrary constants, $\alpha$ is an arbitrary function.

If $c_3\ne0$ or $c_4\ne0$, then $a_5$ and $a_6$ are constants which can be eliminated by the transformation $x\to x+a_5t$, $u\to u+a_6t$. As a result we get  equation (\ref{list9}). If $c_3=c_4=0$, then the equation reads as
$$
u_{t}=u_{xxx}-\frac{3\,u_{xx}^{2}}{4\,u_x} +\alpha (x)u_x+c_1+c_2u.
$$
If here $c_2=0$, then the transformation $u\to u+c_1t$ eliminates the constant $c_1$. If $c_2\ne0$, then by the shift $u\to u-c_1/c_2$ we vanish $c_1$, and then by the transformation $u\to u\exp(c_2t)$ we eliminate $c_2$. Thus, in all cases we arrive at   equation (\ref{list10}).

{\bf II.2}. In view of $a_2\ne0$ we set $a_2=(3/\sqrt{2})\exp(\psi /2)$, then  equations (\ref{a1a2}) are reduced to $a_1=3/2\psi_u$ and to  Liouville equation
\begin{equation}\label{Liu}
\psi _{xu}+e^\psi =0.
\end{equation}
Then from (\ref{a2-1}) and (\ref{a2-3}) we get
\begin{equation}\label{key0}
a_3=0,\quad a_{6}=a_6(u),\quad a_4=\frac{3\,e^{\psi/2}}{\sqrt{2}}\frac{\p\psi}{\p x},\quad \frac{\p a_5}{\p u}=\frac{3\,e^\psi} {2}\frac{\p \psi}{\p x},
\end{equation}
and equation (\ref{a2-2}) is reduced to (\ref{Liu}).
Besides the aforementioned relations,   conditions (\ref{cond1})~--~(\ref{cond4}) imply exactly one more equation,
\begin{equation}\label{key}
\frac{\p }{\p u}(a_2^2a_6)+\frac{\p}{\p x}\left[\left(\frac{\p a_2}{\p x}\right)^2-2\,a_2\frac{\p^2a_2}{\p x^2}-a_5a_2^2\right]=0.
\end{equation}

The form of   equation (\ref{Liu}) slightly differs from the standard one, and this is why we provide its solution
$$
\psi=\ln\frac{2\alpha' \nu' }{(\alpha- \nu)^2},\qquad \alpha'\nu'\ne0.
$$
It yields the formula for $a_2$, which we write as
$$
a_2=3\frac{\sqrt{\alpha'(x)\nu'(u)}}{\alpha-\nu}.
$$
Now it is easy to find $a_1$, $a_4$, and $a_5$,
$$
a_5=\frac{3\,\alpha''}{\alpha -\nu }-\frac{3\,\alpha'^2}{(\alpha -\nu)^2}+q(x),
$$
where $q$ is an arbitrary function. Substituting $a_2$ and $a_5$ into   equation (\ref{key}), we find the functions $a_6$ and $q$,
$$
a_6=\frac{c_0+c_1\nu+c_2\nu^2}{\nu'},\qquad q=\frac{3\,\alpha''^2}{\alpha'^2}-\frac{\alpha'''}{\alpha'}-\frac{c_0+c_1\alpha+c_2\alpha^2}{\alpha'},
$$
where $c_i$ are the constants in the separation of variables.

Since $\nu'\ne0$, by the point transformation $\t u=\nu(u)$ the result can be somewhat simplified and we arrive at  equation (\ref{list11a}).

{\bf Case III.} As $B=1$, it follows from (\ref{A2}) that $A_{2}=0$. Then the total derivative of $\rho_0$ w.r.t. $t$ casts into the form
$$
\frac{d}{dt}\rho_0\sim u_2^3\frac{\p^3 A_1}{\p u_1^3}
+u_2^2\left(3\frac{\p^3A_1}{\p u\p u_1^2}\,u_1+3\frac{\p^3A_1}{\p x\p u_1^2}-2\,A_1\frac{\p^2A_1 }{\p u_1^2}\right)+O(1).
$$
Equating the expressions at $u_2^3$ and $u_2^2$ to zero, we obtain respectively $${A_1=a_0(x,u)+a_1(x,u) u_1+a_2(x,u)u_1^2}$$ and
$$
3\frac{\p a_2}{\p u}\,u_1+3\frac{\p a_2}{\p x}-2\,a_2(a_0+a_1 u_1+a_2u_1^2)=0.
$$
Splitting the last relation w.r.t. $u_1$, we find that $a_2=0$.

Thus,  equation (\ref{eq0_3}) becomes
\begin{equation}\label{eq111}
u_{t}=u_{xxx}+\big(a_0(x,u)+a_1(x,u)\,u_x\big)\, u_{xx}+A_{0}(u_{x},u,x).
\end{equation}
For this equation by a transformation $u\rightarrow \phi(x,u)$ we can reduce the issue to $a_1=0$. After this simplification one can get easily that
$$
\frac{d}{dt}\rho_1\sim u_2^3\frac{\p^4 A_0}{\p u_1^4}
+u_2^2\left(3\frac{\p^4A_0}{\p u\p u_1^3}\,u_1+3\frac{\p^4A_0}{\p x\p u_1^3}-2\,a_0\frac{\p^3A_0 }{\p u_1^3}\right)+O(1).
$$
By this, equating the coefficient at $u_2^3$ to zero, we find
$$
A_0= a_2 u_1^{3}+a_3 u_1^{2}+a_4 u_1+a_5,
$$
where $a_i=a_i(x,u)$. Then the coefficient at $u_2^2$ gives an equation, which can be  splitted w.r.t. $u_1$ into two following ones,
\begin{equation}\label{a0a2}
\frac{\p a_2}{\p u}=0,\qquad 2\,a_0a_2=3\frac{\p a_2}{\p x}.
\end{equation}

Returning back to the analysis of conditions (\ref{cond1})--(\ref{cond3}), we can split now also w.r.t. $u_1$. In view of (\ref{a0a2}) it allows us, in particular, to obtain
\begin{equation}\label{a0a3}
\frac{\p^3 a_0}{\p u^3}=0,\qquad \frac{\p^2 a_3}{\p u^2}=0,\qquad a_2\left(3\frac{\p a_2}{\p x}-2\frac{\p a_3}{\p u}\right)=0.
\end{equation}
Consider alternative cases {\bf III.1} $a_2\ne0$ or {\bf III.2} $a_2=0$.

{\bf III.1}. By the transformation $u\to u\,\mu(x)$ we can normalize $a_2(x)$, $a_2=-1/2$, which yields $a_0=0,\ a_3=\alpha (x)$. Then by  condition (\ref{cond2}) we find
$$
a_4=f_0(x)+c_1e^{2u}+c_2e^{-2u},\qquad a_5=f_1(x)-\frac{u}{6}(4\,\alpha \alpha'+3\,f_0'),
$$
where $c_1$ and $c_2$ are constant and, in addition, $c_i\alpha =c_if_1=c_if_0'=0,\ i=1,2$.

If $c_1\ne0$ or $c_2\ne0$, we obtain the equation differing from (\ref{list4}) by the Galilean transformation. If $c_1=c_2=0$, then  condition (\ref{cond4}) determines $f_0$ and $f_1$,
$$
f_0=k_1-\frac{2}{3}\,\alpha^2,\qquad f_1=k_2 -\frac{2}{3}(k_1\alpha +\alpha'')+\frac{4}{27}\,\alpha ^3,
$$
where $k_1$ and $k_2$ are constant. Making the transformation ${u\to\ds u+\frac{2}{3}\int \alpha(x) dx}$ in the obtained equation, we reduce it to the form $u_t=u_3-u_1^3/2+k_1u_1+k_2$ being equivalent to a partial case of   equation (\ref{list4}).

{\bf III.2}. Equations (\ref{a0a3}) imply $a_0=b_1(x)u^2+b_2(x)u+b_3(x),\ a_3=b_4(x)u+b_5(x)$. In this case the equation (\ref{eq111}) is simplified by the transformation $u\to u\,f_1(x)+f_2(x)$. There are three non-equivalent cases {\bf III.2.a} $a_0=3u^2+b(x)$, {\bf III.2.b} $a_0=3u$ and {\bf III.2.c} $a_0=0$.

In the first two cases a simple check of  conditions
(\ref{cond1})--(\ref{cond4}) leads us to  equations (\ref{list13}) and (\ref{list14}), respectively.

In the case {\bf III.2.c} the form of the equation is determined by three conditions (\ref{cond1})--(\ref{cond3}),
$$
u_t=u_{xxx}+a_3(x) u_x^{2}+a_4(x,u) u_x+a_5(x,u),
$$
where $a_4=b_1u^2+b_2u+b_3,\ a_5=b_4u^3+b_5u^2+b_6u+b_7,\ b_i=b_i(x),\ 1\le i\le7$. Integrability conditions (\ref{cond1})--(\ref{cond4}) yield a cumbersome system of
equations for the functions $b_i$, studying of which yields several forks.

1. If $a_3\ne0$, then $a_3=c_0$, $b_1=b_2=b_4=b_5=b_6=0$, $b_7=c_1x+c_2+\frac12b_3''+\frac14b_3^2$. The transformation $u\to\ds u-\frac12\int b_3\,dx$ yields the
equation $u_t=u_3+c_0u_x^2+c_1x+c_2$. The sixth integrability condition implies $c_1=0$; then by the transformation $u\to u+c_2t$ we eliminate $c_2$ and obtain (\ref{list3}).

2. If $a_3=0$, then $b_1=\const$, $b_4=0$. Then again forks appear;

2.1. If $b_1\ne0$, then $b_3$, $b_5$, $b_6$, and $b_7$ are expressed in terms of $b_2(x)$ so that the transformation $u\to u-b_2/(2b_1)$ leads to the equation equivalent to (\ref{list2}).

2.2. If $b_1=0$,  we obtain $b_2=\const,\ b_2(b_6-b_3')=0,\ b_2(b_3'''+b_3b_3'-b_2b_7)=0$. If $b_2\ne0$, the transformation $u\to u- b_3/b_2$ leads us to the equation equivalent to (\ref{list1}). Otherwise we obtain linear equation (\ref{list15}).
\end{proof}

\subsection{Comments to the list of integrable equations}

It was shown in the previous section that each integrable equation (\ref{kdv}) is reduced to one of  equations (\ref{list1})--(\ref{list15}) by a chain of point transformations.
Although the answer as the list is not invariant w.r.t. point transformations,   integrability conditions (\ref{law1}), (\ref{rekkur_sc}) are so. This is why to check the integrability of a given equation it is not necessary to reduce the equation to one in the list. According to the proof of Theorem~1, it is sufficient to check four conditions (\ref{cond1})--(\ref{cond4}) if the equation belongs to the classes
{\bf I} or {\bf II}, and six conditions (\ref{rekkur_sc}) if the equation belongs to the class {\bf III}. It can be shown that if the right hand side of  equation (\ref{kdv})
is independent explicitly of $x$, then for the equations of the class {\bf III} it is sufficient to check   conditions (\ref{cond1})--(\ref{cond4}).

The orders of  canonical conservation laws (\ref{law1}), (\ref{rekkur_sc}) are discrete invariants of the group of point transformations. The analysis of the structure of these conservation laws shows that these equations are divided into two groups. For the first group of the equations (let us call them $S$-integrable\footnote{The terminology is due to F. Calogero.})
the canonical series contains conservation laws of arbitrarily high order. We observe that this property is more restrictive than just the condition of the existence of an infinite series of conservation laws for the equation. For instance, the linear
equation $u_t=u_{xxx}$ possesses an infinite series of the conservation laws with the densities $u_k^2,\, k\in \N$. However, all its canonical laws are trivial.

For the equations of the second group ($C$-integrable equations) among the canonical conservation laws
there are only a few non-trivial ones. Equations (\ref{list10})--(\ref{list15}) are $S$-integrable. The equations (\ref{list10}) and (\ref{list15}) have no non-trivial canonical conservation laws. Equation (\ref{list11a})
has just one nontrivial canonical conservation law of the first order $\ds\rho_0\sim \frac{\sqrt{\alpha' u_1}+\alpha'}{u-\alpha(x)}$. Each of  equations (\ref{list13}) and (\ref{list14}) has one canonical conservation law of zero order, $\rho_0\sim u^2$ and $\rho_0\sim u$, respectively.

All canonical conservation laws with even indexes of $S$-integrable equations are trivial \cite{ss}, and the orders of odd laws increase by  one, but the initial orders in these sequences are different. In Table 1 we provide the order of first four odd canonical conservation laws for all $S$-integrable equations.

For   equation (\ref{list9}) we provide the orders of the densities in the case of generic constants. In the case $c_1=0$ we have $\rho_1\sim0$, and other orders remain to be unchanged. If $c_2=0$,  the orders are equal to $1$, $0$ ($\rho_3\sim0$), $2$, and $3$.

\rem Equations (\ref{list1})--(\ref{list9}) are integrable by the method of the inverse scattering problem while (\ref{list10})--(\ref{list14}) are linearizable by the differential substitutions (see Section 2.4).

\medskip

{{\small Table 1.
Orders of canonical conservation laws. For zero order conservation laws we indicate in the brackets to what the density is equivalent \label{tab1}}}
\begin{center}
\begin{tabular}{|l|l|l|l|c|c|c|c|c|c|}
\hline
$\rho_i$&(\ref{list1})& (\ref{list2}) & (\ref{list3}) & (\ref{list4})& (\ref{list5})& (\ref{list6})& (\ref{list7})& (\ref{list8})& (\ref{list9})\\
\hline
$\rho_1$ & 0, ($\sim u$) & 0, ($\sim u^2$) & 0, ($\sim0$) & 1 & 2 & 2 & 2 & 2 & 1 \\
\hline
$\rho_3$ & 0, ($\sim u^2$) & 1 & 1 & 2 & 3 & 3 & 3 & 3 & 2 \\
\hline
$\rho_5$ & 1 & 2 & 2 & 3 & 4 & 4 & 4 & 4 & 3 \\
\hline
$\rho_7$ & 2 & 3 & 3 & 4 & 5 & 5 & 5 & 5 & 4 \\
\hline
\end{tabular}
\end{center}

If in the formulation of the original classification problem one assumes that the right hand side of  equation (\ref{kdv}) is independent explicitly of $x$, then the answer changes just for $C$-integrable equations. For equations (\ref{list10}), (\ref{list13}), (\ref{list15}) arbitrary functions are replaced by arbitrary constants and then these constants can be eliminated by point transformations.

Formula (\ref{list11a}) contains two $C$-integrable equations independent of $x$  explicitly,
\begin{align}
&\begin{aligned}\label{list11}
u_t&=u_{xxx}-\frac34\,\frac{u_{xx}^{2}}{{u_{x}}+1}-3\,u_{xx}u^{-1}(\sqrt{u_{x}+1}+u_{x}+1)\\
&\quad+6 \,u^{-2} u_{x}(u_{x}+1)^{3/2}+3 \,u^{-2} u_{x}\,(u_{x}+1)(u_{x}+2),
\end{aligned} \\[2mm]
&\begin{aligned} \label{list12}
u_t&=u_{xxx}-\frac34\,\frac {u_{xx}^{2}}{{u_{x}}+1}-3\,\frac{u_{xx}\,(u_{x}+1)\cosh {u}}{\sinh{u}}+3\,\frac {u_{xx}\sqrt {u_{x}+1}}{\sinh{u}}\\
&-6\,\frac {u_{x}(u_{x}+1)^{3/2}\cosh{u}}{\sinh^{2}u}+3\,\frac {u_{x}\,(u_{x}+1)(u_{x}+2)}{\sinh^{2}u}+u_{x}^2(u_{x}+3).
\end{aligned}
\end{align}
Equation (\ref{list11}) can be obtained from (\ref{list11a}) with $\alpha =x,\ c_0=c_1=c_2=0$ by the transformation $u\to u+x$. Equation (\ref{list12}) is obtained in the case $\alpha =e^{2x}$, $c_0=c_1=c_2=0$ by the transformation
$ u\to e^{2(u+x)}$. It is impossible to obtain from (\ref{list11a}) any other equation independent on $x$. It follows, for instance, from the results of independent classification of  equations (\ref{kdv}) without explicit dependence on $x$.

The situation with  equation (\ref{list14}) is rather instructive. Let the functions $\gamma$ and $\beta$ be constant. Then the constant $\gamma$ is eliminated by the Galilean transformation. It is easy to check that if $\beta\ne 0$, then the integrability conditions hold true but the canonical
conservation laws depend explicitly on $x$. It is impossible if in the classification of equations possessing generalized symmetries we restrict the generalized symmetries not to depend explicitly on $x$. In this case $\beta=0$ and  equation (\ref{list14}) is just a third order symmetry for the Burgers equation. If the dependence on $x$ for the symmetries is admitted, then the constant $\beta$ in the answer should be preserved.

\subsection{Differential substitutions relating the equations in the list}

We say that a differential substitution
\begin{equation}\label{miura}
\tilde u=\Phi(x,u,u_1,\dots,u_k)
\end{equation}
acts from the equation
\begin{equation}\label{eq1}
u_t=u_{n}+g(x,u, u_x,\dots, u_{n-1})
\end{equation}
into the equation
\begin{equation}\label{eq2}
\t u_t=\t u_{n}+f(x,\t u,\t u_x,\dots,\t u_{n-1})
\end{equation}
if for each solution $u(x,t)$ of   equation (\ref{eq1})  formula (\ref{miura}) provides a solution for   equation (\ref{eq2}). The number $k$ is called an order of the substitution. Since for $k>0$  transformation (\ref{miura}) has no inverse transformation of the same form, equations (\ref{eq1}) and (\ref{eq2}) in this definition are not equal in rights. If a differential substitution reads as (\ref{miura}), where $\t u$ satisfies an equation (A) and $u$ does   equation (B), we express it as $(B)\to(A)$.

The most known  differential substitution is the Miura transformation $\t u=u_x-u^2$ relating the Korteweg-de Vries equation
$$
\t u_t=\t u_{xxx}+6\,\t u\,\t u_x
$$
and the modified Korteweg-de Vries equation
$$
u_t=u_{xxx}-6\,u^2\,u_x.
$$
Other substitutions relating the main equations of the list were found in \cite{cvsokyam}. The issue on the invertibility of the differential substitutions was treated in
 \cite{ds3}.

The
orders of possible differential substitutions relating $S$-integrable equations can be found by Table 1. Namely, if  equations (\ref{eq1}) and (\ref{eq2}) are related by the substitution (\ref{miura}), then the orders of the canonical conservation laws with sufficiently large indices
for (\ref{eq1}) are greater by $k$ than the orders of the canonical conservation laws with the same indices for (\ref{eq2}). For instance, if  equations (\ref{list6}) and (\ref{list1})
are related by a differential substitution, then it acts from (\ref{list6}) into (\ref{list1}) and is of the third order.

Below we provide the information on differential substitutions relating various integrable equations in the list. Since the superposition of
differential substitution (\ref{miura}) and the Galilean transformation leads out of the class of the substitutions of the form (\ref{miura}), sometimes in order to find a substitution it is necessary to add a term $c\,u_x$ to the right hand side of   equation (\ref{eq1}). All such cases are mentioned below in the text.

{\bf I. $S$-integrable equations.} It happens that for all $S$-integrable equations except the Krichever-Novikov equation (\ref{list7}) there exist substitutions acting in the Korteweg-de Vries equation. For   equation (\ref{list7}) such substitution exists only when the Weierstrass function degenerates or, which is the same, the polynomial $Q$ in   formula (\ref{kdv4}) has multiple roots.

Let us provide all substitutions in the Korteweg-de Vries equation. If from a given equation there exists several substitutions into (\ref{list1}), we provide all of them. We replace
equations (\ref{list7}) and (\ref{list6}) by (\ref{kdv4}) and (\ref{kdv5}), respectively, since after that the substitutions look more compact.

\noindent {\bf(\ref{list2})$\to$(\ref{list1})}: $\t u=\pm i\sqrt{6}\,u_1+u^2+\lambda $; at that $u_t=u_{xxx}+u^2u_x+\lambda u_x$.
\\
{\bf(\ref{list3})$\to$(\ref{list1})}: $\t u=2\,u_1$.\\
{\bf(\ref{list4})$\to$(\ref{list1})}: ${\ds\t u=3\,u_2-\frac{3}{2}\,u_1^2+2\sqrt{-6\,c_2}\,u_1\,e^{-u}+c_1e^{2\,u}+c_2e^{-2\,u}}$.\\[4mm]
{\bf(\ref{list5})$\to$(\ref{list1})}:
$\ds\t u=\frac{3\,u_3}{\sqrt{u_1^2+1}}-\frac{3\,u_1\,u_2^2}{(u_1^2+1)^{3/2}}-\frac{3\,u_2^2}{2(u_1^2+1)}
-\frac{6\,c_0\,u_1\,u_2}{\sqrt{u_1^2+1}}+6\,c_0\,u_2+3\,a_1\,u_1^2+$\\
\mbox{}\hspace{7em} $\ds+3\,a_1\,u_1\sqrt{u_1^2+1}$, where $c_0=\sqrt{(a_1-a_2)/2}\,.$
\\[4mm]
{\bf(\ref{kdv5})$\to$(\ref{list1})}:
$\ds\t u=\frac{d}{dx}\left(6\,\frac{u_1+\sqrt{Q+u_1^2}}{u-a}-\frac{3}{2}\,\frac{Q'+2\,u_2}{\sqrt{Q+u_1^2}}\right)
- \frac{3}{8}\frac{\big((Q+u_x^2)_{x}\big)^2}{u_1^2\,(Q+u_x^2)}+\frac{1}{2} Q''$, where $Q(a)=0.$
\\[3mm]
{\bf(\ref{list9})$\to$(\ref{list1})}: $\ds\t u=\sqrt{-3\,c_2}\,\frac{u_2}{\sqrt{u_1}}+2c_2\,u_1+\frac{3}{2}\,c_1\sqrt{u_1}$.

The above substitutions of higher orders are the superpositions of first order substitutions. These substitutions relate  some of $S$-integrable equations. The first order substitutions are drawn on the graph (Fig. 1).

The arrows of the graph correspond to the following substitutions,

\noindent {\bf(\ref{list5})$\to$(\ref{list4})}: $\ds\t u=\ln\Big(u_1+\sqrt{1+u_1^2}\,\Big)$. At that, in the equation (\ref{list5}) there should be an additional term
$\ds \frac32\,a_2u_1$. The constants in the equations are related by the formulas $\ds c_1=\frac34(a_1+a_2),\ c_2=\frac34(a_2-a_1)$.

\noindent
{\bf(\ref{kdv5})$\to$(\ref{list4})}: $\ds\t u=\ln\Big(u_1+\sqrt{Q+u_1^2}\,\Big)-\ln(a_0+2\,a_1\,u+a_2\,u^2)$. At that, the polynomial $Q$ is written in the factorized form, ${Q(u)=(a_0+2\,a_1\,u+a_2\,u^2)(k_0+2\,k_1\,u+k_2\,u^2)}$. Moreover, in  equation (\ref{kdv5}) there should be an additional term ${\ds\frac12(a_0k_2+a_2k_0-2\,a_1k_1)u_1}$, and the constants $c_1$ and $c_2$ in   equation
(\ref{list4}) are determined by the formulas ${\ds c_1=\frac32(a_0a_2-a_1^2)}$, ${\ds c_2=\frac32(k_0k_2-k_1^2)}$.

\vskip .3ex
\noindent
\begin{minipage}[t]{78mm}
\baselineskip=6mm
\parindent 4ex \noindent {\bf(\ref{list4})$\to$(\ref{list2})}: $\ds\t u=\pm \frac{i}{2}\sqrt{6}\,u_1+\sqrt{c_1}\,e^u+\sqrt{c_2}\,e^{-u}$, at that, in equation (\ref{list4}) there should be an additional term $2\sqrt{c_1c_2}\,u_1$.

\parindent 4ex
\noindent
{\bf(\ref{list9})$\to$(\ref{list2})}: $\ds \t u=a+b\sqrt{u_1}\,, \ b\ne0$, where $\ds c_1=\frac{4}{3}\,ab,\ c_2=\frac{1}{2}\,b^2$. At that, in   equation (\ref{list9}) there should be an additional term $a^2u_1$.

\parindent 4ex
If in  equation (\ref{kdv4}) the polynomial $Q$ has multiple roots, then there exists the following substitutions from this equation not indicated in the graph,

\parindent 4ex
\noindent
{\bf(\ref{kdv4})$\to$(\ref{list1})}:

\noindent 1) $\ds\t u=\frac{d}{dx}\left(-3\frac{u_2}{u_1}+\frac{12\,u_1}{u-a}\right)
-\frac{3\,u_2^2}{2\,u_1^2}-\frac{Q}{u_1^2}$,\qquad $\ds Q =(u-a)^2\big(c_0+c_1u+c_2u^2\big)$.\\[1mm]

\end{minipage}
\hfill{
\begin{minipage}[t]{77 mm}
\unitlength=1mm
\special{em:linewidth 0.4pt}
\linethickness{0.4pt}
\begin{picture}(40.0,82.0)(0.0,76.00)
\put(10.33,150.00){\circle{10.00}}
\put(35.00,150.00){\circle{10.00}}
\put(60.00,150.00){\circle{10.00}}
\put(35.00,130.00){\circle{10.00}}
\put(60.00,130.00){\circle{10.00}}
\put(60.00,110.00){\circle{10.00}}
\put(35.00,110.00){\circle{10.00}}
\put(47.33,95.67){\circle{10}}
\put(14.0,146.67){\vector(4,-3){17.67}}
\put(35.0,145.00){\vector(0,-1){10.00}}
\put(35.00,125.00){\vector(0,-1){10.00}}
\put(56.53,126.43){\vector(-4,-3){17.73}}
\put(10.60,150.00){\makebox(0,0)[cc]{\ref{list5}}}
\put(35.30,150.00){\makebox(0,0)[cc]{\ref{kdv5}}}
\put(60.00,150.00){\makebox(0,0)[cc]{\ref{kdv4}}}
\put(35.20,130.00){\makebox(0,0)[cc]{\ref{list4}}}
\put(60.20,130.00){\makebox(0,0)[cc]{\ref{list9}}}
\put(60.20,110.00){\makebox(0,0)[cc]{\ref{list3}}}
\put(35.20,110.00){\makebox(0,0)[cc]{\ref{list2}}}
\put(47.53,95.67){\makebox(0,0)[cc]{\ref{list1}}}
\put(38.5,106.50){\vector(2,-3){5.10}}
\put(57.9,105.63){\vector(-4,-3){7.73}}
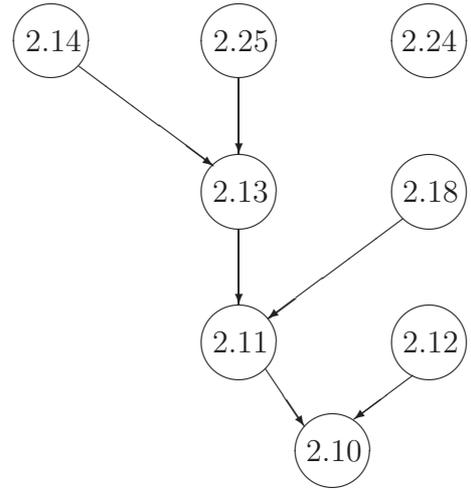
\put(0,82){\small Figure 1. Graph of the substitutions for }
\put(13,77){\small $S$-integrable third order equations}
\end{picture}
\end{minipage}
}

\noindent 2) $\ds\t u=\frac{d}{dx}\left(3\frac{u_2}{u_1}-\frac{12\,h}{u_1}\right)
-\frac{3\,u_2^2}{2\,u_1^2}-\frac{Q}{u_1^2}$,\qquad $h=c_0+c_1u+c_2u^2, \ Q=6\,h^2$.

\noindent {\bf(\ref{kdv4})$\to$(\ref{list4})}: $\ds \t u=\ln u_1-\ln h$, \qquad $\ds h=a_0+a_1u+a_2u^2, \ \ Q=-c_2h^2,\
c_1=\frac32(4\,a_0a_2-a_1^2)$.

In the case $Q=0$  equation (\ref{kdv4}) coincides with  Schwarz-Korteweg-de Vries equation (\ref{list8}). Equation (\ref{list8}) is related with the Korteweg-de Vries equations by three different substitutions,

\noindent {\bf(\ref{list8})$\to$(\ref{list1})}: 1) $\ds\t u=3\,\frac{u_3}{u_1}-\frac{9\,u_2^2}{2\,u_1^2}$; 2) $\ds\t u=-3\,\frac{u_3}{u_1}+\frac{3\,u_2^2}{2\,u_1^2}$;
3) $\ds\t u=-3\,\frac{u_3}{u_1}+\frac{3\,u_2^2}{2\,u_1^2}+12 \left(\frac{u_2}{u}-\frac{u_1^2}{u^2}\right)$.

\noindent All of them are superpositions of first order substitutions. Besides aforementioned, in these superpositions the following substitutions from  equation (\ref{list8}) are involved,

\noindent
{\bf(\ref{list8})$\to$(\ref{list4})}: $1)\ \ds \t u=\ln(u_1),\ \ c_1=c_2=0,\ \ 2)\ \t u=\ln(u_1)-\ln(u^2+c_1/6),\ c_2=0$. One more substitution is obtained from 2) by the replacement  $\t u\to -\t u,\ c_2\leftrightarrow c_1$.

{\bf II. $C$-integrable equations.} \rule{10cm}{0pt}

\noindent {\bf(\ref{list10})$\to$(\ref{list15})}: $\t u=\sqrt{u_1}$, at that, in   equation (\ref{list15}) $\beta=\dfrac12\alpha'$.

\noindent {\bf(\ref{list10})$\to$(\ref{list13})}: $\t u=\sqrt{u_1/(2\,u)}$.

\noindent {\bf(\ref{list15})$\to$(\ref{list14})}: $\t u={u_1/u}$, at that, in  equation (\ref{list15}) $\alpha=\gamma$.

\noindent {\bf(\ref{list11a})$\to$(\ref{list14})}: $\ds \t u=\frac{1}{\xi}(\alpha'+\sqrt{ u_x\alpha'}\,)-\frac{1}{2}\,\alpha''(\alpha')^{-1}$, where $\xi =\alpha (x)- u$. At that, in   equation (\ref{list14})
$$
\begin{aligned}
&\gamma =\frac{1}{2}\,\alpha'''(\alpha')^{-1}-\frac{3}{4}\,(\alpha''/\alpha')^2-(c_0+c_1\alpha +c_2\alpha ^2)/\alpha',\\
&\beta =\frac{1}{2}\,\alpha^{(4)}(\alpha')^{-1}-2\,\alpha''\alpha'''(\alpha')^{-2}+\frac{3}{2}\,(\alpha''/\alpha')^3-\frac{1}{2}\,\alpha''(\alpha')^{-2}(c_0+c_1\alpha +c_2\alpha ^2).
\end{aligned}
$$

\section{Fifth order equations}

In this section we find all equations (\ref{eq0}) possessing infinite sequences of local generalized symmetries. In the process of classification we employed necessary integrability conditions which follow from the existence of formal symmetry \cite{ibshab,ss} and are written as canonical
conservation laws. It is easy to check that each third order integrable equation (\ref{list1})--(\ref{list15}) has a fifth order symmetry (\ref{eq0}). It turns out that if one exclude these symmetries from the consideration, then the list of other integrable equations coincides (up to the equivalency) with the list obtained in  works \cite{DSS,MSS}, where the equations possessing generalized conservation laws were studied.

In Subsection 3.1 we provide the complete list of integrable equations (\ref{eq0}) not being symmetries of lower order equations. The equations in the list differ slightly by the form from equivalent equations in \cite{DSS,MSS}. Subsection 3.2 contains a new recurrent formula for the integrability conditions. We note that in the works \cite{DSS,MSS} only several simplest conditions were provided. For a given equation (\ref{eq0}) the integrability conditions can be easily checked one by one by computer.

In Subsection 3.3 we adduce a schematic proof of the classification theorem. It contains an algorithm for reducing an integrable equation (\ref{eq0}) to one of the canonical forms from Subsection 3.1 by point transformations (\ref{t1})--(\ref{t6}). In other words, once we employ point transformations, we indicate what we normalize by them. We hope that following this algorithm and guidelines given in the text, the interested reader can easily recover all the details of rather hard calculations. Post factum it follows from the proof that if equation (\ref{eq0}) satisfies first ten integrability conditions, then it is integrable. We observe that the equation $u_t=u_{5}+ u u_1$ satisfies fist nine integrability condition, but does not tenth.

\subsection{List of integrable equations}\rule{5cm}{0pt}

\theo {\it Suppose nonlinear equation} (\ref{eq0}) {\it satisfies two conditions},

\noindent 1) {\it there exists an infinite sequence of generalized symmetries}
\begin{equation} \label{sym}
u_{\tau_{i}}=G_i(u,...,u_{n_{i}}), \qquad i=1,2,... , \qquad n_{i+1}>n_{i}>\cdots>5;
\end{equation}

\noindent 2) {\it there exist no symmetries} (\ref{sym}) {\it of orders} $1<n_i<5.$

\noindent {\it Then the equation is equivalent to one in the  list}
\begin{align}
&u_t=u_{5}+5 u u_{3}+5 u_1 u_2+5 u^2 u_1, \label{tst1}
\\
\label{tst2}
&u_t=u_{5}+5 u u_{3}+{25 \over 2} u_1 u_2+5 u^2 u_1, \\
\label{tst3}
&u_t=u_{5}+5 u_1 u_{3}+{5 \over 3} u_1^3, \\
\label{tst4}
&u_t=u_{5}+5 u_1 u_{3}+{15\over 4}u_2^2 + {5 \over 3} u_1^3, \\
\label{tst5}
&u_t=u_{5}+5 (u_1-u^2) u_{3}+5 u_2^2-20 u u_1 u_2-5 u_1^3+5 u^4 u_1, \\
\label{tst6}
&u_t=u_{5}+5 (u_2-u_1^2) u_{3}-5 u_1 u_2^2+u_1^5, \\
\label{tst7}
&\begin{aligned}
u_t&=u_{5}+5 (u_2-u_1^2+\lambda_1 e^{2u}-\lambda_2^2 e^{-4u}) \, u_{3}-5 u_1 u_2^2+15 (\lambda_1 e^{2u}+4 \lambda_2^2 e^{-4u})\, u_1 u_2 \\
&+u_1^5-90 \lambda_2^2 e^{-4u}\, u_1^3 +5(\lambda_1 e^{2u}-\lambda_2^2 e^{-4u})^2\, u_1,
\end{aligned}
\\
 \label{tst8}
&\begin{aligned}
u_t&=u_{5}+5 (u_2-u_1^2-\lambda_1^2 e^{2u}+\lambda_2 e^{-u}) \, u_{3}-5 u_1 u_2^2-15 \lambda_1^2 e^{2u} \, u_1 u_2 \\
&+u_1^5+5(\lambda_1^2 e^{2u}-\lambda_2 e^{-u})^2 \, u_1, \quad \lambda_2\ne 0,
\end{aligned}
\\
\label{tst9}
&
\begin{aligned}
u_t&=u_{5}-5\frac{u_2 u_{4} }{ u_1}+ 5\frac{u_2^2 u_{3}}{ u_1^2}+5\left(\frac{ \mu_1}{ u_1}+\mu_2 u_1^2\right)u_{3}-5\left(\frac{ \mu_1}{ u_1^2}+\mu_2 u_1\right)u_2^2
\\
&-5 \frac{\mu_1^2}{ u_1}+ 5 \mu_1\mu_2 u_1^2 +\mu_2^2 u_1^5,
\end{aligned}
\\
\label{tst10}
&\begin{aligned}
u_t&=u_{5}-5\frac{u_2 u_{4}}{u_1}-\frac{15}{4 }\,\frac{u_{3}^2}{u_1}+\frac{ 65}{4}\,\frac{u_2^2 u_{3}}{ u_1^2} +5\left(\frac{\mu_1}{u_1}+\mu_2 u_1^2\right)\, u_{3}
-\frac{135}{16 }\frac{u_2^4}{u_1^3}
\\
&-5\left(\frac{7 \mu_1}{4 u_1^2}-\frac{\mu_2 u_1}{2}\right) u_2^2-5 \frac{\mu_1^2}{u_1}+ 5 \mu_1\mu_2 u_1^2 +\mu_2^2 u_1^5,
\end{aligned}
\\
&
\begin{aligned}
u_t&= u_{5}-\frac{ 5 }{2}\,\frac{ u_2u_{4} }{ u_1}-\frac{5}{4}\,\frac{u_{3}^2}{ u_1}+5\frac{u_2^2 u_{3}}{u_1^2} +\frac{5\, u_2 u_{3}}{2 \sqrt{u_1}}
-5 (u_1-2 \mu u_1^{1/2}+\mu^2) \, u_{3} -\frac{ 35}{16}\,\frac{u_2^4}{ u_1^3} \\
 & -\frac{5}{3}\,\frac{u_2^3}{u_1^{3/2}} +5\Big (\frac{ 3 \mu^2}{ 4 u_1} -\frac{\mu }{\sqrt{u_1}}+\frac{1}{4}\Big)\, u_2^2
+ \frac{ 5}{ 3}\, u_1^3
 - 8 \mu u_1^{5/2}+15 \mu^2 u_1^2-\frac{40}{3}\,\mu^3 u_1^{3/2},
\end{aligned} \label{tst11}
\\
\label{tst12m}
&\begin{aligned}
u_t&=u_{{{ 5}}}+\frac52\,{\frac { f-u_{1}}{{f}^{2}}}\,u_{{2}}u_{{{ 4}}}+\frac54\,{\frac {2\,f -u_{1}}{{f}^{2}}}\,u_3^{2} +5\,\mu\, ( u_{1}+f ) ^{2}u_{{{3}}}\\
&+\frac54\,{\frac {4\,{u_{1}}^{2}-8\,u_{1}f+{f}^{2}}{{f}^{4}}}\,u_2^{2}u_{{{ 3}}} +{\frac {5}{16}}\,{\frac {2-9\,u_{1}^{3}+18\,u_{1}^{2}f}{{f}^{6}}}\,u_2^{4}\\
&+\frac {5\mu}{4}\,{\frac {( 4\,f-3\,u_{1} )( u_{1}+f )^{2}}{{f}^{2}}}\,u_2^{2}+{\mu}^{2} ( u_{1}+f ) ^{2}\big( 2\,f ( u_{1}+f ) ^{2}-1\big),
\end{aligned}
\\
\label{tst13m}
&\begin{aligned}
u_t&=u_{{{ 5}}}+\frac52\,\frac {f- u_{1}}{f^{2}}\, u_2u_4+\frac54\,\frac {2\,f- u_1}{f^2}\, u_3^{2}-5\,\omega\, ( {f}^{2}+u_1^{2} ) u_3 \\
&+\frac54\,{\frac {4\,u_1^{2}-8\,u_{1}f+{f}^{2}}{{f}^{4}}}\,u_2^{2}u_3+{\frac {5}{16}}\,\frac {2-9\,u_1^{3}+18\,u_1^{2}f}{f^6}\,u_2^{4}
\\
&+\frac54\,\omega\,\frac {5\,u_1^{3} -2\,u_1^{2}f-11\,u_{1}{f}^{2}-2}{f^2}\, u_2^{2}-\frac52\,{\omega'}\, ( u_1^{2}-2\,u_{1}f+5\,{f}^{2} )u_{1}u_2 \\
&+5\,{\omega}^{2}u_{1}{f}^{2} ( 3\,u_{1}+f ) ( f-u_{1}),
\end{aligned}
\\
\label{tst14m}
&\begin{aligned}
u_{{t}}&=u_5+\frac52\,{\frac {f-u_{1}}{{f}^{2}}}\,u_2u_4+\frac54\,{\frac {2\,f -u_{1}}{{f}^{2}}}\,u_3^{2}
+\frac54\,{\frac {4\,u_1^{2}-8\,u_{1}f+{f}^{2}}{{f}^{4}}}\,u_2^{2}u_3
\\
&+{\frac {5}{16}}\,{\frac {2-9\,{u_{1}}^{3}+18\,{u_{1}}^{2}f}{{f}^{6}}}\,u_2^{4}+5\,\omega\,{\frac {2\,{u_{1}}^{3}+{u_{1}}^{2}f-2\,u_{1}{f}^{2}+1}{{f}^{2}}}\,u_2^{2}
\\
& -10\,\omega\,u_{{{3}}} ( 3\,u_{1}f+2\,u_1^{2}+2\,{f}^{2} )-10\,{\omega'} ( 2\,{f}^{2}+u_{1}f+{u_{1}}^{2} )\,u_{1}u_{{2}}\\
& +20\,{\omega}^{2}u_{1} ( {u_{1}}^{3}-1 )( u_{1}+2\,f ),
\end{aligned}
\\
\label{tst15m}
&\begin{aligned}
u_t&=u_5+\frac52\,\frac {f-u_1}{f^2}\,u_2u_4+\frac54\,{\frac {2\,f-u_{1}}{{f}^{2}}}\,u_3^{2}-5\,c\frac {{f}^2+u_1^2}{{\omega}^{2}}\,u_3
\\
&+\frac54\,{\frac {4\,u_1^{2}-8\,u_{1}f+{f}^{2}}{{f}^{4}}}\,u_2^{2}u_3 +\frac {5}{16}\,\frac {2-9\,{u_{1}}^{3}+18\,u_1^{2}f }{{f}^{6}}\,u_2^{4}
\\
&-10\,\omega\, ( 3\,u_{1}f+2\,u_1^{2}+2\,{f}^{2} )\,u_3-\frac54\,c\,\frac { 11\,u_{1}{f}^{2}+2\,u_1^{2}f +2-5\,u_1^{3} }{{\omega}^{2}{f}^2}\,u_2^{2}
\\
&+5\,\omega\,\frac {2\,u_1^{3}+u_1^{2}f-2\,u_{1}{f}^{2}+1}{f^2}\,u_2^{2}+5\,c\,{\omega'}\,\frac {u_1^{2}+5\,{f}^{2}-2\,u_{1}f }{{\omega}^3}\,u_{1}u_2
\\
&-10\,{\omega'}\,(2\,{f}^{2}+u_{1}f+{u_{1}}^{2})\,u_1u_2 +20\,{\omega}^{2}u_{1} ( u_1^{3}-1 ) ( u_{1}+2\,f )
\\
&+40\,{\frac {c\,u_{1}{f}^{3} ( 2\,u_{1}+f ) }{\omega}}+5\,{\frac {{c}^{2}u_{1}{f}^{2} ( 3\,u_{1}+f )( f-u_{1} ) }{{\omega}^{4}}},\ \ c\ne0.
\end{aligned}
\end{align}
{\it Here} $\lambda_1$, $\lambda_2$, $\mu$, $\mu_1$, $\mu_2$, {\it and} $c$ {\it are parameters},
{\it the function} $f(u_1)$ {\it solves the algebraic equation}
\begin{equation}\label{alg1}
(f+u_1)^2(2f-u_1)+1=0,
\end{equation}
{\it and} $\omega(u)$ {\it is any nonconstant solution to the differential equation}
\begin{equation}\label{Waier}
\omega'^2=4\, \omega^3+c. \qquad
\end{equation}

\rem All the equations in the list of Theorem 2 are $S$-integrable. In the proof of the theorem it has been established that each $C$-integrable equation
(\ref{eq0}) is a symmetry of some third order $C$-integrable equation in   list (\ref{list1})~--~(\ref{list15}).

\rem If $\lambda_1=\lambda_2=0$ in  equation (\ref{tst7}), it coincides with (\ref{tst6}). If $\lambda_1=\lambda_2=0$ in  equation (\ref{tst8}),  it also coincides with (\ref{tst6}), and if $\lambda_2=0$, then (\ref{tst8}) coincides with (\ref{tst7}) as $\lambda_2=0$  up to the change $\lambda_1\to - \lambda_1^2$ in the latter.

\rem If $\mu _2\ne0$ in   equation (\ref{tst9}),   the substitution $u=c^{-1}\ln v$ reduces it to
\begin{equation}\label{tst9a}
v_t=v_5-5\frac{v_2v_4}{v_1}+5\frac{v_2^2v_3}{v_1^2} +5\mu\left(\frac{vv_3-v_1v_2}{v_1}-\frac{vv_2^2}{v_1^2}\right) -5\mu^2\frac{v^2}{v_1},
\end{equation}
where $\mu=\mu_1c,\ c=\sqrt{-\mu_2}$.

\rem If $\mu _2\ne0$ in  equation (\ref{tst10}),  the transformation $u=c^{-1}\ln v$ reduces it to
\begin{equation}\label{tst10a}
v_t=v_5-5\frac{v_2v_4}{v_1}-\frac{15v_3^2}{4v_1} +65\frac{v_2^2v_3}{4v_1^2}-\frac{135v_2^4}{16v_1^3}
+5\mu\left(\frac{vv_3}{v_1}+\frac12v_2 -\frac74\frac{vv_2^2}{v_1^2}\right)-5\mu^2\frac{v^2}{v_1},
\end{equation}
where $\mu=\mu_1c,\ c=2\sqrt{-\mu_2}$.

\subsection{Integrability conditions}

The following conditions can be borrowed from the works \cite{MSS,DSS}.

\lem {\it For nonlinear integrable equations (\ref{eq0}) the first four integrability conditions can be written as}
\begin{align}
&\frac{d}{dt}\frac{\p F}{\p u_4}=\frac{d}{dx}\sigma _0, \label{c0}\\
&\frac{d}{dt}\left(2 \left(\frac{\p F}{\p u_4}\right)^2-5 \frac{\p F}{\p u_3}\right)=\frac{d}{dx} \sigma _1, \label{c1}\\
&\frac{d}{dt}\left(15\frac{\p F}{\p u_3} \frac{\p F}{\p u_4}- 25 \frac{\p F}{\p u_2}- 4\left(\frac{\p F}{\p u_4}\right)^3\right)=\frac{d}{dx}\sigma _2, \label{c2}\\
&\frac{d}{dt}\left[25\left(\frac{d}{dx}\frac{\p F}{\p u_4}\right)^2+5\frac{\p F}{\p u_4}\left(5\frac{d}{dx}\frac{\p F}{\p u_3}
+10\frac{\p F}{\p u_2}-{7}\frac{\p F}{\p u_3}\frac{\p F}{\p u_4}\right)+\right. \label{c3}\\
&\hspace{50mm}\left.\vphantom{\frac{d}{d}}+7\left(\frac{\p F}{\p u_4}\right)^4+{25}\left(\frac{\p F}{\p u_3}\right)^2-125\frac{\p F}{\p u_1}\right]=\frac{d}{dx}\sigma_3,\n
\end{align}
{\it where $\dfrac{d}{dx}$ is the operator of the total derivative w.r.t. $x$, and $\dfrac{d}{dt}$ is the evolution derivation w.r.t. $t$ by virtue of
equation (\ref{eq0}).}

The given conditions follow from the existence of a formal symmetry. But technically it is more convenient to employ
the described in Appendix 3 method of calculating the densities of canonical conservation laws
\begin{equation}\label{laws}
\frac{d}{dt}\rho_n=\frac{d}{dx}\theta_n,\ \ n=0,1,\dots
\end{equation}
by the logarithmic derivative of a formal eigenfunction of the
linearization operator for  equation (\ref{eq0}). This approach allows us to obtain the recurrent formula
\begin{align}
\rho_{n+4}=&\frac{1}{5}\theta_n -\frac{1}{5}\left[F_{u_0}\delta_{n,0} +F_{u_1}\delta_{n,-1}+F_{u_2}\delta_{n,-2} +F_{u_3}\delta_{n,-3}+F_{u_4}\delta_{n,-4}+F_{u_1}\rho_n\right]
\nonumber\\
&-2\sum_{0}^{n+3} \rho_{i}\rho_j-2\sum_{0}^{n+2} \rho_i\rho_j\rho_{k}-\frac{1}{5}\sum_{0}^{n} \rho_i \rho_j \rho_k \rho_l\rho_{m}
+\sum_{0}^{n+1}\left(\frac{d}{dx}\rho_{i}\right)\frac{d}{dx}\rho_j
\nonumber\\
&+\sum_{0}^{n}\rho_i \left(\frac{d}{dx}\rho_j\right)\frac{d}{dx} \rho_k-\sum_{0}^{n+1} \rho_i \rho_j \rho_k \rho_l
-\frac{1}{5}F_{u_2}\left[\frac{d}{dx}\rho_n+2\rho_{n+1}+\sum_{0}^{n} \rho_{i}\rho_j\right]
\nonumber\\
&-\frac{1}{5}F_{u_3}\left[\frac{d^2}{dx^2}\rho_n +3\frac{d}{dx}\rho_{n+1}+3\rho_{n+2} +\frac{3}{2}\frac{d}{dx}\sum_{0}^{n} \rho_{i}\rho_j
+3\sum_{0}^{n+1} \rho_{i}\rho_j+\sum_{0}^{n} \rho_{i}\rho_i\rho_k\right] \label{rec}
\\[3mm]
&-\frac{1}{5}F_{u_4}\left[\frac{d^3}{dx^3}\rho_n +4\frac{d^2}{dx^2}\rho_{n+1}+6\frac{d}{dx}\rho_{n+2} +4\rho_{n+3}+2\frac{d^2}{dx^2}\sum_{0}^{n}\rho_{i} \rho_j
+6\frac{d}{dx}\sum_{0}^{n+1}\rho_{i} \rho_j+ \right.
\nonumber\\
&\left.\quad +6\sum_{0}^{n+2}\rho_{i} \rho_j +4\sum_{0}^{n+1}\rho_i \rho_j \rho_k -\sum_{0}^{n}\left(\frac{d}{dx}\rho_{i}\right)\frac{d}{dx}\rho_j
+2\frac{d}{dx}\sum_{0}^{n}\rho_i \rho_j \rho_k+\sum_{0}^{n}\rho_i \rho_j \rho_k\rho_l \right]\nonumber-\\[2mm]
&-\frac{d}{dx}\left[\frac{1}{5}\frac{d^3}{dx^3}\rho _{n}+\frac{d^2}{dx^2}\rho _{n+1}+2\frac{d}{dx}\rho _{n+2}+\sum_{0}^{n} \rho_{i}\frac{d^2}{dx^2}\rho_j
+2\frac{d}{dx}\sum_{0}^{n+1}\rho_{i} \rho_j+3\sum_{0}^{n+2}\rho_{i} \rho_j\right.
\nonumber\\
&\left.\quad+2\rho_{n+3}+\frac{2}{3}\frac{d}{dx}\sum_{0}^{n} \rho_i \rho_j \rho_k +\frac{1}{2}\sum_{0}^{n}\left(\frac{d}{dx}\rho_{i}\right) \frac{d}{dx}\rho_j
+2\sum_{0}^{n+1}\rho_i \rho_j \rho_k+\frac{1}{2} \sum_{0}^{n}\rho_i \rho_j \rho_k\rho_l\right],\nonumber
\end{align}
where $n\ge -4$, $\rho_k=0$, $\forall k<0$, $\delta_{ij}$ is the Kronecker delta, $F_{u_i}=\p F/\p u_i$. In  formula (\ref{rec}) the notation
$$
\sum_{m}^{n}a_i b_j\dots p_z=\sum_{\substack{i+j+\dots+z=n\\ i\ge m,\, j\ge m,\dots,\, z\ge m}}a_i b_j\dots p_z
$$\label{summ}
was employed for multiple sums. The summation indices in  formula (\ref{rec}) are non-negative.

Recurrent formula (\ref{rec}) is published at the first time. It is easy to check that the first four integrability conditions in sequence (\ref{laws}) are equivalent to  conditions (\ref{c0})~--~(\ref{c3}).

Conditions (\ref{rec}), (\ref{laws}) can be used more efficiently for the classification if one first studies the structure of the densities of local conservation laws for equations (\ref{eq0}). We recall that the symbol $O(n)$ indicates a function of the differential order at most $n$. Moreover, we employ the symbol $P_n(u_k)$ to denote a polynomial of the degree $n$ of the variable $u_k$, whose coefficients have differential order less than $k$. In what follows we employ regularly the equivalence $f\dfrac{d}{dx}g\sim -g\dfrac{d}{dx}f$ implied by $\dfrac{d}{dx}(fg)\sim0$. For instance, we have
$$
u_{n+1}f(u,\dots,u_n) \sim-\sum_{i=0}^{n-1}u_{i+1} \frac{\p }{\p u_i}\int fd\,u_n,
$$
which yields, in particular, $u_{n+1}O(n)\sim O(n)$.

\lem {\it If a density $\rho$ of a local conservation
law for   equation (\ref{eq0}) has differential order $ n\ge3$,
then the equation
\begin{equation}\label{den_n}
 \frac{d}{dx}\, \frac{\p^2\rho }{\p u_n^2}=\frac{2}{5} \,\frac{\p^2\rho }{\p u_{n}^2}\,\frac{\p F}{\p u_4}
\end{equation}
holds true.}

\begin{proof} By definition we have
$$
\frac{d}{dt}\rho =\sum_{k=0}^{n} \frac{\p\rho }{\p u_k}\left(u_{k+5}+\frac{d^k}{dx^k}\,F\right).
$$
Using the equivalence, we can lower the order of this expression up to $n+2$. First we show that
$$
\sum_{k=0}^{n-2} \frac{\p\rho }{\p u_k}\left(u_{k+5}+\frac{d^k}{dx^k}\,F\right)\sim O(n+1).
$$
For this it is sufficient to convert the highest order term. Assuming $n\ge 3$, we get
$$
\begin{aligned}
\frac{\p\rho }{\p u_{n-2}}u_{n+3}&\sim -u_{n+2}\frac{d}{dx}\, \frac{\p\rho }{\p u_{n-2}}=-u_{n+2}u_{n+1}\frac{\p^2\rho }{\p u_n\p u_{n-2}}+u_{n+2}O(n) \\
&\sim \frac{1}{2}\,u_{n+1}^2\frac{d}{dx}\frac{\p^2\rho }{\p u_n\p u_{n-2}}-u_{n+1}\frac{d}{dx}O(n)=O(n+1).
\end{aligned}
$$
Thus,
\begin{equation}\label{dok1}
\frac{d}{dt}\rho\sim \frac{\p\rho }{\p u_n}u_{n+5}+\frac{\p\rho }{\p u_{n-1}}u_{n+4}+\frac{\p\rho }{\p u_n}\,\frac{d^n}{dx^n}\,F+
\frac{\p\rho }{\p u_{n-1}}\,\frac{d^{n-1}}{dx^{n-1}}\,F+O(n+1).
\end{equation}

Let us convert the first term,
\begin{align*}
a_1&\eqdef\frac{\p\rho }{\p u_n}u_{n+5}\sim u_{n+3}\frac{d^2}{dx^2}\,\frac{\p\rho }{\p u_n}=u_{n+3}\frac{d}{dx}\,\sum_{i=0}^{n} \frac{\p^2\rho }{\p u_n\p u_i}u_{i+1}\\
&=u_{n+3}\left(\sum_{i=0}^{n} \frac{\p^2\rho }{\p u_n\p u_i}u_{i+2}+\sum_{i,j=0}^{n} \frac{\p^3\rho }{\p u_n\p u_i\p u_j}u_{i+1}u_{j+1}\right)
\\
&\sim -\frac{1}{2}\,u_{n+2}^2\frac{d}{dx}\, \frac{\p^2\rho }{\p u_n^2}-u_{n+2}\frac{d}{dx}\left(\sum_{i=0}^{n-1} \frac{\p^2\rho }{\p u_n\p u_i}u_{i+2}+
\sum_{i,j=0}^{n} \frac{\p^3\rho }{\p u_n\p u_i\p u_j}u_{i+1}u_{j+1}\right) \\
&\sim -u_{n+2}^2\left(\frac{1}{2}\,\frac{d}{dx}\, \frac{\p^2\rho }{\p u_n^2}+ \frac{\p^2\rho }{\p u_n\p u_{n-1}}+2
\sum_{i=0}^{n} \frac{\p^3\rho }{\p u_n^2\p u_i}u_{i+1}\right)+u_{n+2}O(n+1) \\
&\sim -\frac{5}{2}\,u_{n+2}^2\frac{d}{dx}\, \frac{\p^2\rho }{\p u_n^2}-\frac{\p^2\rho }{\p u_n\p u_{n-1}}\,u_{n+2}^2+O(n+1).
\end{align*}

The second term in (\ref{dok1}) is converted in the same way,
$$
\begin{aligned}
a_2&\eqdef\frac{\p\rho }{\p u_{n-1}}u_{n+4}
\sim u_{n+2}\left(\sum_{i=0}^{n} \frac{\p^2\rho }{\p u_{n-1}\p u_i}u_{i+2}+\sum_{i,j=0}^{n} \frac{\p^3\rho }{\p u_{n-1}\p u_i\p u_j}u_{i+1}u_{j+1}\right)\\
&=\frac{\p^2\rho }{\p u_n\p u_{n-1}}u_{n+2}^2+u_{n+2}O(n+1)\sim \frac{\p^2\rho }{\p u_n\p u_{n-1}}u_{n+2}^2+O(n+1).
\end{aligned}
$$

The above calculations are correct if $n\ge 2$. To convert the remaining two terms in (\ref{dok1}), it is important that $n\ge3$,
$$
\begin{aligned}
a_3&\eqdef\frac{\p\rho }{\p u_{n}}\,\frac{d^n}{dx^n}\,F\sim \left(\frac{d^2}{dx^2}\,\frac{\p\rho }{\p u_{n}}\right) \frac{d^{n-2}}{dx^{n-2}}\,F\\
&\hspace{2cm}=\left(\frac{\p^2\rho }{\p u_{n}^2}\,u_{n+2}+O(n+1)\right)\left(\frac{\p F}{\p u_4}\,u_{n+2}+O(n+1)\right)\\
&\hspace{2cm}=\frac{\p^2\rho }{\p u_{n}^2}\,\frac{\p F}{\p u_4}\,u_{n+2}^2+u_{n+2}O(n+1)+O(n+1)\sim \frac{\p^2\rho }{\p u_{n}^2}\,\frac{\p F}{\p u_4}\,u_{n+2}^2+O(n+1),\\
a_4&\eqdef\frac{\p\rho }{\p u_{n-1}}\,\frac{d^{n-1}}{dx^{n-1}}\,F\sim-\left(\frac{d}{dx}\frac{\p\rho }{\p u_{n-1}}\right)\frac{d^{n-2}}{dx^{n-2}}\,F=
-\left(\frac{d}{dx}\frac{\p\rho }{\p u_{n-1}}\right)\left(\frac{\p F}{\p u_4}\,u_{n+2}+O(n+1)\right)\\
&\hspace{2cm}=u_{n+2}O(n+1)+O(n+1)\sim O(n+1).
\end{aligned}
$$

Summing up the obtained expressions for $a_1$,\ldots, $a_4$, we find
$$
\frac{d}{dt}\rho\sim u_{n+2}^2\left(\frac{\p^2\rho }{\p u_{n}^2}\,\frac{\p F}{\p u_4}-\frac{5}{2}\,\frac{d}{dx}\, \frac{\p^2\rho }{\p u_n^2}\right)+O(n+1).
$$
Since a quadratic in higher derivative expression can not be a total derivative of any function, we obtain (\ref{den_n}). \end{proof}

{\bf Corollary.} {\it If we have $n>3$ in Lemma 4, then the density $\rho$ is quadratic in $u_n$. Indeed, the left hand side of  equation (\ref{den_n}) contains the term $\rho_{u_nu_nu_n}u_{n+1}$ whose differential order is greater than $4$ if $n>3$. The order of other terms is less and hence
$\rho_{u_nu_nu_n}=0$.}

We apply the obtained result to the classification of equations (\ref{eq0}).

\lem {\it Assume equation (\ref{eq0}) satisfies condition (\ref{c0}). Then the function $F$ is quadratic in $u_4$.}

\begin{proof}
Applying Corollary of Lemma 4 to the canonical density $\rho=F_{u_4}$, we obtain
$$
\frac{\p F}{\p u_4}=f_1+f_2u_4+f_3u_4^2,
$$
where the functions $f_1\,, f_2$ and $f_3$ are independent of $u_4$. Substituting this expression into (\ref{den_n}), we find
$$
\frac{d}{dx}f_3=\frac{2}{5}\,f_3(f_1+f_2u_4+f_3u_4^2).
$$
The left hand side of this equation is linear in $u_4$, and the right hand side is quadratic; thus, $f_3=0$. It yields $\ds F=f_0+f_1u_4+\frac12\,f_2u_4^2$, where
the functions $f_i$ are independent of $u_4$.
\end{proof}

\subsection{Scheme of proof of main theorem}
\rule{5cm}{0pt}

\lem {\it Suppose  equation (\ref{eq0}) satisfies integrability conditions (\ref{c0}), (\ref{c1}), and (\ref{c2}). Then the function $F$ is linear in $u_4$.}

\begin{proof}
According to Lemma 5, the function $F$ is quadratic in $u_4$, ${\ds F=f_0+f_1u_4+\frac12\,f_2u_4^2}$. It can be  easily verified that
$$
\rho_2\sim u_4^3\,f_2\left(16\,f_2^2-15\,\frac{\p f_2}{\p u_3}\right)+Z_1u_4^2+O(3).
$$
In accordance with Corollary of Lemma 4, a cubic in $u_4$ term should vanish and therefore
$$
\frac{\p f_2}{\p u_3}=\frac{16}{15}\,f_2^2.
$$
In view of this equation we find
$$
\rho_1\sim u_4^2\,f_2^2+O(3).
$$
For this density   relation (\ref{den_n}) read as follows,
$$
\frac{d}{dx}\,f_2=\frac{1}{5}\,f_2(f_1+f_2\,u_4).
$$
It implies the equation
$$
\frac{\p f_2}{\p u_3}=\frac{1}{5}\,f_2^2,
$$
which together with the previous one yields $f_2=0$.
\end{proof}

Thus, if  integrability conditions (\ref{c0})~--~(\ref{c2}) hold true, then   equation (\ref{eq0}) reads as
\begin{equation}\label{eq01}
u_t=u_5+u_4f_1(u,u_1,u_2,u_3)+f_0(u,u_1,u_2,u_3).
\end{equation}

\lem {\it If a function of third differential order $\rho(u,u_1,u_2,u_3)$ is a density of a conservation law for equation (\ref{eq01}), then it is at most quadratic in $u_3$.}

\begin{proof}
 Letting $n=3$ and $F=f_0+f_1u_4$ in (\ref{den_n}) $n=3$, we obtain
\begin{equation}\label{den_3}
\frac{d}{dx}\frac{\p^2 \rho}{\p u_3^2}=\frac{2}{5}\,\frac{\p^2 \rho}{\p u_3^2}f_1.
\end{equation}
Taking into account that $f_1$ is independent of $u_4$, we find that $\rho_{u_3u_3u_3}=0$.
\end{proof}
{\bf Corollary 1.} {\it The function $f_1$ in (\ref{eq01}) is linear in $u_3$.}

\begin{proof}
Indeed, it follows from $F=f_0+f_1u_4$ and (\ref{c0}) that $f_1$ is the density of a conservation law for
equation (\ref{eq01}). Thus, as it was proven above, this function reads as $f_1=g_1+g_2u_3+g_3u_3^2$, where $g_i=g_i(u,u_1,u_2)$. Substituting this expression into (\ref{den_3}) instead of $\rho$, we obtain $g_3=0$.
\end{proof}

{\bf Corollary 2.} {\it If $f_1=g_1+g_2u_3$, $g_i=g_i(u,u_1,u_2)$ in equation (\ref{eq01}) and this equation possesses a conservation law with a density $\rho$ of second
differential order, then the equation
\begin{equation}\label{den_2}
\frac{d}{dx}\frac{\p^2 \rho}{\p u_2^2}=\frac{2}{5}\,\frac{\p^2 \rho}{\p u_2^2}(g_1+g_2u_3)
\end{equation}
holds true.}

\begin{proof}
The statement can be easily checked by straightforward calculations.
\end{proof}

\prop {\it If  integrability conditions (\ref{c0})--(\ref{c2}) hold true, then  equation (\ref{eq0}) reads as
\begin{equation}\label{eq1bis}
u_t=u_5+A_1u_2u_4+A_2u_4+A_3u_3^2+ (A_4u_2^2+A_5u_2+A_6)u_3+A_7u_2^4+A_8u_2^3+A_9u_2^2+A_{10}u_2+A_{11},
\end{equation}
where $A_i=A_i(u,u_1)$. }

\begin{proof}
 By Corollary 1 of Lemma 7, equation (\ref{eq0}) reads as (\ref{eq01}), where $f_1=g_1(u,u_1,u_2)+g_2(u,u_1,u_2)u_3$. Consider then a
density $\rho_1$ of  conservation law (\ref{c1}). It is easy to check that
$$
\rho_1\sim \frac25\,f_1^2+\frac{\p f_1}{\p u_0}u_1+\frac{\p f_1}{\p u_1}u_2+\frac{\p f_1}{\p u_2}u_3-\frac{\p f_0}{\p u_3}.
$$
In this expression all terms except the last one are at most quadratic in $u_3$. By Lemma 7 the considered density must be quadratic in $u_3$ and therefore, the function $f_{0}$ is cubic in $u_3$,
$$
f_0=g_4+g_5u_3+g_6u_3^2+g_7u_3^3,\qquad g_i=g_i(u,u_1,u_2).
$$

Due to the obtained results, densities of  conservation laws (\ref{c1}) and (\ref{c2}) are equivalent to the  expressions
$$
\begin{aligned}
&\rho _1\sim u_3^2\left(5\frac{\p g_2}{\p u_2}+2\,g_2^2-15\,g_7\right)+O(2),\\
&\rho_2\sim u_3^3\left(50\frac{\p g_7}{\p u_2}-25\frac{\p^2 g_2}{\p u_2^2}+30\,g_2\frac{\p g_2}{\p u_2}+8\,g_2^3-90\,g_2g_7\right)+P_2(u_3).
\end{aligned}
$$
According to Lemma 7, the coefficient at $u_3^3$ in the second formula should vanish. Moreover, the condition $\ds \frac{d}{dt}\rho _1\sim 0$ leads to extra four
equations relating the functions $g_2$, $g_7$ and their derivatives w.r.t. $u_2$. By these equations it is easy to obtain $g_2=g_7=0$.

Thus, $F=g_1u_4+g_4+g_5u_3+g_6u_3^2$. Now the density of conservation law (\ref{c0}) equals $g_1$ and we can substitute
$\rho =g_1$ and $g_2=0$ into (\ref{den_2}),
$$
\frac{d}{dx} \frac{\p^2 g_1}{\p u_2^2}=\frac25\frac{\p^2 g_1}{\p u_2^2}g_1.
$$
As above, by this we obtain the linear in higher derivative function $g_1=A_1(u,u_1) u_2+A_2(u,u_1)$.

In view of the obtain results,   condition (\ref{c2}) yields $g_6=A_3(u,u_1)$. Then by  condition (\ref{c1}) we get $\ds \frac{\p^3g_5}{\p u_2^3}=0$. And finally, bearing in mind all the obtained results, we find by   condition (\ref{c2}) that $\ds \frac{\p^5g_4}{\p u_2^5}=0$.
\end{proof}

In studying   equation (\ref{eq1bis}) the following lemma will be useful.

\lem {\it Equation (\ref{eq1bis}) preserves its form under point transformations $u=\phi(v)$.}

Some of the formulas for the conversion of the coefficients
$A_i$ look simple,
\begin{align}\label{coef}
&\tilde A_1(v)=\phi'A_1(u),\qquad \tilde A_2(v)=A_2(u)+\phi''v_1^2A_1(u)+5\phi''(\phi')^{-1}v_1,\nonumber\\
&\tilde A_3(v)=\phi'A_3(u),\qquad \tilde A_4(v)={\phi'}^2A_4(u), \qquad \tilde A_7(v)={\phi'}^3A_7(u).\
\end{align}
Other formulas are much more bulky and we omit them.

It can be checked that the first six densities of the canonical conservation laws for   equation (\ref{eq1bis}) are equivalent to
\begin{align}
&\rho_0=-\frac{1}{5}(A_1u_2+A_2),\qquad \rho_1\sim R_1=\psi _1u_2^2+\psi_2u_2+\psi_3,\label{r1}\\
&\rho_2\sim R_2=\psi _4u_2^3+\psi_5u_2^2+\psi_6u_2+ \psi_7, \qquad\rho_3\sim R_3=\psi _8u_3^2+\psi_9u_2^4+ \psi_{10}u_2^3+\dots,\label{r2-3}\\
&\rho_4\sim R_4=\psi _{11}u_2u_3^2+\psi_{12}u_3^2+\psi_{13}u_2^5+ \dots,\label{r4}\\
&\rho_5\sim R_5=\psi _{1}u_4^2+\psi_{15}u_3^3+(\psi_{16}u_2^2+\psi_{17}u_2+\psi_{18})u_3^2+\psi_{19}u_2^6+ \dots, \label{r5}
\end{align}
where the coefficients $\psi_k$ are expressed in terms of the functions $A_i$ and their derivatives. For instance, $\psi_1$ in (\ref{r1}) and (\ref{r5}) reads as
$$
\psi_1=\frac{1}{25}\left(2A_1^2-5A_4+10\frac{\p A_3}{\p u_1}\right).
$$

\lem {\it If  equation (\ref{eq1bis}) has a conservation law with a density $\rho(u,u_1,\dots,u_n)$ of differential order $n\ge2$, then
\begin{equation}\label{ron}
\rho\sim \alpha_1(u,u_1) u_n^2+\alpha_2(u,u_1,\dots,u_{n-1}),
\end{equation}
at that,
\begin{equation}\label{coef_eq}
5\frac{\p\alpha _{1}}{\p u_1}=2\alpha _1A_1,\qquad 5\frac{\p\alpha _{1}}{\p u_0}\,u_1=2\alpha _1A_2.
\end{equation}}

For $n=3$ and $n=4$ the structure of the densities can be easily specified. If $n=3$, then
$$
\rho\sim \alpha_1 u_3^2+\alpha_2u_2^4+\alpha_3u_2^3+\alpha_4u_2^2+\alpha_5,\quad \alpha_i=\alpha_i(u,u_1),
$$
and $\alpha_1(A_1-2A_3)=0$.
If $n=4$, then
$$
\rho \sim \alpha_1 u_4^2+\alpha_2u_3^3+(\alpha_3u_2^2+\alpha_4u_2+\alpha_5)u_3^2+\beta(u,u_1,u_2) ,\quad \alpha_i=\alpha_i(u,u_1),
$$
where $\beta $ is a polynomial of sixth degree in $u_2$.

{\bf Corollary.} {\it The coefficients $\psi_4$ in (\ref{r2-3}), $\psi_{11}$ and $\psi _{13}$ in (\ref{r4}) are zero.}

The form of  equation (\ref{eq1bis}) depends essentially on the orders of its canonical conservation laws. Among integrable equations (\ref{eq1bis}), there can be equations of the two following types,

{\bf I.} Equations possessing no generalized canonical conservation laws. In other words, all canonical densities
for the equations of the first type are equivalent to densities of zero or first differential order.

{\bf II.} Equations possessing generalized canonical conservation laws of orders $\ge2$.

In the case {\bf I} one should equate all nontrivial terms of higher order in the densities of the canonical conservation laws to zero. This is why in   expressions (\ref{r1})--(\ref{r5}) there should be $\psi_1=\psi_4=\psi_5=\psi_8=\psi_9=\psi_{10}=\dots=0$.
In particular,
$$
A_4=\frac{2}{5}A_1^2+2\frac{\p A_3}{\p u_1}.
$$
By the equation $\psi_4=0$ one can express $A_7$ in terms of $A_1$ and $A_3$, and by $\psi_5=0$ one can express $A_8$ in terms of $A_1,A_2,A_3$ and $A_5$. From six conditions $\rho_i\sim h_i(u,u_1), \ i=1,\dots,6$ one can extract a cumbersome system of differential equations for the remaining functions $A_i$. In this system there is the following closed subsystem of the equations for $A_1$ and $A_3$,
\begin{equation}\label{br1}
A_3=\frac{1}{2}A_1,\qquad \frac{\p A_1}{\p u_1}=\frac{2}{5}A_1^2.
\end{equation}
The latter of these equations has two solutions $A_1=0$ and $A_1=-\frac{5}{2}(u_1+a(u))^{-1}$. If $a(u)\ne0$, then by point transformation $u\to \phi(u)$ one can normalize $a=1$. Thus, there are three possible cases,
$$
{\bf I.a.} \ \ A_1=0; \qquad {\bf I.b.}\ \ A_1=-\frac{5}{2}u_1^{-1};\qquad {\bf I.c.} \ \ A_1=-\frac{5}{2}(u_1+1)^{-1}.
$$

{\bf Case I.a.} It follows from the equations $\psi_i=0$ that $A_2=g_1(u)+g_2(u)u_1$. Employing the point transformation $u\to \phi(u)$, we can assume that $g_2=0$ (see (\ref{coef})). After that all remaining functions $A_i(u,u_1)$ are happened to be polynomials with constant coefficients. To determine these coefficients we check 10 integrability conditions (\ref{laws}). We find out that there exist only three integrable equations of the considered type,
$$
\begin{aligned}
u_t&= u_5+u_4c_1+c_2u_3+c_3u_2+c_4u_1+c_5u+c_6,\\
u_t&= u_5+5u^2u_4+10u\,u_3(u^3+4u_1)+25u \,u_2^2+10u_2(5u_1^2+12u^3u_1+u^6)\\
&+140u^2u_1^3+70u^5u_1^2+5u^8u_1,\\
u_t&= u_5+5u\,u_4+10u^2u_3+15u_1u_3+10u_2^2+10u^3u_2 +50u\,u_1u_2+5u^4u_1+30u^2u_1^2+15u_1^3.
\end{aligned}
$$

The second of these equations is a symmetry for   equation (\ref{list13}), where $\alpha=0$. The third equation is the symmetry of the Burgers equation $u_t= u_2+2uu_1$
(as well as a symmetry of   equation (\ref{list14}), where $\beta=\gamma=0$).

{\bf Case I.b.} There exists only one integrable equation in this class,
$$
u_t=u_5-\frac{5u_2u_4}{2u_1}+5\frac{u^2_2u_3}{u_1^2}-\frac{5u_3^2}{4u_1}-\frac{35u_2^4}{16u_1^3}+k\,u.
$$
It is a symmetry of   equation (\ref{list10}) as $\alpha(x)=c$.

{\bf Case I.c.} By the integrability conditions one can find $A_2=f(u)(u_1+1)+g(u)\sqrt{u_1+1}$, where $f$ and $g$ are arbitrary functions. If $g=0$, all the functions $A_i$ are independent of $u$, and the transformation $u\to u-x$ is admissible. It reduces this case to the case {\bf I.b.}

If $g\ne0$,  there exist two very cumbersome $C$-integrable equations being the symmetries of  equations (\ref{list11}) and (\ref{list12}), respectively.

In the case {\bf II}  equation (\ref{eq1bis}) possesses at least one generalized conservation law, and thus, in accordance with (\ref{coef_eq}), we can write $A_1$ and $A_2$
as
\begin{equation}\label{case2}
A_1=\frac{5}{2f_0}\frac{\p f_0}{\p u_1},\qquad A_2=\frac{5}{2f_0}\frac{\p f_0}{\p u}u_1,\quad f_0=f_0(u,u_1).
\end{equation}
As a result,   equation (\ref{eq1bis}) casts into the form
\begin{equation}\label{eq1bb}
u_t=u_5+\frac{5}{2}(\ln f_0)_x u_4+A_3u_3^2+(A_4u_2^2+A_5u_2+A_6)u_3+A_7u_2^4+A_8u_2^3+A_9u_2^2+A_{10}u_2+A_{11},
\end{equation}
where $A_i=A_i(u,u_1)$. The first canonical conservation law for this equation is trivial,
$$
\rho_0=-\frac{1}{2}\,\frac{d}{dx} \ln f_0,\qquad \theta_0=-\frac{1}{2}\,\frac{d}{dt} \ln f_0.
$$

Second integrability condition (\ref{c1}) is reduced to
$$
\frac{d}{dt}\rho _1\sim u_4^2f_0\frac{d}{dx}\left(\frac{A_4}{f_0}-\frac{2}{f_0}\frac{\p A_3}{\p u_1}-\frac{5}{2\,f_0^3}\left(\frac{\p f_0}{\p u_1}\right)^2\right)
+u_3^3\,Z_1+u_3^2\,Z_2+O(2)\sim 0.
$$
Equating the coefficient at $u_4^2$ to zero, we obtain
\begin{equation}\label{A4}
A_4= 2\,\frac{\p A_3}{\p u_1}+\frac{5}{2f_0^2}\left(\frac{\p f_0}{\p u_1}\right)^2+c_1f_0,
\end{equation}
where $c_1$ is an integration constant. The function $Z_1$ is linear in $u_2$, this is why the identity $Z_1=0$ implies two equations,
\begin{align}
&c_1\left[25\,f_0\frac{\p^2 f_0}{\p u_1^2}-45\left(\frac{\p f_0}{\p u_1}\right)^2+10\,A_3\,f_0\frac{\p f_0}{\p u_1}-14\,f_0^2\frac{\p A_3}{\p u_1}-6\,c_1\,f_0^3\right]=0,
\label{usl2-1}\\
&c_1\left[25\,f_0\,u_1\frac{\p^2 f_0}{\p u_1\p u}-30\,u_1\frac{\p f_0}{\p u_1}\frac{\p f_0}{\p u}+5\,f_0\frac{\p f_0}{\p u}(3+2\,u_1A_3)-2\,f_0^2\,u_1\frac{\p A_3}{\p u}
-3\,f_0^2A_5\right]=0.\label{usl2-2}
\end{align}

The function $Z_2$ is cubic in $u_2$ and therefore the identity $Z_2=0$ implies four equations involving also the factor $c_1$. This is why it is natural to consider two cases, $c_1=0$ and $c_1\ne0$. Moreover, in view of Lemma 9, there appears one more fork, $A_1-2A_3=0$ or $A_1-2A_3\ne0$. Thus, we have the four cases
$$
\begin{aligned}
&{\bf II.a.} \ \ c_1=0,\ \ A_3=\frac{1}{2}A_1; \qquad &&{\bf II.c.}\ \ c_1=0,\ \ A_3=\frac{1}{2}A_1+f_1;\\
&{\bf II.b.} \ \ c_1\ne0,\ \ A_3=\frac{1}{2}A_1,\qquad &&{\bf II.d.}\ \ c_1\ne0,\ \ A_3=\frac{1}{2}A_1+f_1,
\end{aligned}
$$
where $f_1=f_1(u,u_1),\ f_1\ne0$.

{\bf Case II.a.} In this case the density in  condition (\ref{c2}) can be written as
$$
\rho _2\sim u_2^3\,f_0^{-3}\left[5\,f_0^2\frac{\p^3f_0}{\p u_1^3}+5\,f_0\frac{\p f_0}{\p u_1}\frac{\p^2f_0}{\p u_1^2}-5\left(\frac{\p f_0}{\p u_1}\right)^3-16\,A_7\,f_0^3\right]+P_2(u_2).
$$
By Lemma 9 the coefficient at $u_2^3$ should vanish, and it determines the function $A_7$,
\begin{equation}\label{A7}
A_7=\frac{5}{16\,f_0^3}\left[ f_0^2\frac{\p^3f_0}{\p u_1^3}+f_0\frac{\p f_0}{\p u_1}\frac{\p^2f_0}{\p u_1^2}-\left(\frac{\p f_0}{\p u_1}\right)^3\right].
\end{equation}

In view of (\ref{A7}),  forth integrability condition (\ref{c3}) is reduced to
$$
\frac{d}{dt}\rho _3\sim u_5^2f_0\frac{d}{dx}\left[f_0^{-2}\frac{\p^2 f_0}{\p u_1^2}-2\,f_0^{-3}\left(\frac{\p f_0}{\p u_1}\right)^2\right]+P_1u_4^3+
u_4^2u_3(P_2u_2+P_3)+u_4^2O(2)+O(3)\sim 0,
$$
where $P_i$ are some functions of first differential order.
Equating the term at $u_5^2$ to zero, we obtain $f_0=\big(c\,u_1^2+\alpha(u)u_1+\beta (u)\big)^{-1}$, where $c$ is a constant, and $\alpha $ and $\beta $ are arbitrary functions. As a result, the equation $P_1=0$ holds automatically, and $P_2=0$ yields $c=0$. Thus,
$$
f_0=\big(\alpha(u)u_1+\beta (u)\big)^{-1},\ \ A_1=-\frac{5}{2}\frac{\alpha }{\alpha u_1+\beta },\ \ A_2=-\frac{5}{2}\frac{\alpha' u_1^2+\beta' u_1 }{\alpha u_1+\beta }.
$$

In view of   transformation formulas (\ref{coef}) for $A_1$ and $A_2$ one can see that the change $u\to\phi(u)$ allows one to simplify $f_0$. If $\alpha=0$, without loss of generality we put $f_0=1$; if $\beta =0$,  without loss of generality we can let
$\alpha =1$; if $\alpha \beta\ne0$, we may assume that $\beta =\alpha $.

Thus, there appear the following three non-equivalent cases,
$$
{\bf II.a.1.} \ \ f_0=1;\qquad {\bf II.a.2.}\ \ f_0=\frac{1}{u_1};\qquad {\bf II.a.3.}\ \ f_0=\frac{a(u)}{u_1+1}.
$$

{\bf Case II.a.1.} The identities $c_1=0,\,f_0=1$ lead to the relations $A_1=A_2=A_3=$ $=A_4=A_7=0$. Third integrability condition (\ref{c2}) reads as
$$
\frac{d}{dt}\rho _2\sim u_4^2\frac{d}{dx}\left(3\,A_8-\frac{\p A_5}{\p u_1}\right)+\frac{1}{5}\,u_3^3\,A_5\left(3\,A_8-\frac{\p A_5}{\p u_1}\right)+P_2(u_3)\sim 0.
$$
In this expression the coefficients at $u_4^2$ and $u_3^3$ should be equated to zero. At the same time, the density in  condition (\ref{c3}) reads as
$$
\rho _3\sim u_2^3\left(2\frac{\p A_8}{\p u_1}-\frac{\p^2 A_5}{\p u_1^2}\right)+P_2(u_2).
$$
The coefficient at $u_2^3$ should vanish by Lemma 9. The mentioned three identities imply $A_8=A_8(u),\ A_5=3\,(A_8+c_2)u_1+q_1(u)$;\
$c_2\,(A_8+c_2)=0,\ c_2\,q_1=0$, where $c_2$ is a constant.

In view of the above results we find
$$
\rho _4\sim u_2^3\,A_8'+P_2(u_2).
$$
By Lemma 9 it yields $A_8'=0$, and we hence get
$$
A_8=c_3,\ \ A_5=3\,(c_2+c_3)u_1+q_1(u);\ \ \ c_2\,(c_2+c_3)=0,\ \ c_2\,q_1=0.
$$

Now   conditions (\ref{c1}) and (\ref{c3}) are written as
$$
\begin{aligned}
&\frac{d}{dt}\rho _1\sim u_3^2\frac{d}{dx}\left(\frac{\p^2A_6}{\p u_1^2}-2\,q_1'\right)+P_5(u_2)\sim 0, \\
&\frac{d}{dt}\rho _3\sim u_4^2\frac{d}{dx}\left(\frac{\p A_9}{\p u_1}-\frac{\p^2A_6}{\p u_1^2}+\frac{9}{5}(c_2^2-c_3^2)u_1^2-\frac{6}{5}\,c_3\,q_1\,u_1-\frac{1}{5}\,q_1^2\right)
+P_3(u_3)\sim 0.
\end{aligned}
$$
Equating the expressions at $u_3^2$ and $u_4^2$ to zero, we find $A_6$ and $A_9$,
$$
A_6=(q_1'+c_4)u_1^2+q_2\,u_1+q_3,\ \ \ A_9=\frac{3}{5}(c_3^2-c_2^2)u_1^3+\frac{3}{5}c_3\,q_1u_1^2 +\frac{1}{5}(c_5+10\,q_1'+q_1^2)u_1+q_4,
$$
where $q_i=q_i(u)$ are arbitrary functions.

Then it follows from the third and fifth integrability conditions that $c_3=c_2=0$. The third integrability condition determines the
function $A_{10}$ as a polynomial of third degree in $u_1$, and the forth integrability condition determines the function $A_{11}$ as a polynomial of fifth degree in $u_1$. In order to determine the coefficients of the polynomials $A_6$, $A_9$, $A_{10}$, and $A_{11}$ we check 10 integrability conditions.
This work, being technically not difficult, requires the examination of a great number of options while solving equations. The result is $S$-integrable equations (\ref{tst1})~--~(\ref{tst8}), as well as integrable equations  being symmetries of the equations
 (\ref{list1})~--~(\ref{list4}).

{\bf Case II.a.2.} The above obtained formulas for $A_1$, $A_2$, $A_3$, and also (\ref{A4}) and (\ref{A7}) remain to be true. Substituting there $c_1=0$ and $f_0=u_1^{-1}$, we obtain
$$
A_1=-\frac{5}{2\,u_1},\quad A_2=0,\quad A_3=-\frac{5}{4\,u_1},\quad A_4=\frac{5}{u_1^2},\quad A_7=-\frac{35}{16\,u_1^3}.
$$

It is easy to check that  integrability condition (\ref{c2}) can be written as
$$
\begin{aligned}
\frac{d}{dt}\rho_2&\sim u_4^2\left[u_2\,u_1^{-1}\left(6\,A_8+6\,u_1\frac{\p A_8}{\p u_1}+\frac{\p A_5}{\p u_1}-2\,u_1\frac{\p^2 A_5}{\p u_1^2}\right)+
6\,u_1\frac{\p A_8}{\p u}+3\frac{\p A_5}{\p u}-2\,u_1\frac{\p^2 A_5}{\p u_1\p u}\right]+\\
&+\frac{1}{6}u_3^3u_2u_1^{-2}\left(78\,A_8-42\,u_1\frac{\p A_8}{\p u_1}-60\,u_1^2\frac{\p^2 A_8}{\p u_1^2}
+20\,u_1^2\frac{\p^3 A_5}{\p u_1^3}-16\,u_1\frac{\p^2 A_5}{\p u_1^2}+13\frac{\p A_5}{\p u_1}\right)+\\
&+u_3^3\,Q(u,u_1)+P_3(u_3)\sim0.
\end{aligned}
$$
Moreover,
$$
\rho _3\sim u_2^3u_1^{-2}\left(4\,u_1^2\frac{\p A_8}{\p u_1}+12\,A_8\,u_1-2\,u_1^2\frac{\p^2 A_5}{\p u_1^2}-3\,u_1\frac{\p A_5}{\p u_1}+4\,A_5\right)+P_2(u_2).
$$
The coefficients at $u_4^2$ and $u_3^3\,u_2$ in (\ref{c2}) as well as the coefficient at $u_2^3$ in $\rho_3$ should vanish. It gives us four equations, whose solution reads as
$$
A_5=q_1+\frac{q_2}{\sqrt{u_1}},\quad A_8=-\frac{q_1}{2\,u_1}-\frac{2\,q_2}{3\,u_1^{3/2}},
$$
where $q_i=q_i(u)$. A slightly more cumbersome integrability condition (\ref{c1}) yields
$$
A_6=c_2+q_3u_1+2\,c_3\sqrt{u_1}+2\,q_2'u_1^{3/2}+q_1'u_1^2.
$$

Due to these results,   conservation law (\ref{c3}) becomes
$$
\begin{aligned}
\frac{d}{dt}\rho_3&\sim u_4^2u_2\left(\frac{\p^2A_9}{\p u_1^2}+\frac{3}{u_1}\,\frac{\p A_9}{\p u_1}-\frac{q_1^2+15\,q_1'}{5\,u_1}-\frac{5\,q_2'+q_1q_2}{5\,u_1^{3/2}}
-\frac{3}{4}(c_2u_1^{-3}+c_3u_1^{-5/2})\right)+\\
&+u_4^2\left(\frac{\p^2A_9}{\p u_1\p u}\,u_1+2\frac{\p A_9}{\p u}-\frac{2}{5}\sqrt{u_1}\,(5\,q_2'+q_1q_2)'-\frac{1}{5}\,u_1(2\,q_1q_1'+15\,q_1'')
-\frac{2}{5}\,q_2q_2'+\frac{1}{4}\,q_3'\right)\\
&+P_3(u_3)\sim0.
\end{aligned}
$$
The terms with $u_4^2$ should vanish that implies
$$
A_9=c_4-\frac{1}{8}\,q_3+\frac{q_2^2}{10} +\frac{q_4}{u_1^2}+\frac{q_1^2}{15}\,u_1+q_1'u_1 +\frac{4}{25}\sqrt{u_1}\,(5\,q_2'+q_1q_2) -\frac{c_3}{\sqrt{u_1}}-\frac{3\,c_2}{4\,u_1}.
$$

Then from conditions (\ref{c1})~--~(\ref{c3}) we find $A_{10}$ and $A_{11}$, but we do not write these expressions because they are bulky.

To specify constant coefficients and the structure of the functions $q_i(u)$ we check ten integrability conditions.
These conditions are satisfied by the
equation (\ref{tst11}) and an equation being a symmetry of  equation (\ref{list9}).

{\bf Case II.a.3.} The way of arguing in this case is exactly the same as in {\bf II.a.2}, but there are small differences in the formulas. General in the case {\bf II.a} formulas become here
$$
A_1=-\frac{5}{2\,\xi },\quad A_2=\frac{5\,a'\,u_1}{2\,a },\quad A_3=-\frac{5}{4\,\xi},\quad A_4=\frac{5}{\xi^2},\quad A_7=-\frac{35}{16\,\xi^3},
$$
where $a=a(u)$ is an arbitrary function, $ \xi=u_1+1$.

Integrability condition (\ref{c2}) reads as
$$
\begin{aligned}
\frac{d}{dt}\rho_2\sim & u_4^2u_2\,\xi^{-1}\left(6\,A_8+6\,\xi\frac{\p A_8}{\p u_1}+\frac{\p A_5}{\p u_1}-2\,\xi\frac{\p^2 A_5}{\p u_1^2}+\frac{15\,a'}{2\,a\,\xi^2}\right)+
u_4^2u_1\left(6\frac{\p A_8}{\p u}+3\,\xi^{-1}\frac{\p A_5}{\p u}\right.\\
&\left.-2\frac{\p^2 A_5}{\p u_1\p u}+\frac{2\,a'}{a}\frac{\p A_5}{\p u_1}-6\frac{a'}{a}A_8-\frac{3\,a'}{a\,\xi}\,A_5+\frac{15}{2\,a^2\xi^2}\,(2\,{a'}^2-aa'')\right)+\\
&+Q_1(u,u_1)u_3^3u_2+Q_2(u,u_1)u_3^3+P_2(u_3)\sim0.
\end{aligned}
$$
This condition together with the formula for the density of  conservation law (\ref{c3})
$$
\rho _3\sim u_2^3\xi^{-2}\left(4\,\xi^2\frac{\p A_8}{\p u_1}+12\,\xi\,A_8-2\,\xi^2\frac{\p^2 A_5}{\p u_1^2}-3\,\xi\frac{\p A_5}{\p u_1}+4\,A_5-\frac{5\,a'}{a\,\xi}
\right)+P_2(u_2)
$$
and by the relation $Q_1=0$ lead us to four equations with the solution
$$
A_5=q_1+\frac{q_2}{\sqrt{\xi}},\quad A_8=-\frac{q_1}{2\,\xi}-\frac{2\,q_2}{3\,\xi^{3/2}}+\frac{4\,a'}{5\,a\,\xi^2},
$$
where $q_i=q_i(u)$. Then, by integrability condition (\ref{c1}) we determine the $A_6$, and by   integrability condition (\ref{c3}) we find the function $A_9$. After that we determine $A_{10}$ and $A_{11}$ by  conditions (\ref{c1})~--~(\ref{c3}). All these expressions involving arbitrary functions of $u$ are rather bulky and we omit them.

It follows from the fifth integrability condition $\ds \frac{d}{dt}\rho_4\sim0$ that $a'=0,\,q_1=0,\,q_2'=0$ and so forth. Only in the $A_{11}$ their remain two arbitrary functions of $u$. The integrability conditions 5~--~7 yield a vast system of algebraic equations for the constants and for the two remaining functions. It follows from this system that all the functions $A_i$ are independent of $u$. This is why we can apply a transformation $u\to u-x$ leading to the case {\bf II.a.2.} Thus, in the considered case there are no new integrable equations.

{\bf Case II.b} differs from the previous ones by that
canonical conservation law (\ref{c1}) has the second order. It follows from   condition (\ref{c1}) that
$$
f_0=-\frac{5}{2c_1}(u_1^2+a(u)u_1+b(u))^{-1},\qquad ab'=2a'b,
$$
and the functions $A_5$, $A_7$, $A_8$, and $A_9$ are expressed in terms of $f_0$,
$$
\begin{aligned}
&A_5=\frac{15}{2\,f_0}\frac{\p^2 f_0}{\p u\p u_1}\,u_1+\frac{5}{f_0^2}\frac{\p f_0}{\p u}\left(f_0-\frac{\p f_0}{\p u_1}\,u_1\right), \\
&A_7=\frac{c_1}{4}\frac{\p f_0}{\p u_1}+\frac{5}{8\,f_0}\frac{\p^3f_0}{\p u_1^3}-\frac{35}{32\,f_0^2}\frac{\p^2f_0}{\p u_1^2}\frac{\p f_0}{\p u_1}
+\frac{5}{8\,f_0^3}\left(\frac{\p f_0}{\p u_1}\right)^3,\\
&A_8=\frac{5}{24\,f_0^2}\left(14\,f_0-3\frac{\p f_0}{\p u_1}\,u_1\right)\frac{\p^2 f_0}{\p u\p u_1}
+\frac{5\,u_1}{12\,f_0^2}\left(5\,f_0\frac{\p^3f_0}{\p u\p u_1^2}-\frac{\p^2f_0}{\p u_1^2}\frac{\p f_0}{\p u}\right)+\\
&\qquad+\frac{c_1}{3}\frac{\p f_0}{\p u}\,u_1 -\frac{5}{24\,f_0^3}\frac{\p f_0}{\p u}\frac{\p f_0}{\p u_1}\left(3\,f_0+8\,\frac{\p f_0}{\p u_1}\,u_1\right).
\end{aligned}
$$
The formula for $A_9$ is omitted since it is bulky.

Taking into account formulas (\ref{coef}) and an explicit form of the functions $A_1$ and $A_2$, one can observe that if functions $a$ and $b$ are nonzero, then we can make them constant by a point transformation $u\to\phi(u)$. If $a=0$, then up to a point transformation we have either
$b=0$ or $b=1$. If $a\ne0$, by letting $a=2$ we get $b'=0$. Thus, there are the following three possible cases,
$$
\begin{aligned}
&{\bf II.b.1.}\ \ f_0=-\frac{5}{2c_1u_1^2};\qquad {\bf II.b.2.}\ \ f_0=-\frac{5}{2c_1(u_1^2+1)};\\
&{\bf II.b.3.}\ \ f_0=-\frac{5}{2c_1}((u_1+1)^2+c)^{-1},
\end{aligned}
$$
where $c$ is a constant.

{\bf Case II.b.1.} It follows from  conditions (\ref{c1}) and (\ref{c3}) that
$$
\begin{aligned}
&A_5=A_8=0,\ \ A_6=c_2+q_1u_1^2+q_2u_1^{-2},\ \ A_7=-\frac{45}{8}\,u_1^{-3},\\
&A_{10}=q_1'u_1^3+q_2'u_1^{-1},\ \ A_9=-\frac{1}{2}\,q_1u_1-\frac{3}{2}\,c_2u_1^{-1}-\frac{5}{2}\,q_2u_1^{-3},\\
&A_{11}=\frac{1}{5}\left(q_1''+\frac{3}{10}\,q_1^2\right)+\frac{c_2}{5}\,q_1u_1^3-\frac{3}{5}\,c_2\,q_2u_1^{-1}-\frac{1}{10}\,q_2^2u_1^{-3}
+\left(\frac{1}{15}\,q_1q_2-\frac{1}{3}\,q_2''\right)u_1+q_3,
\end{aligned}
$$
where $q_i=q_i(u)$, $c_2$ is a constant.

The check of the conditions 6~--~10 shows that there exist only two integrable equations, which are the symmetries of  equations (\ref{list7}) and (\ref{list8}).

{\bf Case II.b.2.} It follows from   conditions (\ref{c1}) and (\ref{c3}) that
$$
\begin{aligned}
&A_5=A_8=0,\ \ A_6=q+c_2u_1\sqrt{u_1^2+1}+(3\,q+c_3)u_1^2,\ \ A_7=\frac{5}{8}\,u_1 \frac{19-9\,u_1^2}{(u_1^2+1)^3},\ \ q=q(u),\\
&A_9=\frac{3}{2}\,u_1\frac{2\,q+c_3}{u_1^2+1} +\frac{c_2}{\sqrt{u_1^2+1}}-\frac{1}{2}\,c_2\sqrt{u_1^2+1} -\frac{1}{2}\,(3\,q+c_3)\,u_1,\ \ A_{10}=q'u_1(3\,u_1^2+2),\\
&A_{11}=\frac{3}{25}\,c_2(3\,q+c_3)(u_1^2+1)^{5/2} -\frac{1}{5}\,c_2(2\,q+c_3)(u_1^2+1)^{3/2} +\frac{1}{10}(3\,q^2+2\,c_4\,q)\,u_1+\\
&\qquad+\frac{3}{50}\big(10\,q'' +(3\,q+c_3)^2+c_2^2\big)u_1^5+\frac{1}{10}\big(5\,q'' +6\,q^2+5\,c_3\,q+c_2^2\big)\,u_1^3+c_5,\ \ c_2\,q'=0.
\end{aligned}
$$

The check of the conditions 6~--~10 shows that there exist only two integrable equations, which are the symmetries of equations (\ref{list5}) and (\ref{list6}).

{\bf Case II.b.3.} Second integrability condition (\ref{c1}) allows us to show that all the functions $A_i$ are independent of $u$. This is why by the transformation $u\to u-x$ equation (\ref{eq1bb}) is reduced to the equations from the cases {\bf II.b.1} if $c=0$ and {\bf II.b.2} if $c\ne0$.
Thus, in the considered case there is no new integrable equations.

{\bf Case II.c.} The density in (\ref{c2}) is equivalent to the cubic in $u_2$ expression (\ref{r2-3}). The condition $\psi_4=0$ allows us to express
$A_7$ in terms $f_0$ and $f_1$. Then we find
$$
\frac{d}{dt}\rho_2\sim (Z_1u_2+Z_2)\,u_4^2+O(3).
$$
From $Z_1=0,\,Z_2=0$ we deduce two equations of the form
$$
\frac{\p A_8}{\p u_1}=F_1(f_0,f_1),\qquad \frac{\p A_8}{\p u}=F_2(f_0,f_1),
$$
which can be integrated explicitly. Substituting $A_7$ and $A_8$ in all the expressions, we get $\rho_3\sim \alpha \,u_3^2+O(2)$. Since $2\,A_3-A_1=2\,f_1\ne0$, by Lemma 9 we have $\alpha =0$ that yields the Riccatti equation
$$
\frac{\p f_1}{\p u_1}=\phi_1(f_0)f_1^2+\phi_2(f_0)f_1+\phi_3(f_0),
$$
where $\phi_2$ and $\phi_3$ depend both on $f_0$ and on the first and second order derivatives of $f_0$ w.r.t. $u_1$.

As above, we obtain $\rho_4\sim u_3^2(Q_1u_2+Q_2)+O(2)$ and consider the equations $Q_1=0,\,Q_2=0$. The second of these equations determines $A_5$, and the first implies an ordinary differential equation with the derivatives of $f_0$ w.r.t. $u_1$ containing $f_1$. By the forth integrability condition
$$
\begin{aligned}
&\frac{d}{dt}\rho_3\sim u_5^2(P_1\,u_2+P_2)+u_4^3\,P_3+u_4^2u_3(P_4\,u_2+P_5) +u_4^2(P_6\,u_2^2+P_7\,u_2+P_8)+u_3^4P_9+
\\
&\qquad+u_3^3(P_{10}\,u_2^3+P_{11}\,u_2^2+P_{12} \,u_2+P_{13})+u_3^2O(2)+O(2)\sim0
\end{aligned}
$$
we obtain equations $P_i=0$, $i=1,\dots,13$, among those there are many equations involving only $f_1$, $f_0$, and the derivatives of $f_0$ w.r.t. $u_1$. Expressing all the derivatives of $f_0$ from some of the equations and substituting them in other equations, we obtain the contradiction $f_0f_1=0$.

It means that under the conditions {\bf II.c} there exist no integrable equations.

{\bf Case II.d.} We recall that in this case the functions $A_1$ and $A_2$ read as (\ref{case2}) that ensures the triviality of the first canonical conservation law. The function $A_4$ is given by formula (\ref{A4}), and since $c_1\ne0$, we have extra two equations (\ref{usl2-1}) and (\ref{usl2-2}). Moreover, $A_3=A_1/2+f_1,\,f_1\ne0$.

After the exclusion of $A_3$,   equation (\ref{usl2-1}) casts into the form
\begin{equation}\label{ed2-1}
15\,f_0\,\frac{\p^2 f_0}{\p u_1^2}-30\left(\frac{\p f_0}{\p u_1}\right)^2+20f_0\,f_1\,\frac{\p f_0}{\p u_1}-28\,f_0^2\frac{\p f_1}{\p u_1}-12\,c_1f_0^3=0,
\end{equation}
and   equation (\ref{usl2-2}) allows us to express $A_5$ in terms of $f_0$ and $f_1$,
\begin{align}\label{ed2-2}
A_5=\frac{15}{2\,f_0}\,u_1\,\frac{\p^2 f_0}{\p u\p u_1}-\frac{5}{3\,f_0^2}\,\frac{\p f_0}{\p u}\left(3\,u_1\frac{\p f_0}{\p u_1}-3\,f_0-2\,u_1\,f_0\,f_1\right)-\frac{2}{3}\,u_1\,\frac{\p f_1}{\p u}.
\end{align}

In view of said above,   second integrability condition (\ref{c1}) becomes
$$
\frac{d}{dt}\rho_1\sim u_3^2(Z_1\,u_2^3+Z_2\,u_2^2+Z_3\,u_2+Z_4)+Z_5\,u_2^7+P_6(u_2)\sim0.
$$
We express $A_7$ from the equation $Z_1=0$,
\begin{align}\label{ed2-3}
A_7=&\frac{55\,f_0^{-1}}{112}\,\frac{\p^3 f_0}{\p u_1^3}-\frac{f_0^{-2}}{1568}\left(185\,\frac{\p f_0}{\p u_1}+84\,f_0\,f_1\right)\frac{\p^2f_0}{\p u_1^2}-\\
&-\frac{f_0^{-3}}{392}\left(205\left(\frac{\p f_0}{\p u_1}\right)^3-230\,f_0\,f_1\left(\frac{\p f_0}{\p u_1}\right)^2-44\,c_1\,f_0^3\frac{\p f_0}{\p u_1}\right),\nonumber
\end{align}
and $A_8$ from the equation $Z_2=0$,
\begin{align}\label{ed2-4}
A_8=&\frac{55}{28}\,f_0^{-1}\,u_1\frac{\p^3 f_0}{\p u\p u_1^2}-\frac{5\,f_0^{-2}}{84}\,u_1\frac{\p f_0}{\p u}\,\frac{\p^2 f_0}{\p u_1^2}
+\frac{f_0^{-1}}{126}(30\,c_1\,u_1\,f_0-7\,f_1)\frac{\p f_0}{\p u}
\\
&-\frac{f_0^{-2}}{168}\left(25\,u_1\frac{\p f_0}{\p u_1}-490\,f_0+36\,u_1\,f_0\,f_1\right)\frac{\p^2 f_0}{\p u\p u_1}
-\frac{55}{21}\,f_0^{-3} u_1\left(\frac{\p f_0}{\p u_1}\right)^2\frac{\p f_0}{\p u} \nonumber\\
&+\frac{5\,f_0^{-2}}{504}(272\,u_1\,f_1-63)\frac{\p f_0}{\p u_1}\frac{\p f_0}{\p u}-\frac{f_0^{-1}}{63}\,\frac{\p f_1}{\p u}\left(31\,u_1\,\frac{\p f_0}{\p u_1}-7\,f_0\right).
\nonumber
\end{align}

By the equations $Z_3=0$ and $Z_4=0$ we can express the functions $A_9$ and $A_{10}$, respectively, in terms of the functions $f_0$, $f_1$, $A_6$ and their derivatives.
We omit these expressions since they are cumbersome.

The equation $Z_5=0$ is an ordinary fifth order differential equation for $f_0$ w.r.t. the variable $u_1$. Other implications of the second integrability condition involve the derivatives of the functions $f_0$, $f_1$, $A_6$, and $A_{11}$ w.r.t. two variables $u_0$ and $u_1$ and are too complicated for the analysis.

Then, by Lemma 9 it follows from  expressions (\ref{r2-3}) for $\rho_2$ and $\rho_3$ that $\psi_4=0$ and $\psi_8=0$. These two equations read as
\begin{align}\label{ed2-5}
&70\,f_0^2\frac{\p^3 f_0}{\p u_1^3}-f_0\left(405\frac{\p f_0}{\p u_1}-28\,f_0\,f_1\right)\frac{\p^2 f_0}{\p u_1^2}\\
&\qquad+6\frac{\p f_0}{\p u_1}\left(65\left(\frac{\p f_0}{\p u_1}\right)^2-6\,f_0\,f_1\frac{\p f_0}{\p u_1}-2\,c_1\,f_0^3\right)=0,\nonumber\\
&25\,f_0\frac{\p^2 f_0}{\p u_1^2}-10\frac{\p f_0}{\p u_1}\left(5\frac{\p f_0}{\p u_1}-f_0\,f_1\right)+f_0^2(15\,c_1\,f_0+28\,f_1^2)=0.\label{ed2-6}
\end{align}

Expressing the second derivative of $f_0$ from (\ref{ed2-6}) and substituting it into (\ref{ed2-5}), in view of (\ref{ed2-1}) we obtain
$f_0=-4/(5c_1)f_1^2$. After the exclusion of $f_0$,  equations (\ref{ed2-1}) and (\ref{ed2-6}) are reduced to the equation
\begin{equation}\label{ed2-7}
25\,f_1\frac{\p^2 f_1}{\p u_1^2}-75\left(\frac{\p f_1}{\p u_1}\right)^2+10\,f_1^2\frac{\p f_1}{\p u_1}+8\,f_1^4=0,
\end{equation}
and   equation (\ref{ed2-5}) together with the mentioned equation $Z_5=0$ are the implications of  equation (\ref{ed2-7}). Substituting $f_1={5}/({4\,f})$ into (\ref{ed2-7}), we obtain the equation
\begin{equation}\label{ed2-8}
2\,\frac{\p}{\p u_1}\left(f\,\frac{\p f}{\p u_1}\right)+\frac{\p f}{\p u_1}=1,
\end{equation}
whose general integral is written as
\begin{equation}\label{E}
(f+u_1+a)^2(2f-u_1-a)+b=0,
\end{equation}
where $a$ and $b$ are arbitrary functions of the variable $u$.

In view of all obtained results including differential consequences of   equation (\ref{E}), it is easy to check that
$$
\frac{d}{dt}\rho_1\sim u_2^6\,f^{-10}(3a+3u_1-5\,f)(3a'b-ab')+P_5(u_2),
$$
where the prime denotes the derivative w.r.t. $u$. Thus, $3\,a'\,b=ab'$ that implies ${a=c\,b^{1/3},\,c=}$~const if $b\ne0$.

This result allows one to convert both these functions into constants by a point transformation $u=\phi(v)$. Indeed, since $2\,f_1 =2\,A_3-A_1$, in accordance with  formulas (\ref{coef}), $\t f_1(v)=\phi'f_1(u)$. Therefore, the
function $f\sim f_1^{-1}$ is transformed by the law $\t f(v)={\phi'}^{-1}f(u)$. Making the transformation in  equation (\ref{E}), we obtain
\begin{equation}\label{ellu}
[\t f+v_1+a(u){\phi'}^{-1} ]^2[2\t f-v_1-a(u){\phi'}^{-1}]+b(u){\phi'}^{-3}=0.
\end{equation}

If $a=b=0$, then no transformation is needed and we have
$$(f+u_1)^2(2f-u_1)=0.$$
If $b=0$ and $a\ne0$, then letting $\phi'=a$, we reduce equation (\ref{E}) to
$$(f+u_1+1)^2(2f-u_1-1)=0.$$

If $b(u)\ne0$, then $a(u)=k\,b^{1/3}(u)$, where $k$ is a constant. Choosing $\phi'=b^{1/3}$, we obtain equation (\ref{E}) in the form
\begin{equation}\label{E2}
(f+u_1+a)^2(2f-u_1-a)+1=0,
\end{equation}
where $a$ is a constant.

Thus, up to a point transformation, the quantities $a$ and $b$ in (\ref{E}) are constants, and the following three cases are possible,
$$
{\bf II.d.1.}\ \ f=-u_1-a;\quad {\bf II.d.2.}\ \ f=\frac12(u_1+a);\quad {\bf II.d.3.}\ \ f(u_1)\ \text{satisfies (\ref{E2})}.
$$
In each of these cases the parameter $a$ takes one of the values, $a=0$ or $a=1$.

Employing equation (\ref{ed2-8}), we can exclude higher derivatives of $f$ from the expressions for the functions $A_i$ found above. It leads us to rather compact expressions,
$$
\begin{aligned}
&A_1=-\frac{5}{f} f',\ \ \ A_2=0,\ \ \ A_3=\frac{5}{4f}(1-2f'),\ \ \ A_4=\frac{5}{4f^2}(16{f'}^2-3),\\
&A_5=0,\ \ \ A_7=-\frac{5}{16}f^{-3}(2f'-1)(28{f'}^2+20f'+1),\ \ \ A_8=0, \\
& A_9=\frac12 f^2 \frac{\p^3 A_6}{\p u_1^3}+\frac14 f (6f'+1)\frac{\p^2 A_6}{\p u_1^2}-\frac14 \frac{\p A_6}{\p u_1}- \frac{3f'}{2f}A_6,\\[2mm]
&A_{10}=u_1f^2 \frac{\p^3 A_6}{\p u\p u_1^2}+\frac12 u_1f(2f'+1) \frac{\p A_6}{\p u\p u_1}-\frac12 (f+u_1+2ff')\frac{\p A_6}{\p u},
\end{aligned}
$$
and these formulas hold true for each of three cases II.d.1, II.d.2 and II.d.3.

{\bf Case II.d.1.} If $a=1$, then it follows from the integrability conditions that ${A_i=A_i(u_1),\forall i}$. Thus, the transformation $u\to u-x$ is admissible, and we arrive at the case $a=0$. In the case $a=0$ numerous forks lead to the only integrable equation (\ref{tst10}).

{\bf Case II.d.2.} If $a=1$, then exactly as in the previous case we arrive at the case $a=0$, and in the case $a=0$ we get equation (\ref{tst9}).

{\bf Case II.d.3.} Let us consider this case in more details. In view of the above results, the second and forth integrability conditions read as
\begin{align}\label{ro1}
\frac{d}{dt}\rho_1\sim& u_2^5Q_1+u_2^4Q_2+u2^3Q_3+u_2^2Q_4+O(1)\sim 0,\\
\frac{d}{dt}\rho_3\sim&u_4^2(P_1+u_2P_2)+u_3^3(P_3+u_2P_4) +\\
&+u_3^2(P_5+u_2P_6+u_2^2P_7+u2^3P_8)+u_2^7P_9+P_6(u_2)\sim 0.\label{ro3}\nonumber
\end{align}
Here the functions $Q_i$ and $P_j$ depend on $u_0$ and $u_1$ only. For the equivalence of these expressions to zero one needs the identities
$Q_i=0$, $P_j=0$ for all $i,j$. The conditions $Q_1=0$, $P_2=0$, $P_4=0$, $P_8=0$, and $P_9=0$ are homogeneous
ordinary differential equations for the functions
$A_6(u_1)$, and $u$ in involved as a parameter. The first two equations are of the fifth order, the orders of the others are 6, 7, and 9, respectively. By excluding higher derivatives from the first two equations, we arrive at the
equation
$$
2f^2f'\frac{\p^2 A_6}{\p u_1^2}+f(f'+1)(2f'-1)\frac{\p A_6}{\p u_1}+(1-3f')A_6=0.
$$
All remaining equations are its differential consequences. The general solution of the above equation is given by
\begin{equation}\label{A6}
A_6=\gamma (u)(f+u_1+a)^2+10\, \omega(u)(u_1+a)f,
\end{equation}
where $\gamma$ and $\omega $ are arbitrary functions.

Substituting solution (\ref{A6}) into the equation $Q_2=0$, we obtain
\begin{equation}\label{EE}
70\,a\omega'(u_1+a) f^3+a\,\gamma'\Big[7f^4+14(u_1+a) f^3+7(u_1+a)^2 f^2+f-u_1-a\Big]=0.
\end{equation}
Calculating the resultant of the polynomials
(\ref{EE}) and (\ref{E2}) w.r.t. the variable $u_1$ yields
\begin{align*}
&a\Big[34300\,\omega'^2(20\,\omega'+9\gamma')f^{12}+980\,\omega'(165\,\omega'\gamma
'+350\,\omega'^2+3\gamma'^2)f^9\\
&\qquad\qquad-7\gamma'(930\,\omega'\gamma'+2100\,\omega'^2-\gamma'^2)f^6+\gamma'^2(210\,\omega'+59\gamma')f^3-\gamma
'^3\Big]=0.
\end{align*}
Since $\omega$ and $\gamma$ are functions of $u$, and $f$ is a non-constant function of $u_1$, all the coefficients of this polynomial should vanish. It implies
$$
a\gamma'=0,\qquad a\omega'=0.
$$

Next we consider the equations involving $A_{11}$. These are $Q_3=0$, $Q_4=0$, $P_5=0$, and $P_6=0$. First two of them have the second order, from the first we can express $\p^2 A_{11}/\p u_1^2$, and $\p^2 A_{11}/\p u\p u_1$ from the second. By $Q_3=0$ we can exclude $A_{11}$ from $P_6=0$, and it implies the equation for the function $\omega$,
$$\quad \omega''=6\omega ^2.$$
Together with $a\omega' =0$ it implies that $a\omega =0$.

Excluding higher derivatives of $A_{11}$ from $P_5=0$, we obtain the equation
$$
 \frac{\p A_{11}}{\p u}=P_1(f,u_1,\gamma',\omega')P_2^{-1}(f,u_1,\gamma',\omega'),
$$
where $P_1$ and $P_2$ are polynomials in the variables $f$ and $u_1$, such that $P_1(f,u_1,0,0)=0$ and $P_2(f,u_1,0,0)\ne0$. It means that if $a\ne0$ and $\gamma'=\omega'=0$, then
$A_6$ and $A_{11}$ are independent of $u$. It leads to the fact that all $A_i$ depend only on $u_1$. Thus, the transformation $u\to u-ax$ is admitted and it eliminates $a$ in  equation (\ref{E2}). Therefore, it is sufficient to consider only the case $a=0$.

As $a=0$, the equations for $A_{11}$ become not very bulky,
$$
\begin{aligned}
\frac{\p^2 A_{11}}{\p u_1^2}=&\frac{1}{2\,f}\,(\gamma''+4\,\gamma\omega) (8\,{f}^{4}+16\,{u_1}\,{f}^{3}+8\,{{u_1}}^{2}{f}^{2}+ 3\,f+{u_1})\\[2mm]
&\qquad+\frac{2\,{\gamma}^{2}}{5\,f} (9\,{f}^{4}+ 18\,{u_1}\,{f}^{3}+9\,{{u_1}}^{2}{f}^{2}+4\,f+{u_1} )\\[2mm]
&\qquad -\,\frac{10}{\,f}{\omega}^{2} (14\,{f}^{4}+23\,{u_1}\,{f}^{3}-31\,{{u_1}}^{2}{f}^{2} +9\,f-3\,{u_1}),\\[4mm]
 \frac{\p A_{11}}{\p u}=&40\,{f}^{2}\frac {{\omega'}\,\omega\,
(2\,{f}^{6}+{f}^{5}{u_1}-{f}^{4}{{u_1}}^{2} -3\,{f}^{3}-7\,{u_1}\,{f}^{2}-2)}{2\,{f}^{3}+{u_1}\,{f}^{2}-{{u_1}}^{2}f+1}\\[2mm]
&-2\,{f}^{2}\frac {( 2\,{f}^{3}+3\,{u_1}\,{f}^{2}+{{u_1}}^{2}f+1)( 2\,\gamma{\omega'}+3\,{\gamma'}\,\omega)}{2\,{f}^{3}+{u_1}\,{f}^{2}-{{u_1}}^{2}f+1}.
\end{aligned}
$$
Integrating the first equation,\footnote{The method of integration is described in Appendix 2.} we obtain
$$
\begin{aligned}
A_{11}=&\frac{1}{10}\,( 8\,{f}^{5}+2\,{f}^{2}-{{u_1}}^{2}+16\,{u_1}\,{f}^{4}+8\,{{u_1}}^{2}{f}^{3})( 4\,\gamma\omega+{\gamma''})\\
&+\frac{1}{25}\,{\gamma}^{2}( 18\,{{u_1}}^{2}{f}^{3}+36\,{u_1}\,{f}^{4}+7\,{f}^{2}+18\,{f}^{5}-{{u_1}}^{2})\\
&-4\,{\omega}^{2} ( 2\,{f}^{5}+3\,{f}^{2}-{u_1}\,{f}^{4}-8\,{{u_1}}^{2}{f}^{3}+{{u_1}}^{2}) +\alpha\,{u_1}+\beta,
\end{aligned}
$$
where $\alpha $ and $\beta $ are arbitrary functions of $u$. It follows from the second equation that $\alpha $ and $\beta $ are constants. Using the Galilean transformation, we can assume that $\alpha =0$.

Substituting the expression for $\p A_{11}/\p u$ into the equation $Q_4=0$, we arrive at extra two equations for $\gamma $ and $\omega$. Finally the system of equations for these functions reads as
\begin{align}
&\omega''=6\omega ^2, \label{ell0}\\
&\gamma'''=8\gamma'\omega +4\gamma \omega',\label{ell1}\\
&(\gamma +15\omega){ \gamma\,}'+10(\gamma +10\omega) \omega'=0, \label{ell2}
\end{align}

Finding all these functions $A_i$ explicitly and employing equations (\ref{ell0})~--~(\ref{ell2}), it is easy to check completely the integrability conditions 1--4. These conditions lead us to the only restriction $\beta \omega'=0$, where $\beta$ is the constant involved in $A_{11}$.

If $\omega'\ne0$, then $\beta=0$. If $\omega' =0$,   it follows from (\ref{ell0}) that $\omega =0$, and from (\ref{ell2}) that ${\gamma=const}$. In this case the coefficients of   equation (\ref{eq1bb}) are independent of $u$, and the transformation $u\to u+{\beta t}$ eliminating the constant $\beta $ in $A_{11}$ is admissible. Thus, $\beta=0$ for all $\omega$ and $\gamma $.

If $\omega =0$,   letting $\gamma=5\mu$, we obtain   equation (\ref{tst12m}).

If $\omega\ne0,$   it follows from (\ref{ell0}) that $\omega' \ne0 $. In this case the order of   equation (\ref{ell0}) lowers and we obtain the equation ${\omega'}^2=4\,\omega ^3+c$ coinciding with (\ref{Waier}). Since $\omega' \ne0 $, then (\ref{ell2}) implies $\gamma +15\,\omega\ne 0$, and hence one can express $\gamma'$ from (\ref{ell2}). It allows us to exclude the derivatives of the functions $\gamma $ and $\omega$ from (\ref{ell1}). As a result we obtain the equation
\begin{equation}\label{eom}
(\gamma +30\,\omega)(\gamma +5\,\omega)(\gamma +20\,\omega)\big[(\gamma+20\,\omega)(\gamma+5\,\omega)^2+125\,c\big]=0,
\end{equation}
where $c$ is a constant in (\ref{Waier}).

If $\gamma =-30\,\omega$, then it follows from (\ref{ell2}) that $\omega=0$, which contradicts the assumption. If $\gamma =-5\,\omega $, then we get the
equation (\ref{tst13m}), and if $\gamma =-20\,\omega $, then we obtain  equation (\ref{tst14m}).

Consider the case
\begin{equation}\label{last}
 (\gamma+20\,\omega)(\gamma+5\,\omega)^2+125\,c=0.
\end{equation}
Cubic curve (\ref{last}) is rational and is parametrized by the substitution
$$
\omega =\t\omega +\t c\,\t\omega ^{-2},\quad \gamma =-5\,\t c\,\t\omega ^{-2}-20\,\t\omega,\quad
$$
where $c=-27\,\t c.$ Substituting these expressions into (\ref{ell0})~--~(\ref{ell2}), we find that $\t\omega $ satisfies equation (\ref{Waier}) with the constant $\t c$ instead of $c$. Substituting the found functions $A_i$ into  equation (\ref{eq1bis}), employing the expressions for $\omega$ and $\gamma$, and redenoting $\t\omega \to\omega$, $\t c\to c$, we obtain  equation (\ref{tst15m}).


\subsection{Differential substitutions relating equations in the list}\rule{1cm}{0pt}

As it was noted in Section 2.4, while calculating differential substitutions, it is useful to know the orders of the canonical conservation laws. In Table 2 we provide the
orders of several canonical conservation laws for   equations in list (\ref{tst1})~--~(\ref{tst15m}).

Even densities are not indicated in Table 2, since they happened to be trivial $\rho_{2n}\sim0$. For equation (\ref{tst11}) the orders of $\rho_1$ and $\rho_9$ are provided for the case of generic constants; if
$\mu=0$, then $\rho_1\sim0$, $\rho_9\sim0$.

The differential substitutions admitted by fifth order $S$-integrable equations are shown on Figure 2.

Below we provide the substitutions for the equations with generic constants.

\noindent
{\bf(\ref{tst12m})$\to$(\ref{tst5})}: $\t u=\ds \frac{u_2}{2f}+\sqrt{-\mu}\,(f+u_1)$.

\noindent
{\bf(\ref{tst14m})$\to$(\ref{tst8})}: $\t u=\ln(f+u_1)-\ln \phi$. At that, $\ds \omega =\frac{\lambda_1^2}{4\,\phi^2}+\frac{1}{2}\,\lambda_2\,\phi$, and the constant $c$ in the
equation (\ref{Waier}) satisfied by $\omega $ equals $\ds c=-\frac{27}{16}\,\lambda_1^2 \lambda_2^2$.

\medskip
\begin{minipage}[t]{150mm}
{\small Table 2.
The orders of canonical conservation laws. For zero order conservation laws we indicate in the brackets to what the density is equivalent \label{tab2}
}
\begin{center}
\begin{tabular}{|c|c|c|c|c|c|c|c|c|}
\hline
$\rho_i$&(\ref{tst1})&(\ref{tst2})& (\ref{tst3}) & (\ref{tst4}) & (\ref{tst5}) &(\ref{tst6}) &(\ref{tst7})&(\ref{tst8}) \\
\hline
$\rho_1$&0, ($\sim u$)&0, ($\sim u$) &0, ($\sim0$)&0, ($\sim0$)&0, ($\sim u^2$)& 1 & 1 & 1 \\
\hline
$\rho_3$&$\sim 0$&$\sim 0$& 1 &$\sim 0$&$\sim 0$&$\sim 0$&$\sim 0$&$\sim 0$ \\
\hline
$\rho_5$ & 1 & 1 & 2 & 2 &2 & 3 &3 & 3 \\
\hline
$\rho_7$ & 2 & 2 & 3 & 3 & 3 & 4 &4 & 4 \\
\hline
$\rho_9$&$\sim 0$&$\sim 0$& 4 &$\sim 0$&$\sim 0$&$\sim 0$&$\sim 0$ &$\sim 0$\\
\hline
$\rho_{11}$& 4 & 4 & 5 & 5 & 5 & 6 &6 & 6 \\
\hline
\hline
$\rho_i$&(\ref{tst9})& (\ref{tst10})&(\ref{tst11})&(\ref{tst12m})&(\ref{tst13m}) &(\ref{tst14m}) &(\ref{tst15m}) &\\
\hline
$\rho_{1}$& 2 & 2 & 1 & 2 & 2 & 2 & 2 & \\
\hline
$\rho_3$&$\sim 0$&$\sim 0$&$\sim 0$&$\sim 0$&$\sim 0$&$\sim 0$&$\sim 0$& \\
\hline
$\rho_{5}$& 4 & 4 & 3 & 4 & 4 & 4 & 4 &\\
\hline
$\rho_{7}$& 5 & 5 & 4 & 5 & 5 & 5 & 5 &\\
\hline
$\rho_9$&$\sim 0$&$\sim 0$& 3 &$\sim 0$&$\sim 0$&$\sim 0$&$\sim 0$& \\
\hline
$\rho_{11}$& 7 & 7 & 6 & 7 & 7 & 7 & 7 &\\
\hline
\end{tabular}\\[1mm]
\end{center}
\end{minipage}\\[5mm]

\noindent
{\bf(\ref{tst9})$\to$(\ref{tst8})}: $\t u=\ln u_1$, \quad $\mu_1=\lambda_2, \quad \mu_2=-\lambda_1^2$.

\noindent
{\bf(\ref{tst15m})$\to$(\ref{tst8})}: $\t u=\ln(f+u_1)-\ln(2\,\omega \lambda_2^{-1}), \quad c=\ds\frac{1}{4}\,\lambda_1^2 \lambda_2^2$.

\noindent
{\bf(\ref{tst15m})$\to$(\ref{tst7})}: $\ds\t u=\ln(f+u_1)+\frac{1}{2}\ln(-4\,\omega \lambda_1^{-1}), \quad c=\frac{1}{4}\,\lambda_1 \lambda_2^2 $.

\noindent
{\bf(\ref{tst15m})$\to$(\ref{tst5})}: $\ds\t u=\frac{u_2}{2f}-\frac{f\omega'}{\omega }+\frac{1}{2\omega }(3\sqrt{c}-\omega')u_1$.

\noindent
{\bf(\ref{tst13m})$\to$(\ref{tst7})}: $\ds\t u=\ln(f+u_1)+\frac{1}{2}\ln \phi$. At that, $\omega =-\lambda_1\,\phi+4\,\lambda_2^2\,\phi^{-2},\ c=-108\,\lambda_1^2 \lambda_2^2$.

\noindent
{\bf(\ref{tst10})$\to$(\ref{tst7})}: $\ds\t u=-\frac{1}{2}\ln u_1, \quad \mu_1=\lambda_1, \quad \mu_2=- \lambda_2^2$.

\noindent
{\bf(\ref{tst6})$\to$(\ref{tst5})}: $\ds\t u=u_1$.

\noindent
{\bf(\ref{tst8})$\to$(\ref{tst1})}: $\ds\t u=-u_2-u_1^2\pm3\,\lambda_1 e^uu_1-\lambda_1^2e^{2\,u}+\lambda_2e^{-u}$.

\noindent
{\bf(\ref{tst11})$\to$(\ref{tst5})}: $\ds\t u=\sqrt{u_1}-\mu$. At that,  equation (\ref{tst11}) should involve an additional term $5\,\mu^4u_1$.

\noindent
{\bf(\ref{tst7})$\to$(\ref{tst2})}: $\ds\t u=2\,u_2-u_1^2\pm6\,\lambda_2e^{-2\,u}u_1+\lambda_1 e^{2\,u}-\lambda_2^2e^{-4\,u}$.

\noindent
{\bf(\ref{tst5})$\to$(\ref{tst1})}: $\ds\t u=-u_1-u^2$.

\noindent
{\bf(\ref{tst5})$\to$(\ref{tst2})}: $\ds\t u=2\,u_1-u^2$.

\noindent
{\bf(\ref{tst3})$\to$(\ref{tst1})}: $\ds\t u=u_1$.

\noindent
{\bf(\ref{tst4})$\to$(\ref{tst2})}: $\ds\t u=u_1$.

Moreover, there also exist the substitutions for special values of the parameters involved in the equations.

\examp {\bf(\ref{tst7})$\to$(\ref{tst5})}: $\ds\t u=u_1+\sqrt{-\lambda_1}\,e^u\pm \lambda_2\,e^{2\,u}, \ \lambda_1 \lambda_2=0$. In each of the cases $\lambda_2=0$ or $\lambda_1=0$, the logarithmic substitution $u\to-\ln u$ or $u\to-\dfrac12\ln u$ leads us to a first order linear equation for $u$. That is,  the function $u$ can be expressed in terms of $\widetilde{u}$ by one quadrature.

If here we let $\lambda_2=0,\ \lambda_1\to-\lambda_1^2$, then we obtain substitution {\bf(\ref{tst8})$\to$(\ref{tst5})} with $\lambda_2=0$.

\examp {\bf(\ref{tst12m})$\to$(\ref{tst7})}: $\t u=\ln(f+u_1),\ \lambda_2=0,\ \lambda_1=\mu $. Here one can also express $u$ in terms of $\t u$ by one quadrature. Indeed, as one can check easily,   third degree curve (\ref{alg1}) has the  parametric representation
$$
u_1=\frac{1}{3}\left(2\,e^v+e^{-2\,v}\right),\qquad f=\frac{1}{3}\left(e^v-e^{-2\,v}\right),
$$
at that, $v=\t u$. Thus, we have $\ds u=\frac{1}{3}\int\left(2\,e^{\t u}+e^{-2\,\t u}\right)dx$.

\vspace{0.5 true cm}
\begin{center}
{
\unitlength 1mm 
\linethickness{0.4pt}
\ifx\plotpoint\undefined\newsavebox{\plotpoint}\fi 
\begin{picture}(140,92)(20,75)

\put(30,166){\circle{10}}
\put(55,166){\circle{10}}
\put(80,166){\circle{10}}
\put(105,166){\circle{10}}
\put(130,166){\circle{10}}
\put(155,166){\circle{10}}
\put(29,141){\circle{10}}
\put(68,141){\circle{10}}
\put(105,141){\circle{10}}
\put(130,141){\circle{10}}
\put(68,116){\circle{10}}
\put(93,116){\circle{10}}
\put(93,92){\circle{10}}
\put(30,166){\makebox(0,0)[cc]{\ref{tst12m}}}
\put(55,166){\makebox(0,0)[cc]{\ref{tst14m}}}
\put(80,165.75){\makebox(0,0)[cc]{\ref{tst9}}}
\put(105,165.75){\makebox(0,0)[cc]{\ref{tst15m}}}
\put(130,166){\makebox(0,0)[cc]{\ref{tst13m}}}
\put(155,165.75){\makebox(0,0)[cc]{\ref{tst10}}}
\put(29,141){\makebox(0,0)[cc]{\ref{tst6}}}
\put(68,140.75){\makebox(0,0)[cc]{\ref{tst8}}}
\put(68,116){\makebox(0,0)[cc]{\ref{tst5}}}
\put(93,116){\makebox(0,0)[cc]{\ref{tst4}}}
\put(93,92){\makebox(0,0)[cc]{\ref{tst2}}}
\put(93,110.75){\vector(0,-1){13.75}}
\put(43,116.25){\circle{10}}
\put(43,92.25){\circle{10}}
\put(43,116.25){\makebox(0,0)[cc]{\ref{tst3}}}
\put(43,92.25){\makebox(0,0)[cc]{\ref{tst1}}}
\put(43,111.25){\vector(0,-1){14}}
\put(126.5,144.25){\vector(1,-1){.07}}\multiput(108.5,162.5)(.0337078652,-.03417603){534}{\line(0,-1){.03417603}}
\put(130,141){\makebox(0,0)[cc]{\ref{tst7}}}
\put(105,141.25){\makebox(0,0)[cc]{\ref{tst11}}}
\put(129.75,161){\vector(0,-1){15}}
\put(133.5,144){\vector(-1,-1){.07}}\multiput(152.25,161.75)(-.0355787476,-.0336812144){527}{\line(-1,0){.0355787476}}
\put(72.25,143.75){\vector(-3,-2){.07}}\multiput(101,163)(-.0503502627,-.0337127846){571}{\line(-1,0){.0503502627}}
\put(70.5,145.5){\vector(-1,-2){.07}}\multiput(77.5,161.5)(-.033653846,-.076923077){208}{\line(0,-1){.076923077}}
\put(65.25,145.75){\vector(1,-2){.27}}\multiput(57,161.5)(.0336734694,-.0642857143){245}{\line(0,-1){.0642857143}}
\put(72.25,118.75){\vector(-3,-2){.07}}\multiput(101.25,138)(-.0507880911,-.0337127846){571}{\line(-1,0){.0507880911}}
\put(70.5,120.25){\vector(-3,-4){.07}}\multiput(101.75,162.25)(-.03371089536,-.04530744337){927}{\line(0,-1){.04530744337}}
\put(71,119.75){\vector(-3,-4){.07}}\multiput(102.25,161.75)(-.03371089536,-.04530744337){927}{\line(0,-1){.04530744337}}
\put(64.25,119.5){\vector(3,-4){.07}}\multiput(32.5,162)(.03370488323,-.04511677282){942}{\line(0,-1){.04511677282}}
\put(65,120){\vector(3,-4){.07}}\multiput(33.25,162.25)(.03370488323,-.04485138004){942}{\line(0,-1){.04485138004}}
\put(63.5,118.25){\vector(3,-2){.07}}\multiput(32.75,138)(.0524744027,-.0337030717){586}{\line(1,0){.0524744027}}
\put(45.2,96.5){\vector(-1,-2){.07}}\multiput(64.95,137)(-.0337030717,-.069112628){586}{\line(0,-1){.069112628}}
\put(46,96){\vector(-1,-2){.07}}\multiput(65.75,136.75)(-.0337030717,-.0695392491){586}{\line(0,-1){.0695392491}}
\put(46.75,95.25){\vector(-1,-1){.07}}\multiput(64.5,112.5)(-.0346679688,-.0336914063){512}{\line(-1,0){.0346679688}}
\put(89.25,95){\vector(1,-1){.07}}\multiput(71,112.5)(.0351637765,-.0337186898){519}{\line(1,0){.0351637765}}
\put(96.25,95.5){\vector(-3,-4){.07}}\multiput(126.75,137.25)(-.03373893805,-.04618362832){904}{\line(0,-1){.04618362832}}
\put(96.75,95){\vector(-3,-4){.07}}\multiput(127.25,136.75)(-.03373893805,-.04618362832){904}{\line(0,-1){.04618362832}}
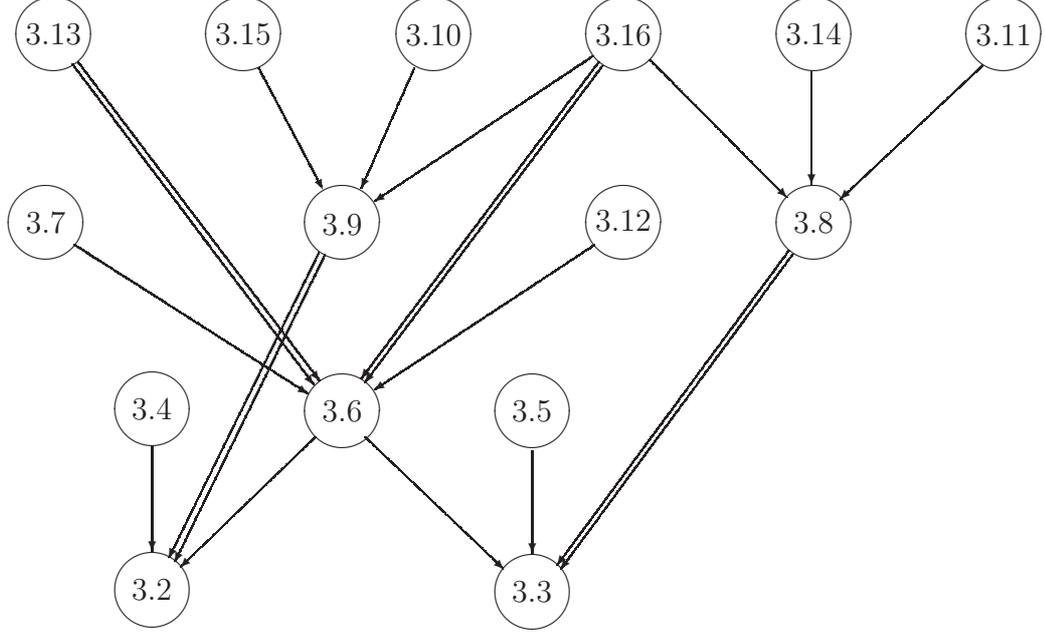
\put(25,78){\small Figure 2. The graph of the substitutions for fifth order $S$-integrable equations}
\end{picture}
}\end{center}%

\section*{Appendix 1. Discrete symmetries of Weierstrass function $\omega$ }

Equations (\ref{tst12m})~--~(\ref{tst15m}) can be written in various ways. We note first that in the paper \cite{MSS} these equations are written in terms of the function $R=f+u_1$ satisfying the equation $2R^3-3u_1R^2+1=0$.

Moreover, there exist transformations preserving the form of  equation (\ref{Waier}). Indeed, consider the functions
$\omega$ and $\t\omega$ satisfying the following equations of the form (\ref{Waier}),
\begin{align}
&\omega'^2=4\,\omega^3 +c, \tag{A1.1} \label{W1} \\
&\tilde\omega'^2=4\,\t\omega^3 +k, \tag{A1.2} \label{W2}
\end{align}
where $c\,k\ne0$. It is easy to check that the  simplest transformations
\begin{align}
&\omega =a\frac{a-\tilde\omega }{a+2\,\tilde\omega },\quad k=c=\frac{1}{2}\,a^3; \tag{$\bsy T_1$}\label{T1} \\
&\omega =\tilde\omega +\frac{k}{\tilde\omega^2},\quad c=-27\,k; \tag{$\bsy T_2$} \label{T2} \\
&\omega =\frac{c+\sqrt{c}\ \tilde\omega'}{2\,\tilde\omega ^2}\,, \quad k=c \tag{$\bsy T_3$}\label{T3}
\end{align}
map a solution of (\ref{W2}) to that of (\ref{W1}).

The transformation $\bsy T_1$ is invertible, and $\t\omega$ is expressed in terms of $\omega$ by the same formula. It is also possible to invert the transformation
$\bsy T_2$, but the problem is reduced to solving a cubic equation. To invert the transformation $\bsy T_3$ one has to solve the Riccatti equation. We note that  formula (\ref{T2}) helps to find the parametrization of  cubic equation (\ref{last}).

The superpositions of the elementary transformations $\bsy T_i$ lead to new transformations preserving the form of   equation (\ref{Waier}). For instance,
\begin{align*}
&\bsy T_1*\bsy T_2:\quad \omega =\frac{3}{2}\,a+\frac{27\,a^2\tilde\omega ^2}{2(2\tilde\omega +a)(\tilde\omega -a)^2}\,, \quad k=\frac{a^3}{2},\ c=-\frac{27}{2}\,a^3;
\\
&\bsy T_2*\bsy T_2:\quad \omega =\tilde\omega +\frac{k}{\tilde\omega ^2}-\frac{27\,k\,\tilde\omega ^4}{(\tilde\omega ^2+k)^2}\,, \quad c=729\,k;
\\
&\bsy T_3*\bsy T_1:\quad \omega =\frac{a(2\tilde\omega +a)^2-{3}\sqrt{2a^3}\,\tilde\omega'}{4(\tilde\omega -a)^2}\,, \quad c=\frac{a^3}{2}\,;
\\
&\bsy T_3*\bsy T_2:\quad \omega =\frac{c\,\tilde\omega ^4+\sqrt{c}\,\tilde\omega (\tilde\omega ^3-2\,k)}{2(\tilde\omega ^3+k)^2}\,, \quad c=-27\,k.
\end{align*}
Moreover, $\bsy T_1*\bsy T_1$ is the identity transformation, and $\bsy T_3*\bsy T_3$ differs from $\bsy T_3$ just by the sign of the root $\sqrt{c}$. Thus, the, equations
(\ref{tst12m})~--~(\ref{tst15m}) can be written in an infinite number of ways different from the first impression.

\section*{Appendix 2. Explicit integration of functions depending on $u_1$ and $f$}

To check the integrability conditions of  equations (\ref{tst12m})~--~(\ref{tst15m}) we need the table of the integrals of rational expression $R(u_1,f)$,
where the function $f$ is defined by   equation (\ref{alg1}). These integrals can be found by the rational parametrization
\begin{equation}\label{param}
f=\frac{w^3-1}{3\,w^2},\qquad u_1=\frac{2\,w^3+1}{3\,w^2}\, \tag{A2.1}
\end{equation}
of  curve (\ref{alg1}). The parametrization allows one to convert the integral of an irrational function $R(u_1,f)$ of the variable $u_1$ into that of a rational function of the variable $w$,
$$
\int R(u_1,f)\,du_1=\int R\left(\frac{2\,w^3+1}{3\,w^2},\frac{w^3-1}{3\,w^2}\right) \left(\frac{2\,w^3+1}{3\,w^2}\right)'dw.
$$
The answer can be written in terms of original variables $u_1$ and $f$ by the formula $w=f+u_1$ implied by (\ref{param}).

To check the integrability conditions we have made use of the integrals
$$
\int u_1^n\,f^m\,du_1,\quad n=0,1,2;\ \ \ -5\le m \le 11.
$$
For instance,
\begin{align*}
&\int f\,du_1=\frac{1}{2}u_1^2-f^2,\qquad \int u_1^2f\,du_1=\frac{1}{20}\big(8f^4+14f^3u_1+f^2u_1^2-u_1\big), \\
&\int u_1f\,du_1=\frac{1}{9}\big(2f^3+f^2u_1+2fu_1^2-2\ln(f+u_1)\big),\\
&\int \frac{du_1}{f}=2\ln(f+u_1),\qquad \int \frac{u_1\,du_1}{f}=2f+u_1, \qquad \int \frac{u_1\,du_1}{f^2}=2\ln(f+u_1)+2\ln f.
\end{align*}
To calculate iterated integrals, one should add to the table also the integrals of logarithms, for instance,
$$
\int \ln(f+u_1)\,du_1=u_1\ln(f+u_1)-\frac{u_1}{2}-f.
$$
In the proof of Theorem 2 we have used around two tens of such
formulas.

To check any of given formulas, it is sufficient to differentiate it, exclude $f'=\dfrac{u_1-f}{2\,f}$, and lower the degree of $u_1$, if needed, by the identities
$$
u_1^3=1+3u_1f^2+2f^3,\quad u_1^4=u_1(1+3u_1f^2+2f^3), \dots
$$
implied by (\ref{alg1}).

\section*{Appendix 3. On recurrent formulas for canonical densities}

Here we discuss the way of obtaining recurrent formulas like (\ref{rekkur_sc}) and (\ref{rec}). The original idea of this method is contained in the work \cite{ZS}, where a simple method for deducing recurrent formulas for the conservation laws of Lax equations was suggested. In the work \cite{CLL} this approach was applied for the linearization of evolution equations and systems.

For the sake of completeness of the content, we first describe briefly the essence of Zakharov-Shabat method.

Suppose  equation (\ref{scalar}) has a Lax representation,
$$
\frac{d L}{dt}=[A,\,L]\Longleftrightarrow u_t=u_n+F(x,u,u_1,\dots,u_{n-1}),
$$
where by square brackets we denote the commutator of linear operators. For simplicity we assume that $A=A(\p_x,\mu,u)$ and $L=L(\p_x,\mu,u)$ are scalar differential operators independent of $\p_t$, $\mu$ is a spectral parameter, $u$ is a solution to equation (\ref{scalar}).

The Lax representation ensures the compatibility of the linear system
\begin{equation}\label{p1}
L\psi =0,\qquad \psi _t=A\psi . \tag{A3.1}
\end{equation}
We introduce notations for the logarithmic derivatives of the function $\psi$,
$$
(\ln \psi)_x=R,\qquad (\ln \psi)_t=T.
$$
It is obvious that the functions $R$ and $T$ are related by the identity
\begin{equation}\label{p2}
R_t=T_x , \tag{A3.2}
\end{equation}
and $R\,dx+T\,dt=d\ln\psi $. This is why up to a multiplicative constant we have
\begin{equation}\label{log}
\psi =\exp\left(\int R\,dx+T\,dt\right), \tag{A3.3}
\end{equation}
where the integral in the exponent is a curvilinear integral with variable upper limit $(x,t)$.

Since $\psi _x=\psi R,\ \psi _t=\psi T$, the operator formulas
$$
\psi ^{-1}\left(\frac{d}{dx}\right)^n\psi =\left(\frac{d}{dx}+R\right)^n,\qquad \psi ^{-1}\left(\frac{d}{dt}\right)^n\psi =\left(\frac{d}{dt}+T\right)^n, \ \ n=0,1,2,\dots
$$
hold true. By these formulas we have
$$\psi ^{-1}L(\p_x,\mu,u)\psi =L(\p_x+R,\mu,u), \quad \psi ^{-1}A(\p_x,\mu,u)\psi =A(\p_x+R,\mu,u).$$
Hence, equation (\ref{p1}) can be rewritten in terms of the functions $R$ and $T$,
\begin{align}
&L(\p_x+R,\mu,u) (1)=0, \tag{A3.4} \label{p3}
\\
&T=A(\p_x+R,\mu,u) (1). \tag{A3.5} \label{p4}
\end{align}
These two equations are nonlinear in $R$. Their solutions are often sought as Laurent series in the parameter $\mu$. Due to (\ref{p2}), the coefficients of these series are the densities of the conservation laws.

\examp For Korteweg-de Vries equation $u_t=u_{xxx}-6uu_x$ the associated linear system can be written as
\begin{align}
&\psi _{xx}-u\psi -\mu^2\psi =0, \tag{A3.6} \label{lax1}\\
&\psi_t=4\psi_{xxx}-6u\psi_x-3u_x\psi . \tag{A3.7} \label{lax2}
\end{align}
Formulas (\ref{p3}),(\ref{p4}) lead us to the equations for $R$ and $T$,
\begin{align}
&R_x+R^2-u-\mu^2=0, \tag{A3.8} \label{ric}\\
&T=4(\p_x+R)^2(R) -6u\,R-3u_x. \tag{A3.9} \label{tmp}
\end{align}
Equation (\ref{tmp}) can be simplified by (\ref{ric}) that yields
\begin{equation}
T=(4\mu^2-2u)R+u_x. \tag{A3.10} \label{tmp1}
\end{equation}

If we substitute the series
\begin{equation}\label{exp}
R=\mu+\sum_{n=0}^{\infty } \rho_n \mu^{-n} \tag{A3.11}
\end{equation}
into equation (\ref{ric}) and equate the coefficients at the equal powers of $\mu$ to zero, we obtain the recurrent formula
\begin{equation}\label{den}
\rho_{n+1}=\frac{1}{2}\left(u\delta_{n0}-\sum_{i=1}^{n-1} \rho_i\rho_{n-i}- \frac{d}{dx}\rho_n \right),\ \ n=0,1,2,\dots,\tag{A3.12}
\end{equation}
where $\delta_{n0}$ is the Kronecker delta. We note that the scale transformation $\rho_n\to\rho_n(-2)^{-n}$ reduces the formula to the form provided in the monograph \cite{Z}. Let us write down first elements of the sequence $\rho_n$,
$$
\rho_0=0,\ \ \ \rho_1=\frac{1}{2} u,\ \ \ \rho_2=-\frac{1}{4}u_1,\ \ \ \rho_3=\frac{1}{8}(u_2-u^2).
$$

Next, we substitute  series (\ref{exp}) into  equation (\ref{tmp1}) to obtain the expansion
\begin{equation}\label{toki}
T=4\mu^3+\sum_{n=1}^{\infty} \theta_n\mu^{-n},\ \tag{A3.13}
\end{equation}
where
\begin{equation}\label{rtoki}
\ \theta_n=4\rho_{n+2}-2u\rho_n,\ \ n>0. \tag{A3.14}
\end{equation}
Since the parameter $\mu$ is arbitrary,  formula (\ref{p2}) defines an infinite sequence of the conservation laws
\begin{equation}\label{canonlaw}
\frac{d}{dt}\rho_n=\frac{d}{dx} \theta_n,\ \ n=1,2,\dots , \tag{A3.15}
\end{equation}

To obtain the canonical densities $\rho_n$, it is sufficient to have (\ref{p2}) and one of  equations (\ref{ric}) or (\ref{tmp}).

If we use   equation (\ref{ric}), we arrive again at  recurrent formula (\ref{den}), but we lose  formula (\ref{rtoki}). The fluxes $\theta_n$ associated with the densities $ \rho_n$ can be found from (\ref{canonlaw}) by inverting the total derivative operator $\ds\frac{d}{dx}$ (the algorithm was discussed in Remark 4 on page \pageref{D-1}).

For further reasoning it is more important to understand how to get the canonical densities from  equations (\ref{tmp}) and (\ref{p2}). Since   equation (\ref{tmp}) does not involve the parameter, we introduce the parameter apriori and we can choose the structure of the expansion for $R$ as we wish. If, for instance, we assume that $R$ is the Taylor series
$$
R=\sum_{n=0}^{\infty } \rho_n \mu^n,
$$
where $\mu$ is the parameter, then
$$
T=\sum_{n=0}^{\infty } \theta_n \mu^n,
$$
where the coefficients $\theta _n$ are determined by  equation (\ref{tmp}). It is easy to check that
$$
\theta_n=4\sum_{0}^{n} \rho_i \rho_j\rho_k-3\,u_1\delta_{n0}-6\,u\rho_n+4\frac{d^2}{dx^2}\rho_n+6\frac{d}{dx}\sum_{0}^{n} \rho_i \rho_j ,
$$
where we have used the notations for the sums introduced on the page \pageref{summ}. Since in the left and right hand sides of this formula the unknown functions $\theta_n$ and $\rho_n$ appear simultaneously, it does not help for calculating the conservation laws.

The situation changes if we postulate the expansion of the function $R$ as the Laurent series
 \begin{equation*}\label{gexpR}
R=\mu^{-1}+\sum_{n=0}^{\infty } \rho_n \mu^n. \tag{A3.16}
\end{equation*}
In this case by   equation (\ref{tmp}) we obtain the expansion for $T$
\begin{equation*}\label{gexpT}
T=4\mu^{-3}+\theta_{-2}\mu^{-2}+\theta_{-1}\mu^{-1}+\sum_{n=0}^{\infty} \theta_n\mu^{n}, \tag{A3.17}
\end{equation*}
and the recurrent formula
\begin{align}\label{rec_z}
\rho_{n+2}&=\frac12\,u\rho_n+\frac14 u_1\delta_{n,0}-\sum_{0}^{n+1} \rho_i\rho_j+\frac{1}{12}\theta_n-\frac{1}{3}\sum_{0}^{n} \rho_i\rho_j\rho_k+\frac{1}{12}\theta_{-2}\delta_{n,-2}-\nonumber\\
&-\frac{d}{dx}\left(\rho_{n+1}+\frac12\sum_{0}^{n}\rho_i\rho_j+\frac{1}{3}\frac{d}{dx}\rho_n \right) +\frac{1}{12}\left(6\,u+\theta _{-1}\right)\delta_{n,-1}, \tag{A3.18}
\end{align}
where $n=-2,-1,0,\dots$ Let us consider the corresponding series of   conservation laws (\ref{canonlaw}), where $ n=-2,-1,0,1,2,\dots $. If the conservation laws with the indices $i\le n+1$ are known, we find $\rho_{n+2}$ by (\ref{rec_z}), and then $\theta_{n+2}$ by (\ref{canonlaw}), and so forth. While finding $\theta_{n+2}$, we have to invert the operator $\ds\frac{d}{dx}$. Under the assumption that the densities and fluxes of conservation laws (\ref{canonlaw}) are explicitly independent of $t$, this procedure is absolutely algorithmic (see page \pageref{D-1}). At that, the function $\theta_{n+2}$ is determined uniquely up to an integration constant.

The beginning of this recurrence is as follows. According to (\ref{gexpR}), we have $\rho _{-2}=\rho _{-1}=0$, and this is why by (\ref{canonlaw}) we obtain that the corresponding fluxes are constant, $\theta_{-1}=12c_{-1},\ \theta_{-2}=12c_{-2}$. Then we find $\rho_0=c_{-2}$. Next two densities read as
$$
 \rho_1=\frac12 u+c_{-1},\ \ \rho_2=\frac{1}{12} \theta_0-\frac{c_{-2}}{2}u-\frac{1}{4}u_1-\frac{c_{-2}^3}{3}-2c_{-1}c_{-2}.
$$
To determine $ \theta_0$, we again have to employ  equation (\ref{canonlaw}) as $n=0$ that implies $ \theta_0=c_0$.

It is important to note that the constants $c_i$ appearing in finding the fluxes $\theta_{i}$ are not essential since they can be eliminated by the change of the parameter $\mu$,
\begin{equation}\label{mumu}
\mu \to \mu+\sum_{i=2}^{\infty} k_i \mu^i. \tag{A3.20}
\end{equation}

Consider now an arbitrary evolution equation with one spatial variable
\begin{equation}\label{eqg}
u_t=K(x,u,u_x,\dots, u_n), \qquad n>1. \tag{A3.20}
\end{equation}
In   case (\ref{scalar}) we have $K=u_n+F(x,u,u_x,\dots, u_{n-1})$. Denote by $K_*$ the Fr\'echet derivative of the function $K$,
$$
K_*=\sum_{i=0}^{n} \frac{\p K}{\p u_i}\frac{d^i}{dx^i}.
$$
The formal series
$$
L=\sum_{k=-\infty }^{1} f_k\frac{d^k}{dx^k},
$$
whose coefficients depend on $x,u,u_x,\dots$ that satisfies the equation
\begin{equation}\label{fsym}
L_t=\left[ K_*,\,L\right], \tag{A3.21}
\end{equation}
is called a formal symmetry (formal recurrence operator) of  equation (\ref{eqg}). It is known that the equation possessing generalized symmetries or
conservation laws has a formal symmetry \cite{ibshab,ss,MSS}.

Equation (\ref{fsym}) ensures the compatibility of the following pair of linear equations,
\begin{equation}
L\psi=\lambda \psi ,\qquad \psi_t=K_*\psi , \tag{A3.23} \label{lax3}
\end{equation}
where $\lambda$ is a spectral parameter. To this system one can apply the procedure of obtaining canonical densities described above. Since the operator $L$ is not known apriori, we employ equation (\ref{p4}),
\begin{equation}
T=\sum_{i=0}^{n} \frac{\p K}{\p u_i}\left(\frac{d}{dx}+R\right)^i (1) . \tag{A3.25} \label{cansym}
\end{equation}
Let
\begin{equation}
R=\rho _{-1}\mu^{-1}+\sum_{k=0}^{\infty } \rho_k \mu^k, \label{seq1} \tag{A3.27}
\end{equation}
then
\begin{equation}
 T=\mu^{-n}+\sum_{i=1}^{n-1} \theta_{-i}\mu^{-i}+\sum_{k=0}^{\infty } \theta_k\mu^k. \label{seq2} \tag{A3.28}
\end{equation}
Indeed, the minimal degree of $\mu $ in the right hand side of identity (\ref{cansym}) is contained in the term
$$
 \frac{\p K}{\p u_n}\left(\frac{d}{dx}+R\right)^n (1)= \frac{\p K}{\p u_n}R^n+\dots=\frac{\p K}{\p u_n}(\rho_{-1}) ^n\mu ^{-n}+\dots,
$$
and hence the series for $T$ should begin with the term $\theta_{-n}\mu ^{-n}$. Since $n>1$, then $\rho_{-n}=0$ and $\theta_{-n}$=const$\ne0$. By scaling the parameter $\mu$ we convert $\theta_{-n}$ into one and obtain (\ref{seq2}).

Substituting   expansions (\ref{seq1}), (\ref{seq2}) into (\ref{cansym}) and equating the terms at $\mu^{-n}$ in the
equation (\ref{cansym}), we obtain the first density
$$
 \rho _{-1}=\left( \frac{\p K}{\p u_n}\right)^{-1/n}.
$$
The formulas for several next canonical densities can be found in \cite{MSS}.

Let us consider now equation (\ref{kdv}) and adduce the deduction of  recurrent formula (\ref{rekkur_sc}) for canonical densities following of the above scheme.

1st step. We write  linearization for  equation (\ref{kdv}),
$$
\left[\left(\frac{d}{dx}\right)^3+\frac{\p F}{\p u_2}\left(\frac{d}{dx}\right)^2+\frac{\p F}{\p u_1}\frac{d}{dx}+\frac{\p F}{\p u}-\frac{d}{dt}\right]\psi =0.
$$

2nd step. By the substitutions
$$
\psi =\exp\left(\int R\,dx+T\,dt\right),\ \ \text{where \ } R_t=T_x,
$$
we obtain the equation with ``extended derivatives'',
$$
\left[\left(\frac{d}{dx}+R\right)^3+\frac{\p F}{\p u_2}\left(\frac{d}{dx}+R\right)^2+\frac{\p F}{\p u_1}\left(\frac{d}{dx}+R\right)+\frac{\p F}{\p u}-\left(\frac{d}{dt}+T\right)\right](1) =0,
$$
which is equivalent to the relation
\begin{align}\label{Ricc}
T=\left(\frac{d^2}{dx^2}+\frac{d}{dx}R+R\frac{d}{dx} +R^2\right)(R)+\frac{\p F}{\p u_2}\left(\frac{d}{dx}+R\right)(R) +\frac{\p F}{\p u_1}R+\frac{\p F}{\p u}. \tag{A3.29}
\end{align}

3rd step. We choose an appropriate expansion for $R$. The simplest choice is to let
\begin{equation}\label{series1}
R=\mu^{-1}+\sum_{n=0}^{\infty } \rho_n\mu^n. \tag{A3.30}
\end{equation}

\rem Our several attempts to find the expansions with the poles of higher order gave nothing new. If, for instance, we assume for  equation (\ref{kdv})
$R=\ds\mu^{-2}+\sum_{n=-1}^{\infty } \rho_n\mu^n$, then after checking several conditions (\ref{canonlaw}) we obtain $\rho_{2n+1}=0,\ \forall n$. It is equivalent to that
$R$ is expanded w.r.t. the parameter $\xi=\mu^2$. Similar results were obtained for some other equations and systems as well (see \cite{IP})

Having chosen  expansion (\ref{series1}), we should accept
\begin{equation}\label{series2}
T=\mu^{-3}+\theta_{-2}\mu^{-2}+\theta_{-1}\mu^{-1}+\sum_{n=0}^{\infty } \theta_{n}\mu^n, \tag{A3.31}
\end{equation}
in order to cancel the terms with $\mu^{-3}$ in  equation (\ref{Ricc}).

For   expansion (\ref{series1}) we have $\rho_{-1}=1, \rho_{-2}=0$, which implies that $\theta_{-2}$ and $\theta_{-1}$ are constants. Since additive integration constants in the fluxes are eliminated by the transformation of  parameter (\ref{mumu}), we let $\theta_{-2}=\theta_{-1}=0$.

Now, as one can easily make sure, substituting   expansions (\ref{series1}) and (\ref{series2}) into equation (\ref{Ricc}), we arrive at  formula (\ref{rekkur_sc}) with indicated there $\rho_0$ and $\rho _1$.


\label{lastpage}

\end{document}